\documentclass[fleqn,usenatbib]{mnras}

\usepackage{newtxtext,newtxmath}

\usepackage[T1]{fontenc}

\DeclareRobustCommand{\VAN}[3]{#2}
\let\VANthebibliography\thebibliography
\def\thebibliography{\DeclareRobustCommand{\VAN}[3]{##3}\VANthebibliography}

\usepackage{graphicx}	
\usepackage{amsmath}	
\usepackage{threeparttable} 

\title[Modelling Stellar Irradiances II]{Modelling Stellar Irradiances II: Correlations of solar irradiances with proxies of activity along a cycle}

\author[E. Deliporanidou et. al.]{
E. Deliporanidou$^{1}$\thanks{E-mail: ed650@cam.ac.uk},
G. Del Zanna$^{1,2}$, 
T.N. Woods$^{3}$, D. Woodraska$^{3}$
\\
$^{1}$DAMTP, Centre for Mathematical Sciences, University of Cambridge, Wilberforce Road, Cambridge, CB3 0WA, UK\\
$^{2}$ School of Physics \& Astronomy, University of Leicester, Leicester  LE1 7RH, UK \\
$^{3}$ Laboratory for Atmospheric and Space Physics, University of Colorado, 1234 Innovation Dr., Boulder, CO 80303, USA
}
\date{Accepted XXX. Received YYY; in original form ZZZ}

\pubyear{2024}

\begin{document}
\label{firstpage}
\pagerange{\pageref{firstpage}--\pageref{lastpage}}
\maketitle

\begin{abstract}
There is a pressing need to model X-ray Ultra-Violet (XUV: 1-30 nm) stellar irradiances, given the scarcity of current measurements. One of the measurable effects of a stellar cycle is the significant (more than one order of magnitude) variation in XUV irradiance. As a first step in modelling stellar irradiances, we present EUV irradiances in a sample of strong spectral lines formed in different layers and regions of the solar atmosphere, obtained from Solar Dynamics Observatory Extreme Ultraviolet Variability Experiment (SDO EVE). These irradiances span half a solar cycle. We present correlations with several proxies of solar activity, such as the Mg II index, sunspot numbers, and cm radio fluxes. Among these, the sunspot number proves to be the poorest proxy, whereas the Mg II index is a very good proxy for coronal lines (hotter temperature lines). We find a relatively strong linear relationship, which enables us to build a model essential for various applications. Additionally, we compare our results with the previous EUV standard solar irradiances reported by Del Zanna and Andretta 2014, derived from Solar and Heliospheric Observatory Coronal Diagnostic Spectrometer (SOHO CDS) data, as well as historical records from the literature. We have also run a DEM analysis on Quiet Sun (QS) and Active Regions (AR) and list the blends and formation temperatures for the strong lines. Finally we provide a simple routine for deriving the irradiances of the strong lines based on proxy values.
\end{abstract}




\section{Introduction}

This paper is the second in a series where we study XUV stellar irradiances, aiming to improve the atomic data and provide accurate modelling. We continue the research that started from the another series of papers that studied the EUV spectrum of the Sun \citep{delzanna_etal:10_cdscal, delzanna_andretta:2011, delzanna_andretta:2015}, using data from the Solar and Heliospheric Observatory (SOHO) Coronal Diagnostic Spectrometer (CDS) \citep{harrison95} and Extreme ultraviolet Variability Experiment (EVE) \citep{Woods2012_EVE}
onboard the Solar Dynamics Observatory (SDO).

It is well known that both stellar and solar cycles drive magnetic flux emergence and result in large variations in XUV irradiances.  Modelling XUV irradiances is critical for both stars and the Sun for several reasons. Solar XUV irradiances significantly impact the Earth's ionosphere and thermosphere, making them essential inputs for general  global circulation atmospheric models \citep[see, e.g.][]{Solomon_Qian_2005}. This is due to the solar variability that affects long-term climatic changes to	telecommunications, as the Sun drives the density in the thermosphere via direct and indirect heating of this layer \citep{vourlidas_2018}. Knowledge of XUV irradiances is also needed for a variety of 
solar studies, for example, to assess the importance of photo 
excitation and photo ionization in the solar corona.

 The importance of XUV irradiances extends beyond the Sun, as stellar XUV irradiances can have profound effects on the atmospheres of exoplanets, influencing their potential habitability \citep{linsky_host_2019}. However, despite advancements in astrophysical research, there remains a substantial gap in our knowledge of stellar EUV emissions. Most of the available data comes from a limited number of instruments such as the Extreme Ultraviolet Explorer (EUVE), the Space Telescope Imaging Spectrograph (STIS), and the Chandra Low Energy Transmission Grating Spectrometer (LETG), which primarily observed nearby stars. These instruments have provided valuable, yet incomplete, data, leading to a reliance on models for the Sun’s EUV irradiances, which are often based on proxies. 

Accurate measurements of solar EUV irradiances have historically been sparse before SDO/EVE, as they were limited to snapshots from sounding rockets or specific 
wavelengths covered by e.g. SOHO CDS and the Thermosphere Ionosphere Mesosphere Energetics Dynamics (TIMED) Solar EUV Experiment (SEE). The EUV irradiances, although a small fraction of the total solar output, show significant variations over the 11-year solar cycle, which are important for understanding solar forcing on the Earth's atmosphere and potentially its climate.

In light of the lack of complete and continuous measurements of XUV irradiances, our aim is to develop methods and models to predict the long-term  variation in irradiances using proxies of solar activity. We have already improved the modelling of UV stellar irradiances \citep{Deliporanidou_2024} using advanced ionisation equilibrium models \citep{dufresne_chianti_2024}. Our current work focuses on correlating solar XUV irradiances, particularly those from EVE, with various proxies of solar activity, including the Mg II index, sunspot numbers, and cm radio fluxes. 
This study is the first step needed to develop a predictive model
for the XUV irradiance of the Sun.

\section{Brief summary of previous observations and models}

\subsection{EUV irradiances}

There is a significant body of literature on historical measurements and models of the solar irradiance in the EUV. Some are described in \cite{delzanna_andretta:2011}, where a comparison table on a few selected spectral lines was provided. It emerged that a large scatter in the irradiances was present, most of which was not due to intrinsic solar variability but rather was due to different radiometric calibrations. It is worth mentioning here that the stronger EUV line, the \ion{He}{ii} 304~\AA, had for about forty years an incorrect calibration of about a factor of two, which was resolved only with the recent CDS and EVE measurements. 

\cite{reeves_parkinson:1970b} produced an extensive list of
irradiances obtained from full-Sun scans with the
Harvard instrument on OSO-IV. It had a low spectral resolution
(about 2~\AA), so only strong isolated lines
were free from significant blending.  The calibration 
procedure, described in \cite{reeves_parkinson:1970a}, mostly
relied on comparing irradiances to earlier measurements, which provided the reference EUV irradiances in the 1960s. The absolute calibration uncertainty was stated to be about a factor of two, which appears to be an over-estimate, by comparisons with later results.

The EUV measurements that have consistently indicated an excellent
agreement with the most recent irradiance measurements (in absolute terms)
and with the values predicted by theory (in relative terms) 
are those produced by Heroux and collaborators, see \cite{malinovsky_heroux:1973,heroux_etal:1974,heroux_higgins:1977,heroux_hinteregger:1978}.
They have a stated accuracy of about 30\% in absolute terms, but e.g.
the \cite{malinovsky_heroux:1973} ones are 
accurate to within  10\% for lines close in wavelength, as 
confirmed by atomic data benchmark studies performed in a series of 
papers by Del Zanna. Such an accuracy has not been achieved since then.

During the 1996 solar minimum, the  Solar \& Heliospheric Observatory (SOHO)  
Solar Ultraviolet Measurements of Emitted Radiation
(SUMER, \citealt{wilhelm95})
made some full-Sun scans \citep{wilhelm_etal:98} [W98],
and provided irradiance measurements in a few strong lines. 
SUMER was well calibrated on the ground and the early data
have a relative uncertainty of about 15\%. The absolute values,
when compared to various measurements, indicated an overall
uncertainty of about 20\%. 
However, the scans suffered from various limitations: nearly all 
radiances were obtained by summing spectral regions on-board, and 
corrections for line blends and continuum contributions had to be estimated,
as well as corrections for detector gain / dead time. 
Such corrections amounted to up to 40\% and carried significant 
uncertainties. On the positive side, these observations allowed for 
limb-brightening estimates and  covered the whole Sun. 
In a later study \cite{dammasch_etal:1999} [D99] provided a set 
of irradiances, obtained in 1997 by sampling the Sun with a small number of slit positions, essentially estimating the limb-brightening but with very poor sampling.  Line radiances were obtained by summing spectral regions
and subtracting a background.

The  Coronal Diagnostics Spectrometer  \citep[CDS;][]{Harrison-etal:95} 
onboard SOHO carried out regular full-Sun scans between 1998 and 2014
in the two wavelength bands of the \textit{Normal Incidence Spectrograph} (NIS),
308--379 and 513--633~\AA\  in first order and 260--315~\AA\ in second order.
The entire solar disk was observed by scanning the 4\arcsec\ slit in 24\arcsec\ steps.
The line radiances were obtained with fitting routines, and irradiances 
obtained with a complex procedure of spatial interpolation, described in 
\citep{delzanna_andretta:2015} and references therein.
The spectral resolution in first order was  about $\sim 0.6$~\AA\ FWHM. 
It took  over 10 years to finally achieve a mission-long in-flight calibration of the NIS with
an accuracy of about 20\%, as described in a series of papers,
the main ones being \cite{delzanna01_cdscal,delzanna02_issi-ADS,delzanna_etal:10_cdscal,delzanna_andretta:2015}. 
It was based on an analysis of the degradation in the cores of the strong lines,
the assumption that the basal quiet Sun chromosphere and transition region
emission does not vary, the use of reliable theoretical intensity ratios of lines
from the same ion to establish the relative calibration for the three
spectral ranges, and finally use a 1997 sounding rocket
\citep{brekke_etal:00} to fix the
absolute calibration of the strongest first order line, from
\ion{He}{i} 584~\AA. The latter carried a NASA/Laboratory for Atmospheric and Space Physics (LASP)
EUV Grating Spectrograph (EGS)  that was calibrated on the ground
against synchrotron emission. Several checks against other simultaneous observations,
including sounding rockets, have been carried out. 
The NIS only degraded smoothly by a factor of two over its
entire mission, hence has been the best EUV spectrometer to date. 

Several sounding rockets carrying EGS instruments have been flown since the 1990s,
see e.g. \cite{1990JGR....95.6227W,woods_etal:1998,2004SPIE.5538...31C}.
They were all calibrated on the ground and obtained very useful datasets, although 
with a lower resolution (about 4~\AA) than more recent ones.

The NASA  Thermosphere Ionosphere Mesosphere Energetics Dynamics (TIMED)
 Solar EUV Experiment (SEE) EGS  
\citep{woods_etal:2005}  has provided irradiance measurements 
since 2002 of some of the strongest EUV lines, but with a lower
spectral resolution (4~\AA\ FWHM) and poorer radiometric calibration
(about 50\% in the 270--450~\AA\ range),
as also suggested by comparison with other measurements \citep{delzanna_andretta:2015}.

\subsection{EVE}

The Solar Dynamics Observatory (SDO) 
 Extreme ultraviolet Variability Experiment (EVE)
 \citep{woods_etal:2012} has provided EUV irradiances since May 2010
 with its Multiple EUV Grating Spectrographs (MEGS) at about 1~\AA\ spectral resolution (FWHM). 
 The MEGS A,  a grazing incidence spectrograph for the 50--380~\AA\ range, operated
 until 2014. The MEGS B,
 a double-pass normal incidence spectrograph for the 330--1050~\AA\ range is still
 operational but does not observe the Sun continuously, to limit its
 degradation.  The EVE channels suffered large in-flight degradations, which are
 partially monitored with the use of redundant filters, partially
 with calibration sounding rockets flying with an identical instrument every few
 years. 
 The first flight of the EVE calibration rocket (PEVE) was flown on  2008 April 14
 during a deep solar minimum  \citep{woods_etal:2009}. 
 The calibration and flight instruments were radiometrically calibrated 
 on the ground before and after the flight
 \citep{chamberlin_etal_07,hock_etal_09,hock_etal:2012} with an accuracy
 in principle better than 20\%.  However, comparisons between an earlier (version 4) flight
 EVE calibration and the PEVE results showed large (more than 20\%) discrepancies for several
 of the strongest lines \citep{delzanna_andretta:2015}. The flight
 EVE data, which have been calibrated with several subsequent PEVE flights,
 generally agreed better with the CDS and historical results.
We use the PEVE irradiances obtained by \cite{delzanna:2019}.

Details of the in-flight EVE calibration will be reported in a separate paper. 
Here, we note that the present analysis has flagged a few anomalies which are to be associated 
with instrumental issues.  The main anomaly occurred on 2014 May 26 with a capacitor short in the MEGS-A CCD camera electronics that prevents valid spectral images from MEGS-A. Power cycling the MEGS-A CCD camera has not been successful to clear the failed capacitor, so the MEGS-B CCD sensor heater setpoint was changed on 2014 July 31 to account for lower power with the MEGS-A camera being powered off.
Other major events are CCD bakeouts, which were carried out in 2010 on 
May 13-14, May 18, June 16-18, and September 23-27, followed by two further bakeouts in 2012,
on March 12-14 and 20-22. Finally, on 2016 August 2 EVE was powered off for safe mode. 
Some of these events have affected the irradiances and the wavelengths of several spectral lines,
the most notable on the strong \ion{Ne}{VII} 465 \AA\ line, as shown below. 
Also, we note that the sounding rocket flights were used to correct for the relative 
degradation of the instrument, but that the absolute scale was fixed using the 2010 
flight.

\subsection{Quiet Sun EUV radiances and conversion to irradiances}

\begin{table*}
\caption{Observed  EUV quiet Sun radiances R and derived irradiances I. 
\label{tab:strong_lines} } 
\begin{center}
\begin{tabular}{@{}lcrrrrlrl@{}}
\hline\hline \noalign{\smallskip}
Ion  & $\lambda$ (\AA) & R OSO-4 & R Skylab & R SUMER & R SUMER  & f & I Skylab \\
&   & DR71 &  VR78  & W05  & W98  &   &  (phot) \\
\noalign{\smallskip}\hline\noalign{\smallskip}
\ion{He}{i} & 584.3  & 325 & 545  & 484 & 445 & 1.0  & 10.9  \\
\ion{O}{v} &  629.7 & 302 & 335  & 416 & 271 & 1.45 & 10.4 \\
 \ion{O}{iii} (sbl) &  702-704  & 67.3 & 57 &  80 & -  & 1.45 & 2.0 \\
\ion{Ne}{viii}  & 770.4 & 32.2  & 54 & 73  & 79  & 1.50 & 2.1 \\
\ion{O}{iv} & 787.7  & 47  & 44 & 58  &  53 & 1.45 & 1.7  \\
\ion{O}{iv} (sbl)&  788--790 & 114  & 127  & 167 &  -  & 1.45 & 4.9 \\
 \ion{O}{ii}, \ion{O}{iii} & 833-835 & 103 & 148 & 159 & 203.5 & 1.45 & 6.1 \\
\ion{C}{iii} &  977.0 & 900 & 963  & 1206& 702 & 1.50 & 48 \\
\ion{H}{i} & 1025.7  & 673 & 747  & 845 & 766 & 1.16 & 31 \\
\ion{O}{vi}  & 1031.9  & 266  & 305  & 358 & 248 & 1.50 & 16 \\
\noalign{\smallskip}\hline 
\end{tabular}
\begin{tablenotes}
    \item[] {The first column indicates the ion and if the 
    line is a self-blend (sbl), while the second column indicates the 
    wavelength (or range) of the transition. 
    The following columns indicate R, the averaged quiet Sun radiances 
    (erg cm$^{-2}$ s$^{-1}$ sr$^{-1}$) as measured by various instruments. 
    f is the correction factor to obtain the mean disk radiances from the centre-disk radiances. The last column lists the irradiances I 
    (10$^8$ photons cm$^{-2}$ s$^{-1}$) corresponding to the Skylab ATM irradiances. }
\end{tablenotes}  
\end{center}
\end{table*}

When comparing the irradiances from different instruments
we found several inconsistencies, mostly between EVE and historical 
measurements. We have therefore resorted to estimate 
irradiances for solar minimum conditions obtained from 
accurate quiet Sun radiances near disk centre and a limb-brightening correction, to see if they are consistent or not with the previous 
records of irradiances. We describe here a selection. 

The Harvard instrument on OSO-IV was the first to provide
spatially resolved spectra of the Sun. A quiet Sun spectrum
was obtained by averaging 27 observations during 1967 October 26-27
(a period of relatively high solar activity) 
over an area of 1\arcmin\ and published by \cite{dupree_reeves:1971}.
The calibration procedure is described in 
\cite{reeves_parkinson:1970a} and \cite{reeves_parkinson:1970b}, as previously mentioned.

EUV radiances of various solar regions were obtained by \cite{vernazza_reeves:1978}
[VR78]  with the Harvard  Skylab ATM  instrument, which had a
spectral resolution of about 1.6~\AA. They were the standard reference for a long time. 
The  quiet Sun values were obtained by averaging near Sun
centre 16 days of observations, during which one of the calibration sounding
rockets (Calroc) was flown. Calroc observed the same region and was
calibrated on the ground before and after, reaching an uncertainty of
about 35\% in the EUV, a major milestone for the time. The degradation over the
9 months of operation was monitored measuring count rates in the
quiet Sun. Details of the calibration procedures are described in \cite{timothy_etal:1975} and \cite{reeves_etal:1977b}.

A few papers have  provided SUMER averaged radiances of the quiet Sun during
nearly minimum conditions. There is a significant scatter in the values, as
discussed in \cite{dufresne_etal:2023}. This can only be due to solar variability
and the fact that the published observations were on small spatial regions.
Furthermore, it took a relatively long time for SUMER to acquire EUV
reference spectra. 
We consider here the \cite{warren:2005} [W05],
obtained by averaging over a small disk-centre region,
and the \cite{wilhelm_etal:1998a} ones, obtained from a single-slit exposure
near Sun centre.

We have selected a few among the strongest lines, listed
in Table~\ref{tab:strong_lines}, produced from low-temperature
ions, as the conversion to quiet Sun fluxes is more accurate and are less variable
with solar activity.
The selected radiances show a scatter of values. 
To obtain the mean disk radiances from the centre-disk radiances 
we have adopted the correction factors f listed in Table~\ref{tab:strong_lines}. 
They are mostly from the analysis of the limb brightening
from full-Sun CDS radiances during solar minimum described in
\cite{andretta_delzanna:2014}, with a
few values from \cite{wilhelm_etal:1998a}, noting that
there is general agreement between these two studies. 
As we regard the Skylab results more reliable, given the large
sample, we list in Table~\ref{tab:strong_lines} the irradiances
obtained from the  ATM  radiances, by multiplying by f $\times\, \pi\, \times\, \left(\frac{R_{\odot}}{d_{\odot}} \right)^2$.

\subsection{Solar cycle variability}

 There has been an extensive literature on solar cycle
 variability in irradiances. Even the total solar irradiance
 (mostly due to visible light), which can
 in principle be measured with great accuracy, showed a huge
 scatter of values.  For a long time, it seemed that the
 last few solar minima differed, which was at odds
 with the accurate measurements carried out by Bill Livingston
 over several cycles of photospheric and chromospheric line widths
 on the quiet Sun, which did not show any significant variations
 \citep[cf.][]{livingston_etal:07,livingston_etal:2010}.
 However, recent revisions confirm that the basal solar minimum
 irradiance does not change 
\citep[see, e.g. discussions in][]{2017GeoRL..44.1196D}.

 Historical records clearly showed that irradiances of
 higher-temperature (2--3 MK) lines, produced by active regions,  
 varied a lot more than  other lower-temperature lines, with the
 exception of chromospheric / transition region
 lines from \ion{He}{i}, \ion{He}{ii}, and \ion{H}{i} 
 \citep[see, e.g.][]{oster:1983,woods_etal:1998}.
 It was only with the mission-long consistent set of CDS measurements
 which covered the large solar maximum in 2002 and the following deep minimum
 \citep{delzanna_andretta:2011,delzanna_andretta:2015} that the variations were quantified. 
It was clear that all the chromospheric / transition region 
lines formed between 20,000 and 600,000 K have a small variation of at most 20-30\%.
The larger variations in the H and He lines are most
likely due to the variability of the  photoionization by
coronal radiation, which affects the intensities of the lines
via subsequent complex recombination processes.

\subsection{Proxies}

Proxies of solar activity are indirect indicators used to estimate the Sun's activity levels. There are several proxies of solar activity that
could be considered, see e.g. the references in the 
review by \cite{vourlidas_2018}. We used the solar sunspot number, that provides a measure of solar activity based on visible sunspots, the F10.7 cm and F30 cm radio fluxes, that are both radio emissions from the Sun, and the Mg II index, which is derived from UV spectral lines, and indicates chromospheric activity. A useful resource is the LASP Interactive Solar Irradiance Datacenter (LISIRD)
website: \url{https://lasp.colorado.edu/lisird/}, from where we got most of our proxies' data. 

The daily  measurements of the solar disc emission at 2800~MHz (10.7cm),
scaled to 1 AU,  have been available since 1949. 
They were made by 
the National Research Council of Canada
(at the Algonquin Radio Observatory, near Ottawa until 1991 and 
at the Dominion Radio Astrophysical 
Observatory, near Penticton, British Columbia). The F10.7 index has been widely used 
as one of the indicators of EUV activity \citep{tapping2013}. While it does not fully encapsulate all sources of EUV irradiance variability \citep{lean_solar_2011}, its record is highly continuous, with only a few weeks of interruptions over its entire data set \citep{vourlidas_2018}. 
The coronal origin of the radio emission was discussed by 
\cite{2015ApJ...808...29S}, where it was pointed  out that this was not an ideal proxy 
for EUV emission. 

Similar measurements at 30cm (and other wavelengths) have been taken (since Nov. 1, 1957) 
and are now available
at \url{https://spaceweather.cls.fr/services/radioflux/}. As we shall show later, the 
30cm emission correlates better with the EUV fluxes.

The Mg II index, is an excellent absolute measurement
of solar activity and is based on the relative core-to-wing ratio of the Mg II: 2796.352 and 2803.530~\AA\ lines, formed in the upper chromosphere. It's available since 1978 and it is derived from the ratio of the sum of the
two spectral lines in the UV spectrum of the Sun, specifically the Mg II h and k lines (ionised magnesium) relative to the sum of the photospheric surrounding blue and red wing regions. A superior high-precision, high-cadence Mg II index is provided by the EXIS EUVS-C instruments on GOES-16, 17, 18, and 19 since early 2017 that are expected to continue through 2034. The noise scatter between EXIS instruments shows the high precision values can be corrected to 0.1\% when accounting for the systematic spacecraft-to-sun distance and doppler effects. (McClintock et al 2025, submitted to JSWSC).
We use the Bremen Mg II composite data, available at
\url{https://www.iup.uni-bremen.de/gome/gomemgii.html}. This index is used to measure solar chromospheric activity and is particularly sensitive to changes in the Sun's UV output, which affects the Earth's upper atmosphere and space weather.

Perhaps the longest historical record of direct solar activity is the
sunspot number, as it goes back to the first measurements by Galileo Galilei
in Italy in the 17th century. The sunspots are observed optically with their number increasing during solar maxima and decreasing during solar minima. 
There is a long history of different types of sunspot numbers and
calibration issues \citep[see, e.g.][]{clette_etal:2015}.
However, it has been noted in the literature that this activity record is not
as good as the previous ones.
We use the AAVSO American Relative Sunspot Number,
see \url{https://www.aavso.org/solar}, although we note that several variants are available.

Obviously, as all the solar atmospheric activity is ultimately
caused by some conversion of magnetic energy, the ultimate proxy
of solar activity is the  magnetic field. Ideally,
one should consider the coronal or the chromospheric magnetic field but 
they are not available. We do have, however,  line-of-sight (LOS) measurements of the
photospheric magnetic field for over 100 years. 
As in any other solar measurements, significant
discrepancies with results obtained by different instruments are present.
As a measure of solar activity, some authors have taken just the
total unsigned flux, others have assumed that the field is radial
everywhere (a strong assumption) and made LOS corrections.
As shown by e.g. \cite{2016usc..confE..28S} with a detailed analysis, the unsigned magnetic flux
correlates extremely well with the F10.7 flux, which we consider here. We plan to extend our analysis to the 
unsigned magnetic flux and other proxies, in a follow-up paper.

 \subsection{Proxy modelling}

It has been known for a long time that irradiances of some spectral
lines correlate with activity proxies.
For example, \cite{timothy_timothy:1970}
showed a correlation (although not linear) between
the \ion{He}{ii} 304~\AA\ irradiance and the F10.7 cm radio flux. The relationship of the EUV irradiances with the F10.7 cm radio flux in general, has been further explored in models aimed at describing the EUV spectrum based on such proxies. For instance, the "EUV81" model by \cite{Hinteregger_1981} was among the first to parametrise EUV fluxes using the F10.7. This was followed by the "EUVAC" model by \cite{Richards_1994}, which improved spectral resolution and updated solar emission features. More recently the "SOLAR2000" model by \cite{Tobiska_2000} extended these efforts by integrating additional proxies and data sets to provide a comprehensive description of solar EUV variability.

\cite{toriumi_universal_2022} also explored correlations between solar activity proxies (i.e. the F10.7 cm radio flux and the sunspot area) and spectral irradiances across a wide wavelength range. They developed a catalogue of power-law scaling indices that connect solar activity proxies to spectral lines of varying formation temperatures. \cite{sanz_forcada_estimation_2011} used scaling relations based on solar proxies to estimate XUV radiation received by exoplanets orbiting active stars and demonstrated that such proxies are critical for reconstructing the XUV spectra of stars. \cite{namekata_reconstructing_2023} extended solar proxy scaling to model XUV spectra for active solar-like stars, combining magnetic flux measurements with scaling relations to synthesise spectral irradiances. 

The 10.7cm flux has been used historically by several early models of the
solar EUV/UV irradiance.
\cite{delzanna_andretta:2011} used the well-calibrated CDS irradiances to
find large  discrepancies in a number of lines between the measured
irradiances and those predicted by some earlier models, not just in the absolute 
values, but also in the cycle variations.
On the other hand, a relatively good linear correlation between the CDS 
irradiances of the helium and coronal lines vs. the  10.7cm flux was found,
for data covering the 1998-2010 period.

\cite{2008AdSpR..42..903D} searched for correlations between the
TIMED SEE EUV irradiances and various proxies but with little success.
\cite{lean_etal:2011} also used the TIMED SEE to provide 10-~\AA\ bin 
models of the EUV solar irradiance based on a combination of 
Mg II and F10.7 proxies.

The Flare Irradiance Spectral Model (FISM) was mainly developed for space
weather applications, but also provides daily spectral irradiances
from 0.1 to 1900~\AA\ at 1~\AA\ resolution. The latest version, 
FISM2, is described in \cite{chamberlin_etal:2020}.
The FISM model is based on SDO EVE and SORCE observations and on a
combination of several proxies, which include the  10.7cm radio flux,
the Mg II index, the H I Lyman $\alpha$ among others.
The method essentially searches for each 1~\AA\ bin
which single daily proxy most accurately represents the observation
and then uses it to predict the irradiances. The FISM2 is a multiple component daily model with a solar minimum reference spectrum component, a solar cycle component, and a solar rotation component. It is more complex since it switches to lower quality proxies when gaps occur in the primary proxies. 

\begin{table}
\caption{Main spectral lines from EVE MEGS-A and MEGS-B. The table includes the temperature of formation of each ion, along with information on active region (AR) and quiet Sun (QS) blends derived from a DEM analysis.}
\label{tab:mainlines}
\setlength{\tabcolsep}{2.5pt}
\begin{center}
\begin{tabular}{@{}lrlll@{}}
\hline\hline \noalign{\smallskip}
Ion & $\lambda$ (\AA) & log & AR blends & QS blends \\
 & & T[K] & & \\
\noalign{\smallskip}\hline\noalign{\smallskip}
\ion{Fe}{ix} & 171.1 & 5.99 & 4\% \ion{Ni}{xiv} & - \\
\ion{Fe}{x} & 174.5 & 6.07 & - & - \\
\ion{Fe}{xi} & 180.4 & 6.14 & 5\% \ion{Fe}{x}, 2\% \ion{S}{x} & 7\% \ion{Fe}{x} \\
\ion{Fe}{xi} & 188.2 & 6.14 & 7\% \ion{Fe}{ix}, 5\% \ion{S}{xi} & 11\% \ion{Fe}{ix} \\
\ion{Fe}{xiii} & 202.0 & 6.27 & 9\% \ion{Fe}{xi} & 23\% \ion{Fe}{xi}, 7\% \ion{S}{viii} \\
\ion{Fe}{xiii} & 203.8 & 6.26 & 11\% \ion{Fe}{xii} & 22\% \ion{Fe}{xii}, 3\% \ion{Fe}{xi} \\
\ion{Fe}{xiv} & 211.3 & 6.3 & - & - \\
\ion{Si}{x} & 261.1 & 6.17 & 2\% \ion{Fe}{ix}, 2\% \ion{Si}{ix}, &  \\
            &       &      & 2\%\ion{Fe}{xiii} & 3\% \ion{Si}{ix} \\
  
\ion{S}{x} & 264.2 & 6.24 & - & - \\
\ion{Fe}{xiv} & 264.8 & 6.33 & 6\% \ion{Fe}{xi} & 3\% \ion{Fe}{xi} \\
\ion{Si}{x} & 271.9 & 6.10 & 2\% \ion{O}{iv}, 4\% \ion{Fe}{xiii} & 7\% \ion{O}{iv} \\
\ion{Fe}{ixv} & 274.2 & 6.32 & 2\% \ion{Si}{vii} & 16\% \ion{Si}{vii} \\
\ion{S}{xi} & 281.4 & 6.32 & 6\% \ion{Fe}{xi} & 15\% \ion{Al}{ix} \\
\ion{Fe}{xv} & 284.2 & 6.39 & - & -\\
\ion{He}{ii} & 303.8 & 5.81 & 49\% \ion{Si}{xi}, 2\% \ion{Fe}{xiii} & - \\
\ion{Fe}{xvi} & 335.5 & 6.42 & 2\% \ion{Mg}{viii}, 2\%\ion{Fe}{xii} & 37\% \ion{Mg}{viii}, \\
               &       &     &                  & 12\% \ion{Fe}{xii}, 8\% \ion{Fe}{ix} \\
\ion{Fe}{xvi} & 360.8 & 6.43 & - & 12\% \ion{Fe}{x} \\
\ion{Mg}{ix} & 368.0 & 6.06 & 6\% \ion{Mg}{vii}, 13\% \ion{Fe}{xiii} & 2\% \ion{Fe}{xiii}\\
\ion{Fe}{xv} & 417.3 & 6.40 & 42\% \ion{S}{xiv}, 3\% \ion{Fe}{xiii} & - \\
\ion{Ne}{vii} & 465.0 & 5.74 & - & - \\
\ion{Si}{xii} & 499.0 & 6.37 & - & - \\
\ion{Si}{xii} & 520.7 & 6.37 & - & - \\
\ion{O}{iv} & 554.5 & 5.17 & - & - \\
\ion{He}{i} & 584.0 & 4.50 & - & \\
\ion{O}{iii} & 599.6 & 4.87 & - & - \\ 
\ion{Mg}{x} & 610.0 & 6.03 & 14\% \ion{O}{iv} & 23\% \ion{O}{iv} \\
\ion{Mg}{x} & 625.0 & 6.21 & 5\% \ion{Si}{x} & 4\% \ion{Si}{x} \\
\ion{O}{v} & 630.0 & 5.36 & - & - \\
\ion{N}{iv} & 765.2 & 5.08 & - & - \\
\ion{Ne}{viii} & 771.0 & 5.37 & - & - \\
\ion{Ne}{viii} & 780.0 & 6.02 & - & -\\
\ion{S}{v} & 786.5 & 5.16 & - & -\\
\ion{O}{iv} & 787.7 & 5.17 & - & -\\
\ion{O}{iv} & 790.1 & 5.15 & - & - \\
\ion{C}{iii} & 977.1 & 4.68 & - & - \\
\ion{H}{i} & 1025.7 & 4.16 & - & -\\
\ion{O}{vi} & 1031.9 & 5.67 & - & -\\
\ion{C}{ii} & 1036.6 & 4.45 & - & - \\
\ion{O}{vi} & 1037.6 & 5.67 & - & -\\
\noalign{\smallskip}\hline
\end{tabular}
\end{center}
\end{table}

\begin{figure*}
    \centerline{
        \includegraphics[width=8cm, keepaspectratio, trim=3cm 1.5cm 3cm 1.5cm, clip]{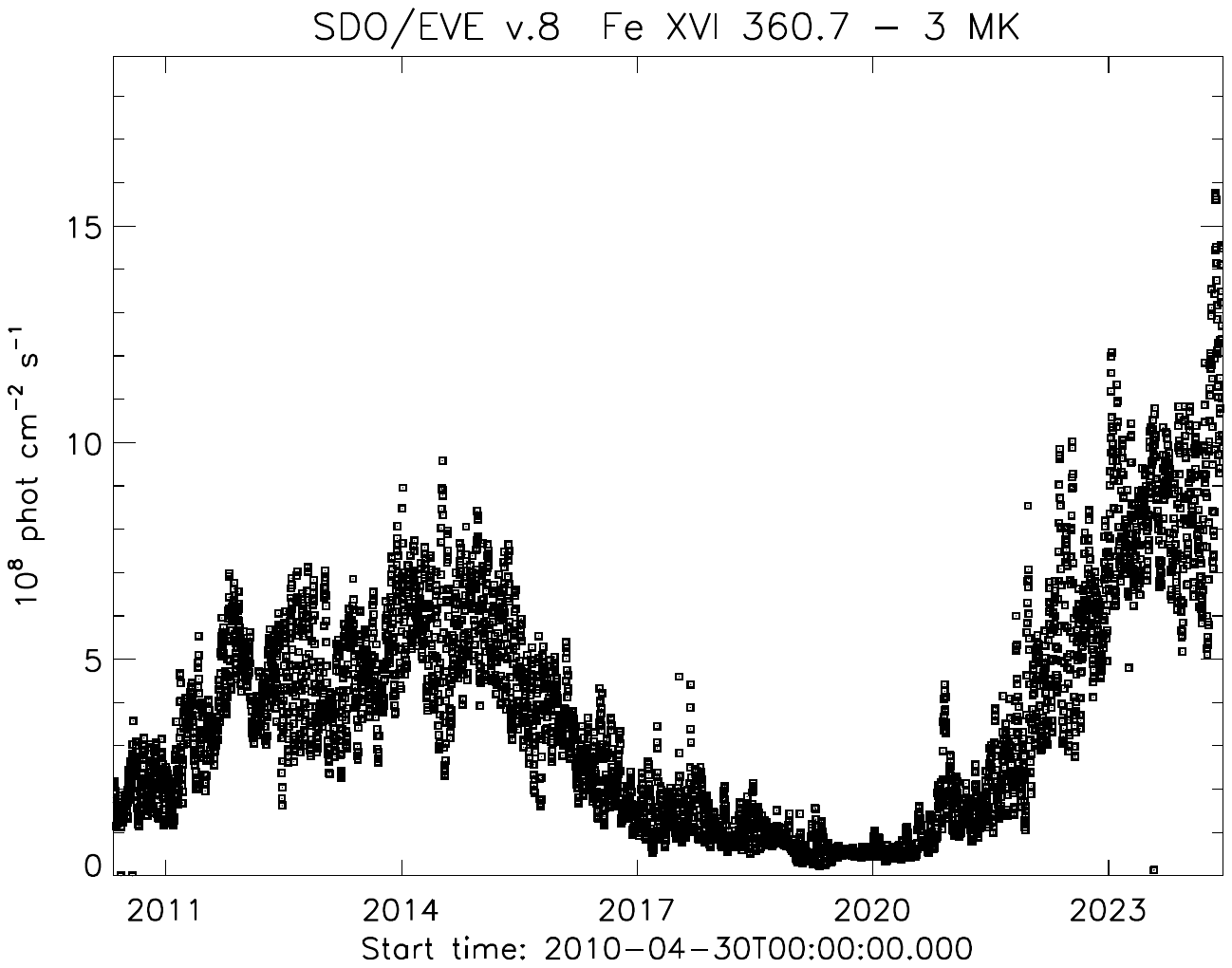}
        \includegraphics[width=8cm, keepaspectratio, trim=3cm 1.5cm 3cm 1.5cm, clip]{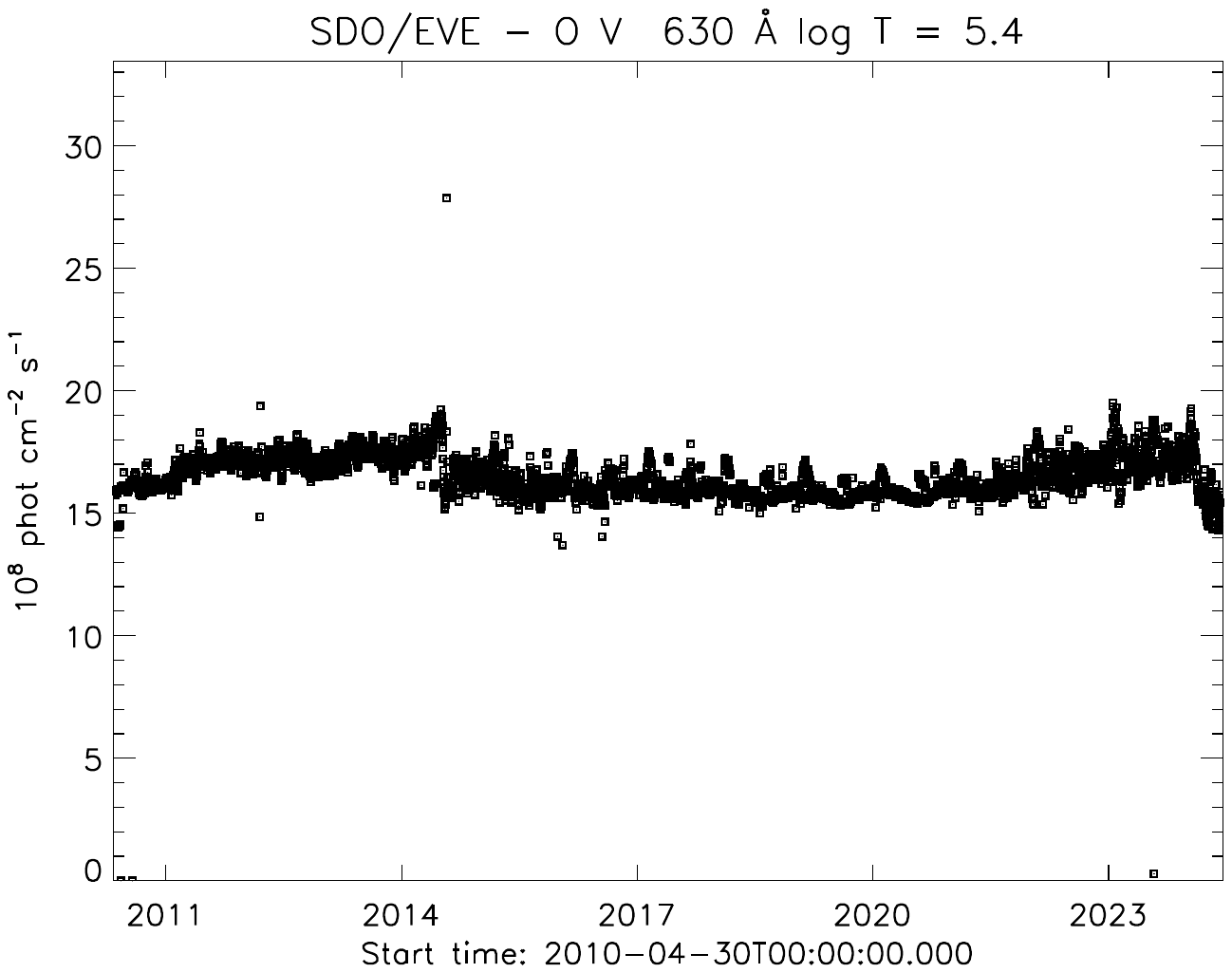}}
    \caption{Variation of SDO/EVE irradiances (10$^8$ phot cm$^{-2}$ s$^{-1}$) of calibration version 8 with the solar cycle, for 360.7\,\AA\ \ion{Fe}{xvi} (left) and 630\,\AA\ \ion{O}{v} (right).}
    \label{fig:irr_vs_cycle}
\end{figure*}

\section{Methods}

We have analysed the SOHO CDS and the version 8 SDO EVE daily-averaged level 3 merged spectra (from 2010 until 17 June, 2024). We computed line irradiances by fitting Gaussian profiles in a few selected spectral ranges and by subtracting a linear background. For SDO EVE irradiances, and in order to be able to analyse the relationship between the irradiances and the proxies, we binned and re-binned the data to reduce noise and calculate robust uncertainties. We removed invalid data points (i.e. negative values for irradiances or proxies), and divided the proxies into bins of fixed intensity ranges. Within each bin, we calculated the mean irradiance and its standard deviation ($\sigma$). We then removed the outliers (data points deviating more than 2$\sigma$ from the mean) to improve the reliability of the bin statistics. We then performed a linear polynomial fit on the re-binned data, yielding coefficients that quantify the relationship between the proxy and the irradiance.

We applied a further renormalisation technique based on the method introduced by \cite{Hinteregger_1981} to improve the comparability of irradiance measurements across different datasets and solar activity levels. In this approach, the proxies (P) and irradiance (I) values are normalised relative to their respective solar minimum values (P$_{\rm min}$ and I$_{\rm min}$). The normalised variable P' and I' are defined as: 

\begin{equation}
\rm P' = \frac{\rm P}{\rm P_{\rm min}} - 1, \quad \rm I' = \frac{\rm I}{\rm I_{\rm min}} - 1
\end{equation}

We then performed a linear fit on the normalised data, where I' = a + b P'. 

The plots presented in this work depict simple linear fits to the un-normalised data where the relationship between the irradiance (I) and the proxies (P) is expressed as I = a + b P. However, in the appendices, we include a table listing the coefficients obtained using the \cite{Hinteregger_1981} renormalisation technique. 

We performed a DEM analysis on the EVE (MEGS-A and MEGS-B) irradiances of two observational dates: one during the solar minimum around 2008 (QS) and the other during the solar maximum around 2014 (AR), using CHIANTI version 11 \citep{dufresne_chianti_2024}. Based on the results of the analysis we determined the formation temperatures of the main spectra lines for the ions discussed in this work, and their associated blends (see table~\ref{tab:mainlines}). For the DEM analysis we used photospheric abundances by \cite{asplund_chemical_2021}, a pressure of 5 $\times 10^{15}$ [cm$^{-3}$ K] for the Active Sun, and a pressure of 5 $\times 10^{14}$ [cm$^{-3}$ K] for the Quiet Sun. 

Finally, we developed and provide a simple routine for a selection of the main and strong lines within the EVE wavelength range. This routine allows the user to input a specific wavelength and a chosen value for any of the four proxies analysed. The routine then outputs the corresponding irradiance value, the average irradiance, the $\pm 1 \sigma$ irradiance range, and the total $\pm$ uncertainty. The outputs are based on the coefficients of the correlations of the irradiances with the proxies.

\begin{figure*}
    \centerline{
        \includegraphics[width=8cm, keepaspectratio, trim=3cm 1.5cm 3cm 1.5cm, clip]{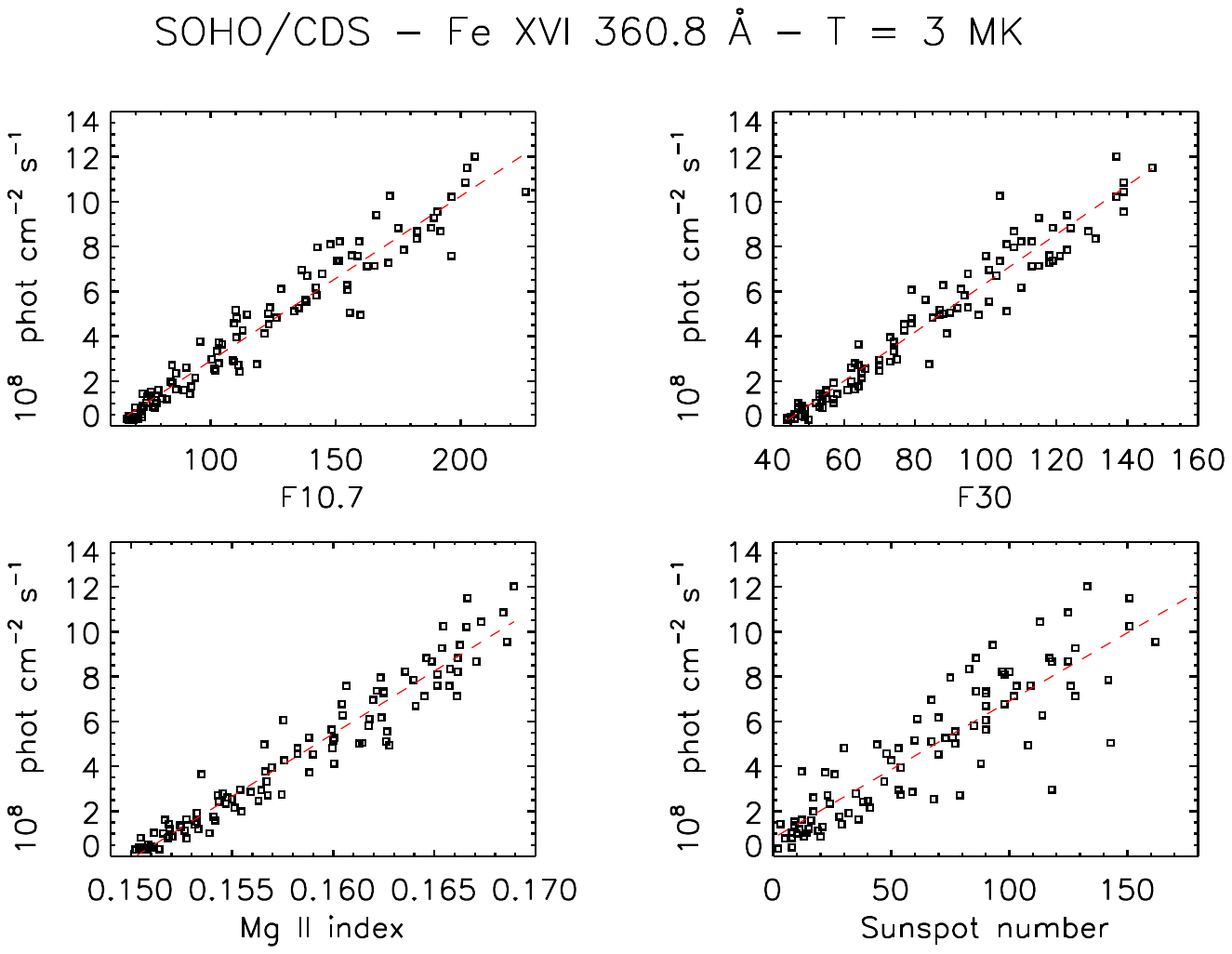}
        \includegraphics[width=8cm, keepaspectratio, trim=3cm 1.5cm 3cm 1.5cm, clip]{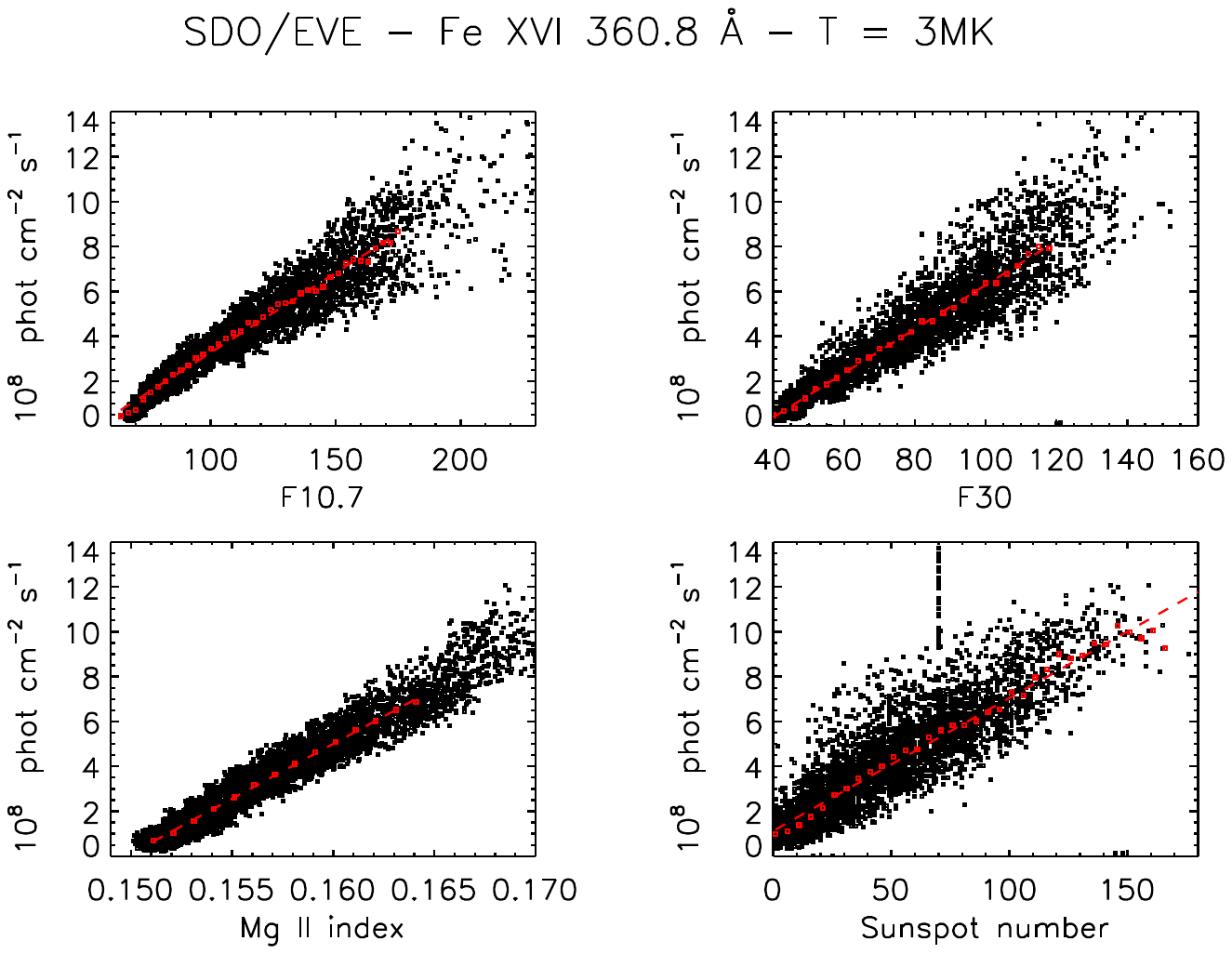}}
    \centerline{
        \includegraphics[width=8cm, keepaspectratio, trim=3cm 1.5cm 3cm 1.5cm, clip]{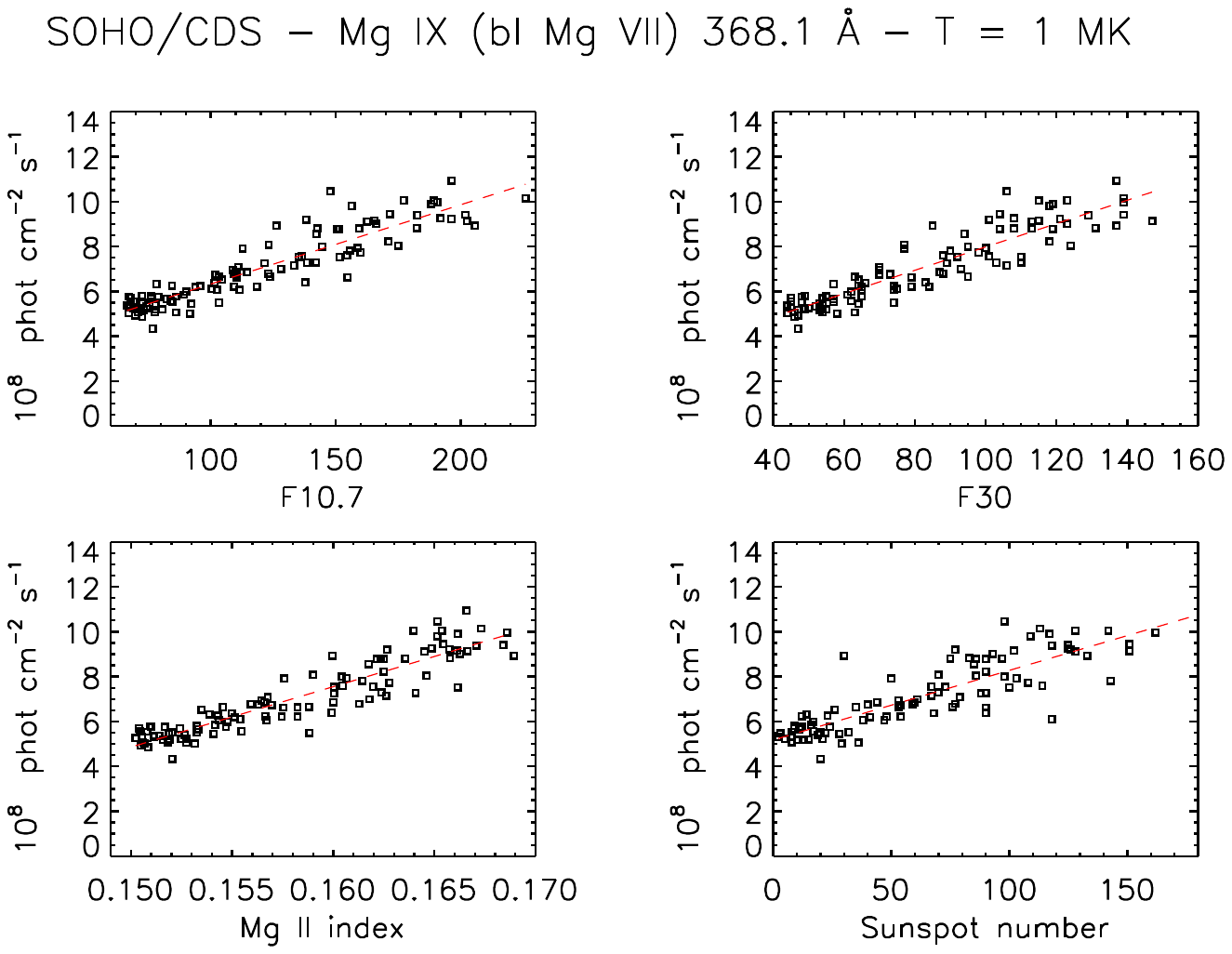}
        \includegraphics[width=8cm, keepaspectratio, trim=3cm 1.5cm 3cm 1.5cm, clip]{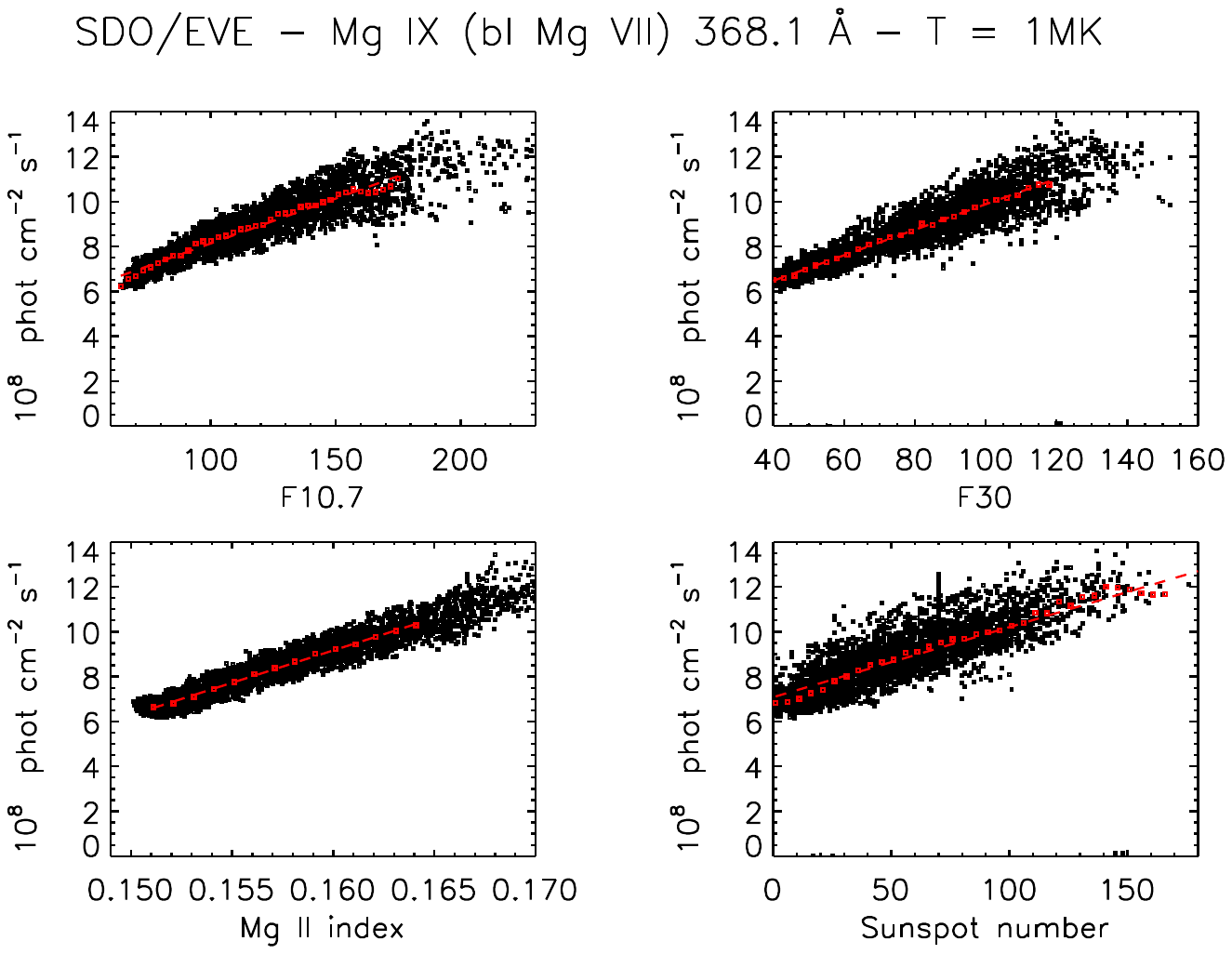}}
    \caption{Correlation between the irradiances of SOHO/CDS and SDO/EVE 360.8\,\AA\ \ion{Fe}{xvi} and 368\,\AA\ \ion{Mg}{ix} hot coronal lines, with the daily F10.7 and F30 cm radio fluxes (\(10^{-22}\,\mathrm{W\,m^{-2}\,Hz^{-1}}\)), the Mg II index, and the Sunspot Number.}
    \label{fig:Coronal_vs_proxy}
\end{figure*}

\begin{figure*}
    \centerline{
        \includegraphics[width=8cm, keepaspectratio, trim=3cm 1.5cm 3cm 1.5cm, clip]{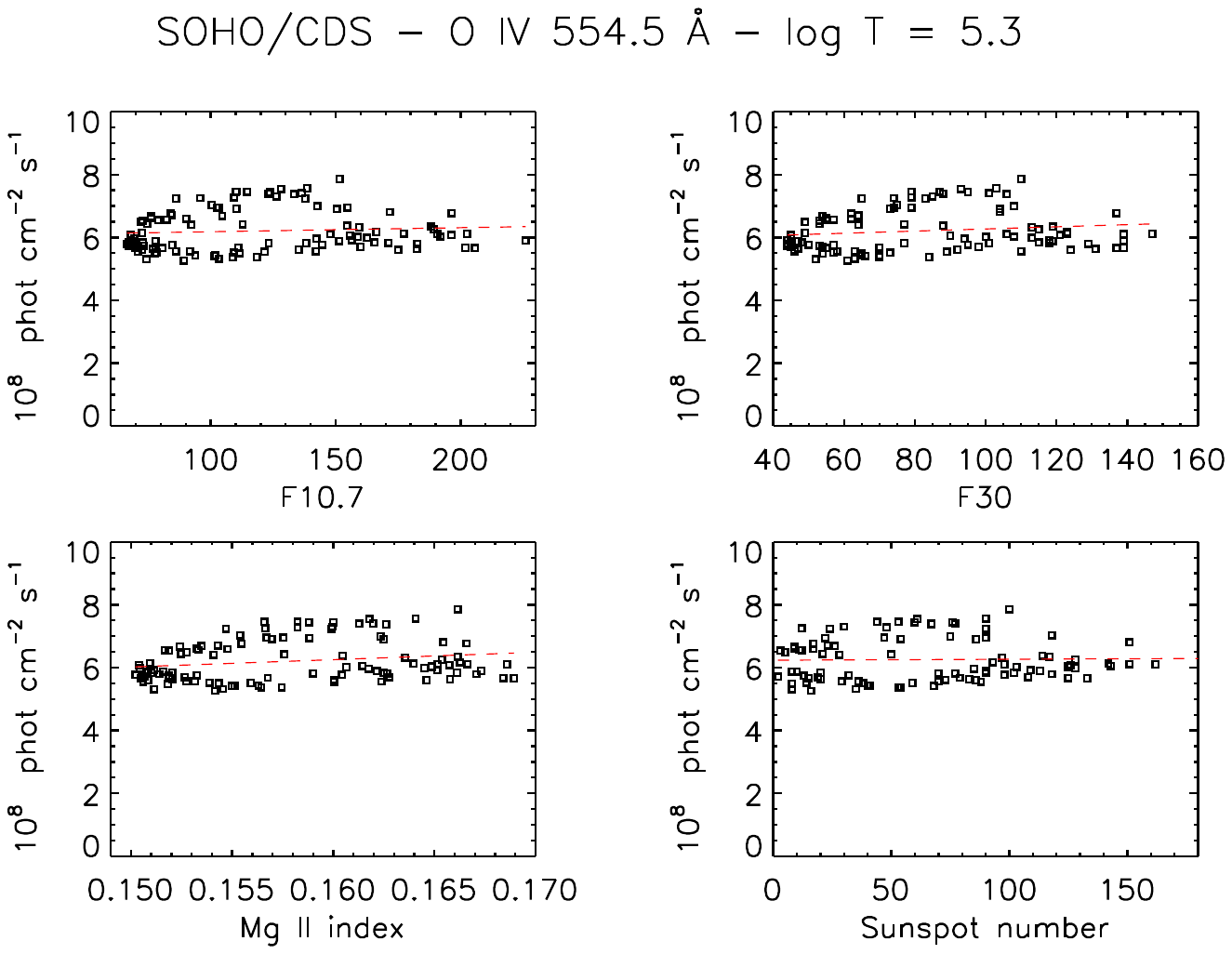}
        \includegraphics[width=8cm, keepaspectratio, trim=3cm 1.5cm 3cm 1.5cm,  clip]{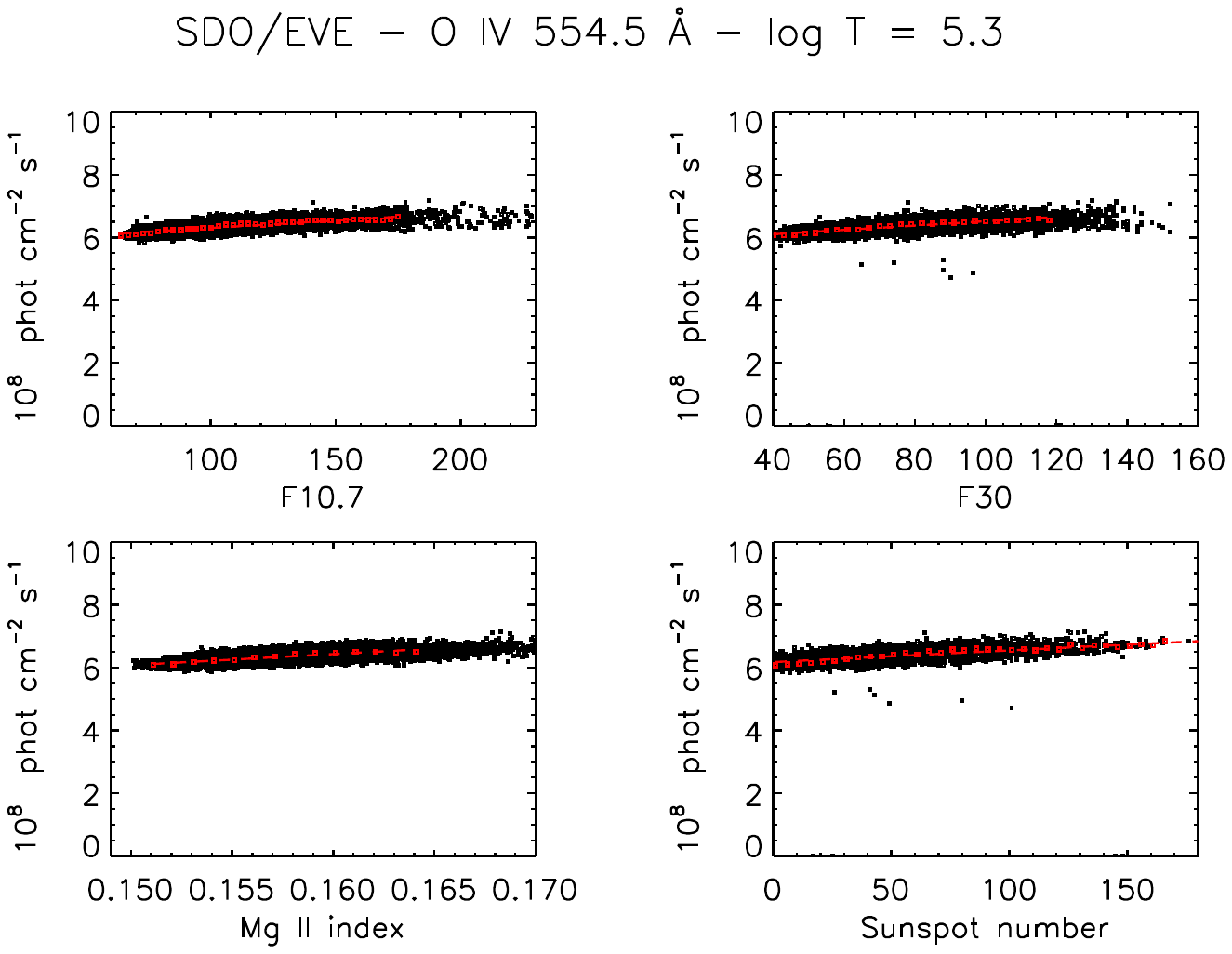}}
    \caption{Correlation between the irradiances of SOHO/CDS and SDO/EVE 554\,\AA\ \ion{O}{iv} transition region line, with the daily F10.7 and F30 cm radio fluxes (\(10^{-22}\,\mathrm{W\,m^{-2}\,Hz^{-1}}\)), the Mg II index, and the Sunspot Number. The 554\,\AA\ \ion{O}{iv} is a self-blended multiplet with 553.34\, \AA\  \ion{O}{iv} and 555.26\,\AA\ \ion{O}{iv}.}
    \label{fig:554_tr_vs_proxy}
\end{figure*}
        
\begin{figure*}
\centerline{
    \includegraphics[width=8cm, keepaspectratio, trim=3cm 1.5cm 3cm 1.5cm, clip]{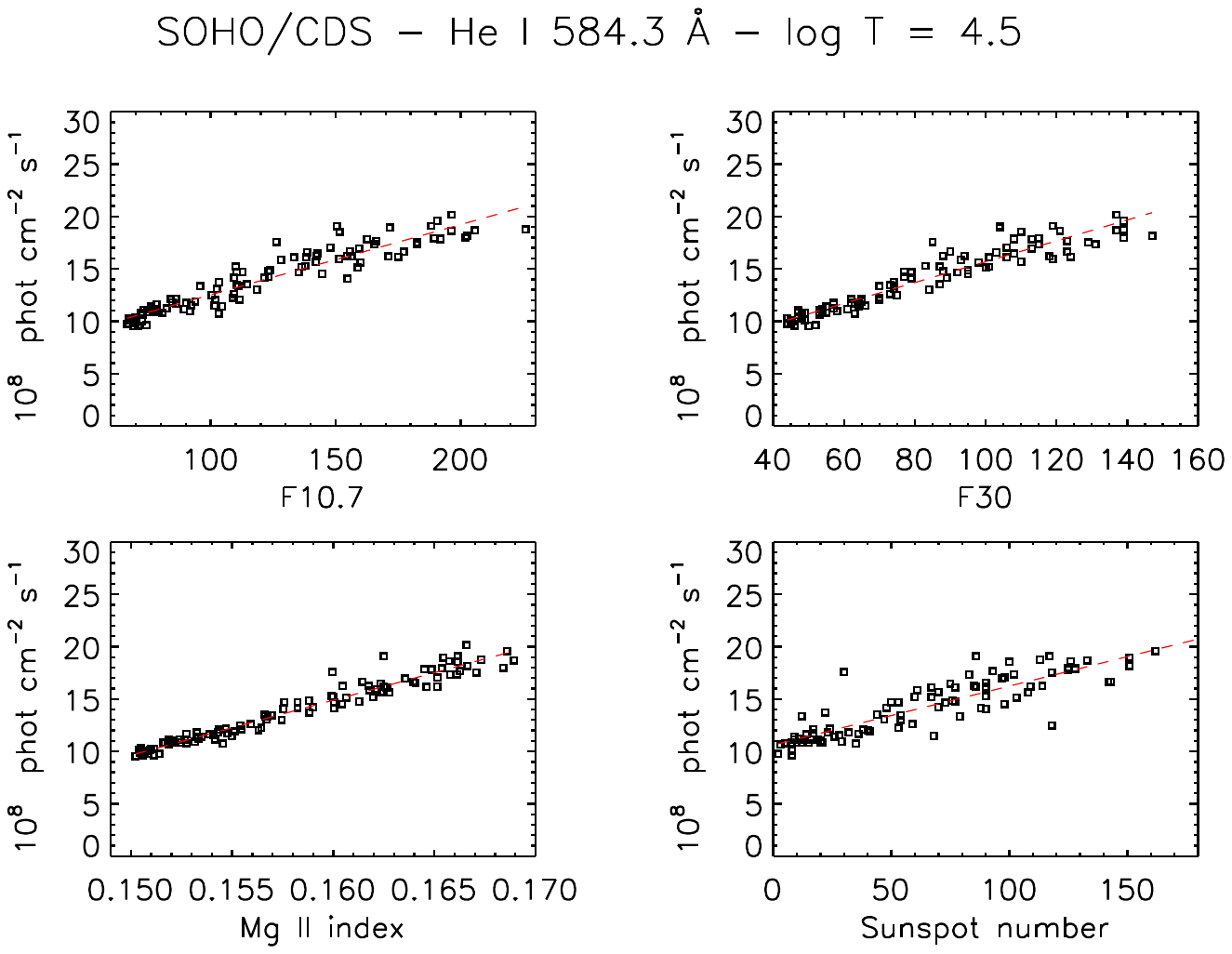}
    \includegraphics[width=8cm, keepaspectratio, trim=3cm 1.5cm 3cm 1.5cm, clip]{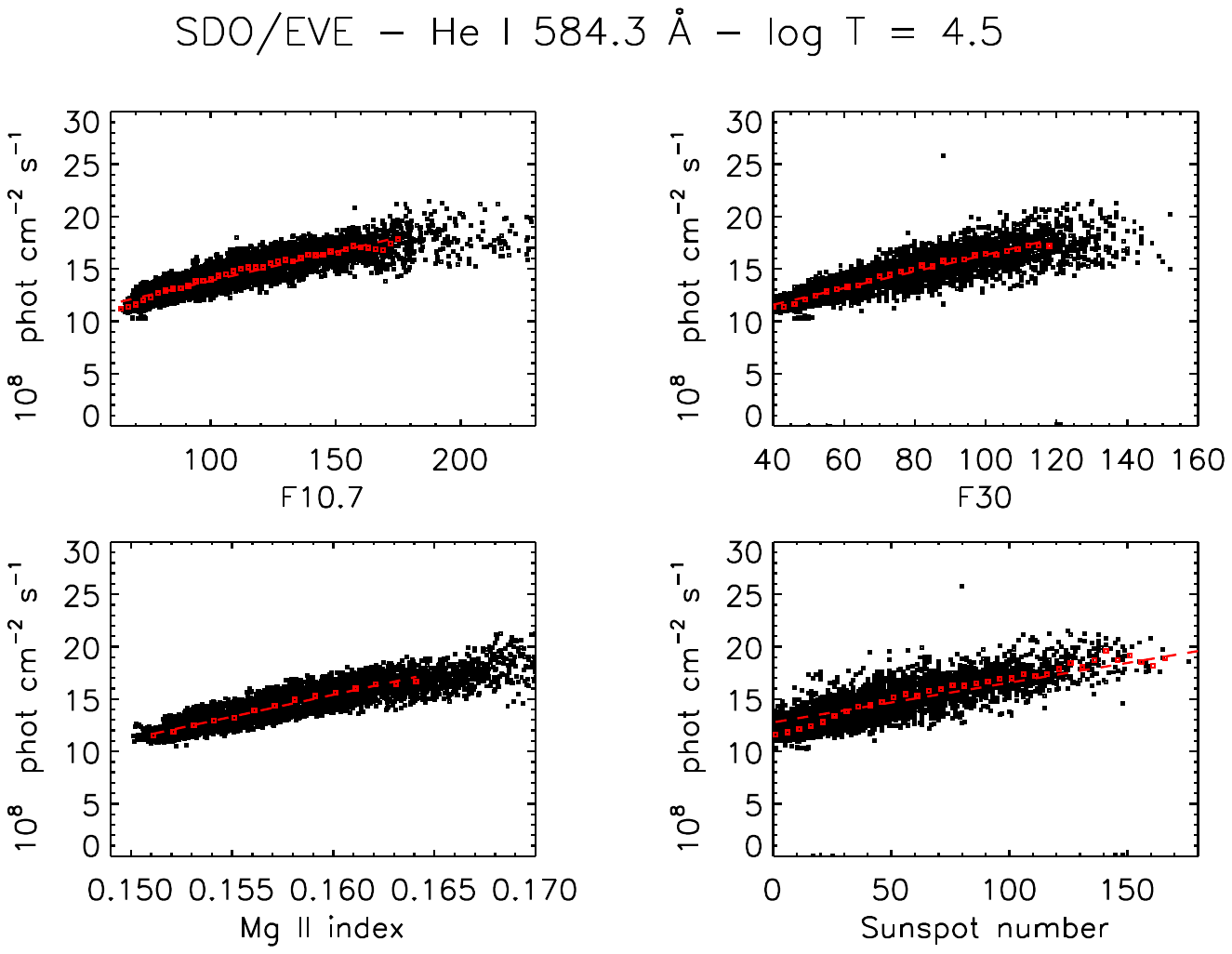}}
    \caption{Correlation between the irradiances of SOHO/CDS and SDO/EVE 584\,\AA\ \ion{He}{i} line, with the daily F10.7 and F30 cm radio fluxes (\(10^{-22}\,\mathrm{W\,m^{-2}\,Hz^{-1}}\)), the Mg II index, and the Sunspot Number.}
    \label{fig:he_vs_proxy}
\end{figure*}

\begin{figure*}
\centerline{
    \includegraphics[width=8cm, keepaspectratio, trim=3cm 1.5cm 3cm 1.5cm, clip]{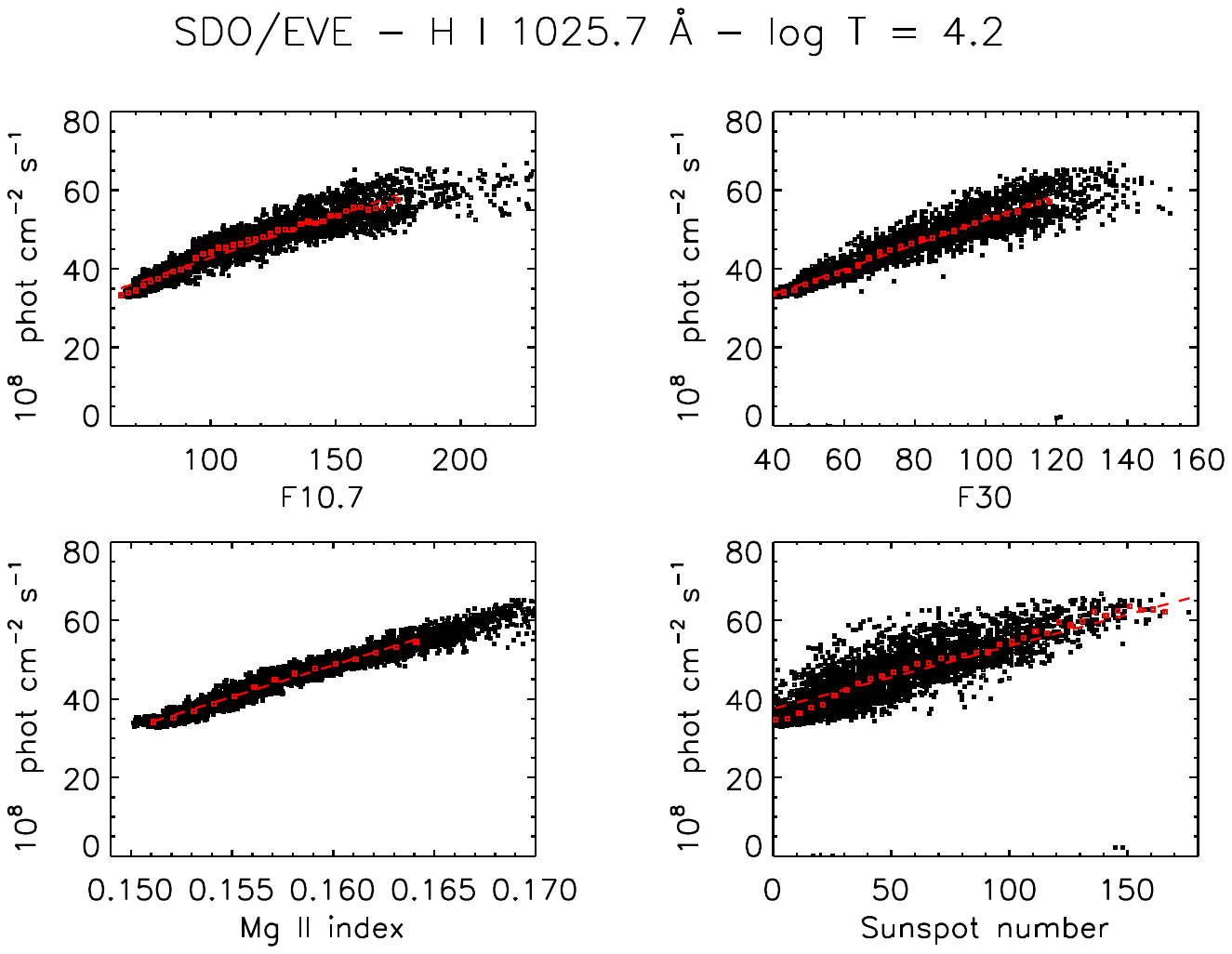}
     \includegraphics[width=8cm, keepaspectratio, trim=3cm 1.5cm 3cm 1.5cm, clip]{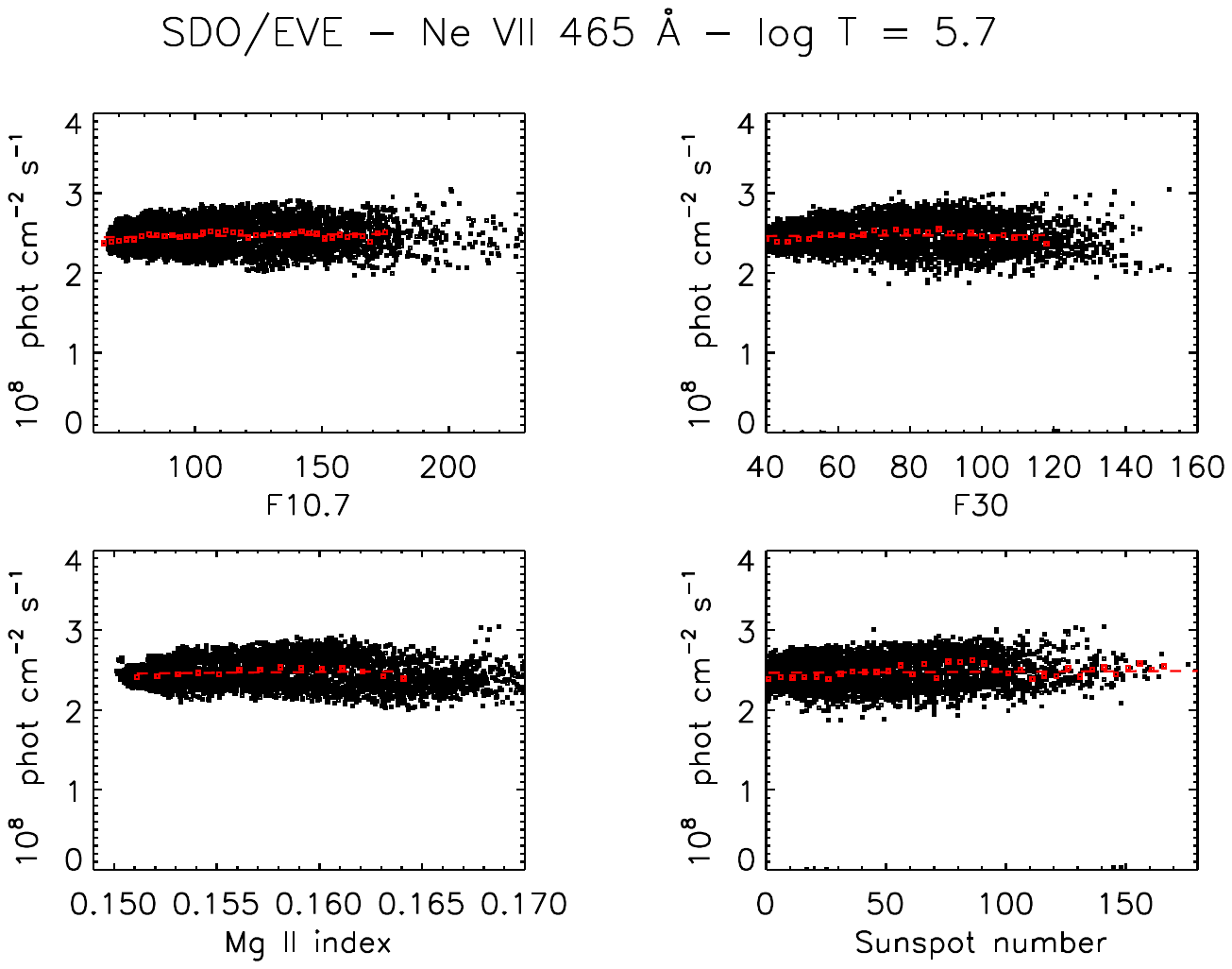}}
    \caption{Correlation between the irradiances of SOHO/CDS and SDO/EVE 1025.7\,\AA\ \ion{H}{i} and 465\,\AA\ \ion{Ne}{viii} lines, with the daily F10.7 and F30 cm radio fluxes (\(10^{-22}\,\mathrm{W\,m^{-2}\,Hz^{-1}}\)), the Mg II index, and the Sunspot Number.}
    \label{fig:tr_h_vs_proxy}
\end{figure*}

\section{Results}

We first investigated the variation of the irradiances with the solar cycle for all the EVE lines. We found that the latest calibration version (8) significantly improved discrepancies found in the previous calibration version (7). We present the results for the 360.7\ \AA\ \ion{Fe}{xvi} which is a very hot coronal lines (T$\approx$ 3MK) and the 630\ \AA\ \ion{O}{v} which is a cooler (log T$\approx$ 5.4) TR line, in figure~\ref{fig:irr_vs_cycle}. The figure shows the variability of the solar irradiances for these lines over a span of approximately 12 years, nearly a full solar activity cycle. The behaviour of the 360.7\ \AA\ \ion{Fe}{xvi} reflects the strong influence of magnetic activity on the hot coronal plasma, with peaks around 2014 and close to 2025 in the solar maxima, while it exhibits a minimum around 2020 during the solar minima. In contrast, the 630\ \AA\ \ion{O}{v} shows a constant behaviour with minor fluctuations over the cycle, as it has been shown in the past that the TR lines are less affected by the cycle.  Note that over- or under-corrected degradation can cause artificial increases or decreases respectively. The quality of the previous version 7 data for recent times was poor because trends were extrapolated using solar minimum trends, under-correcting degradation. 

We then investigated the variation of the irradiances with the proxies of activity that we discussed in the earlier sections. We present results for the main coronal, TR, \ion{He}{i} and \ion{H}{i} lines in  figures~\ref{fig:Coronal_vs_proxy}, \ref{fig:he_vs_proxy}, \ref{fig:554_tr_vs_proxy}, and \ref{fig:tr_h_vs_proxy}. 

The comparison between the SDO/EVE and the SOHO/CDS irradiances shows a good overall agreement, with minor discrepancies likely attributable to differences in radiometric calibration between the instruments. Additionally, all coronal lines exhibit strong correlations with the proxies of activity, as shown in figure~\ref{fig:Coronal_vs_proxy}.  Note that the strong resonance Mg IX line is mostly blended 
with Mg VII, which is (barely) resolvable by CDS. Therefore,
we have added the Mg VII contribution to Mg IX in the CDS case,
for direct comparison with EVE.

In contrast, the TR lines (fig.~\ref{fig:554_tr_vs_proxy} and \ref{fig:tr_h_vs_proxy}) exhibit weaker correlations with the proxies and the solar cycle. The irradiance of the 554.5~\AA\ \ion{O}{iv} line is the total from the  multiplet  (553.329~\AA\, , 554.500~\AA\, and 555.263~\AA\, as 554.5~\AA). We notice from fig.~\ref{fig:554_tr_vs_proxy} that the spread of the scatter for the 554\ \AA\ \ion{O}{iv} is much broader for CDS compared to the EVE irradiances, although there is overall agreement. 

We believe that the smaller variability of TR lines with the solar cycle could in part  be attributed to dark halos. Dark halos are very large regions of reduced emission in TR lines seen all around active regions in XUV observations \citep{Lezzi_2023}. 
So when the Sun is very active the contribution of the active regions
increases, but at the same time the emission in TR lines decreases.
This does not occur in higher-temperature lines.

Finally, the \ion{He}{i} and \ion{H}{i} lines that form in-between the chromosphere and the TR show a bigger variation with the proxies (and the cycle), as already known previously. 
This is most likely due to photo-ionization effects from the 
coronal radiation \citep{andretta_euv_2003}.

Notably, the correlations with the Mg II index and the F30 cm flux are generally more linear across all lines, compared to those with the F10.7 cm flux and the sunspot number. Based on this, we reviewed the literature for accurate historical EUV irradiance records and plotted them against the F30 cm radio flux. We also used a selection of PEVE irradiances from the 2008 solar minimum. These plots can be found in figures~\ref{fig:MEGSA_his_rec},\ref{fig:MEGS_B_his_rec1} and \ref{fig:MEGS_B_his_rec2} in the appendices.  

The best calibrated spectrum in the EUV is that one of \cite{malinovsky_analysis_1973} with typical uncertainties of 10\% for lines close in wavelength, when benchmarked against the best atomic data. The \cite{heroux_etal:1974} spectrum is also excellent and has a stated accuracy of 30\% over a broad wavelength range. Spectra of similar accuracy are those of \cite{higgins_solar_1976} and \cite{heroux_higgins:1977}.

We also found that the EVE irradiances tend to be much more increased compared to historical records for lines with $\lambda>$ 700 \AA. For the same wavelength range we found that the PEVE irradiances are much more increased compared to the EVE irradiances. These results are found in figure~\ref{fig:MEGS_B_his_rec2} in the appendices.

Finally, in table~\ref{tab:mainlines} we present the main lines discussed in the paper, with the log T [K] and the associated blends when found in Active Sun (AR) and Quiet Sun (QS). Blends of 2\% or less are not mentioned in the table. We present the respective plots for all of these lines in the paper and the appendices.

\subsection{Technical Issues}

We found good agreement of EVE irradiances with historical records (see appendix~\ref{append:Historical Records}), especially for hot coronal lines, with some exceptions. 
We also confirm the discrepancies noted (when using 
the earlier EVE calibration version 4) by
\cite{del_zanna_euv_2015} between EVE and PEVE irradiances of several strong lines such as the 465\ \AA\ \ion{Ne}{vii}, the 554.5\ \AA\ \ion{O}{iv}, the 600\ \AA\ \ion{O}{iii} and the 630\ \AA\ \ion{O}{v}.

\begin{figure}
    \centering
    \includegraphics[width=\columnwidth,trim=50 20 50 20, clip]{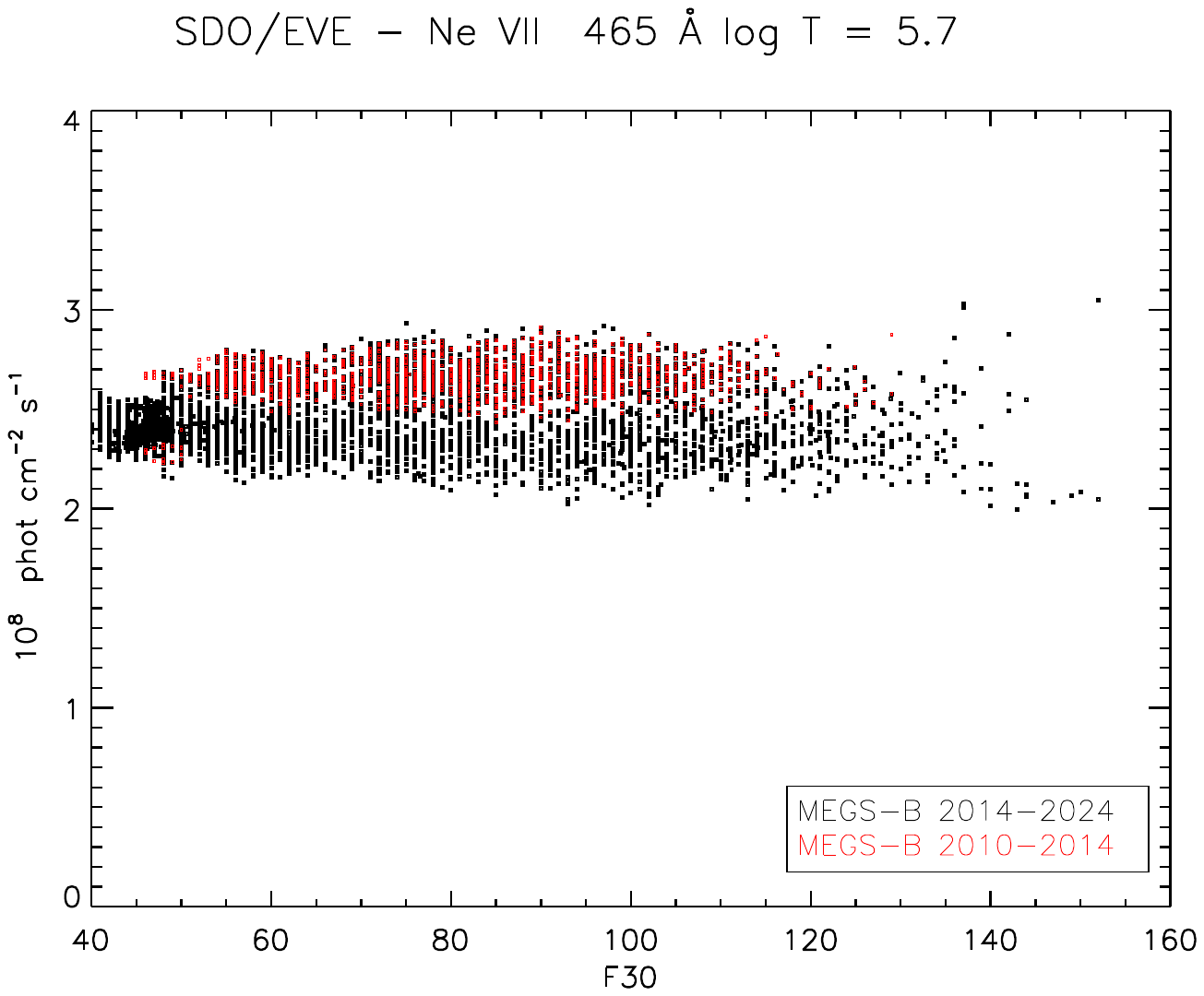}
    \caption{Comparison of SDO/EVE 465\ \AA\ \ion{Ne}{vii} irradiances measured by MEGS-B from 2014 to 2024 (black) and 2010 to 2014 (red). The jump in irradiance observed after 2014 corresponds to the failure of the MEGS-A detector, resulting in changes in wavelength calibration.}
    \label{fig:ne_vii_technical}
\end{figure}

The 465\ \AA\ \ion{Ne}{vii} was significantly affected after the failure of MEGS-A in 2014, as shown in figure~\ref{fig:ne_vii_technical}, where we have plotted the MEGS-B irradiances of 465\ \AA\ \ion{Ne}{vii} line against the F30 cm radio flux from 2010-2014 (in red) and 2014-2024 (in black). We clearly see a wavelength jump that was caused due to a temperature shift in the MEGS-B instrument (as mentioned in the Level 3 Version 8 ReadMe file). In fig.~\ref{fig:tr_h_vs_proxy} in which we show the correlation of this line with all proxies of activity, we've particularly selected MEGS-B data from 2010 to 2014, so the linear coefficients there only correspond to this period. 

There are several more lines that are affected by the failure of MEGS-A, among which are the: 334\ \AA\ \ion{Fe}{xiv}, 345\ \AA\ \ion{Si}{ix}, 349\ \AA\ \ion{Mg}{vi}, 512\ \AA\ \ion{Ca}{ix}, 515\ \AA\ \ion{He}{i}, 537\ \AA\ \ion{He}{i}, 572\ \AA\ \ion{Ne}{v}, 600\ \AA\ \ion{O}{iii}, 630\ \AA\ \ion{O}{v}, 719\ \AA\ \ion{O}{ii}, 765\ \AA\ \ion{N}{iv} and 1036\ \AA\ \ion{C}{ii}.

\subsection{Renormalised Coefficients}
\label{append:coefficients}

We present two tables (CDS and EVE results) for a selection of lines, as an example, summarising the renormalised coefficients associated with the F30 cm radio flux. For each spectral line, we list the minimum F30 cm radio flux value observed (P$_{\rm min}$) and the minimum irradiance value (I$_{\rm min}$).  We then provide the coefficients 'a' and 'b', which describe the linear relationship y = a + bx, based on the renormalised variables (as discussed earlier in the paper), P' = (P / P$_{\rm min}$) - 1 and I' = (I / I$_{\rm min}$) -1, following our modelling approach. Within the associated data, we provide extended tables for all four proxies for the CDS and EVE lines, and in the latter case we also provide a 1$\pm \sigma$ measure of the scatter. 

It is evident from table~\ref{tab:coefficients_EVE} and \ref{tab:coefficients_CDS}, which list the coefficient b-the slope of the irradiance vs. proxy plots—that hotter coronal lines exhibit significantly higher values of b compared to cooler TR or weaker lines. This is again consistent with the expectation that strong coronal lines and influenced by solar activity. The figures presented in this paper, along with those in Appendices~\ref{append:extra plots}, clearly illustrate this behaviour.

\begin{table}
\caption{Correlation of EVE (MEGS-A and MEGS-B) irradiances with the F30 cm radio flux. P$_{\rm min}$ is the minimum value of the F30 observed, I$_{\rm min}$ is the minimum intensity observed for that line and a and b are the renormalised coefficients based on the linear regression fitting. The * next to coefficients indicates that these lines were affected by the failure of MEGS-A and hence the analysis corresponds to irradiances from the period of 2010-2014.}
\label{tab:coefficients_EVE}
\begin{center}
    \centering
    \begin{tabular}{@{}l c | c c c c @{}}
    \hline\hline
    Ion & $\lambda$ (\AA)  
    & \multicolumn{4}{c}{F30 cm} \\ 
    \hline 
   & & P$_{\rm min}$ & I$_{\rm min}$ & a & b  \\
    \hline

\ion{Fe}{ix} & 171.1 & 46 & 3.46 & 0.121 & 0.186 \\
\ion{Fe}{x} & 174.5 & 46 & 2.73 & 0.133 & 0.307 \\
\ion{Fe}{xi} & 180.4 & 46 & 3.13 & 0.148 & 0.455 \\
\ion{Fe}{xi} & 188.2 & 46 & 2.50 & 0.139 & 0.449 \\
\ion{Fe}{xiii} & 202.0 & 46 & 1.46 & 0.352 & 1.08 \\
\ion{Fe}{xiii} & 203.8 & 46 & 0.360 & 0.437 & 3.20 \\
\ion{Fe}{xiv} & 211.3 & 46 & 0.598 & 0.573 & 3.70 \\
\ion{Si}{x} & 261.1 & 46 & 0.552 & 0.145 & 0.314 \\

\ion{Fe}{xiv} & 264.8 & 46 & 0.284 & 1.01 & 4.44 \\
\ion{Si}{x} & 271.9 & 46 & 0.481 & 0.0499 & 1.14 \\
\ion{Fe}{xiv} & 274.2 & 46 & 0.508 & 0.460 & 3.69 \\
\ion{Fe}{xv} & 284.2 & 46 & 1.57 & 0.487 & 5.33 \\
\ion{He}{ii} & 303.8 & 46 & 65.1 & 0.0322 & 0.239 \\

\ion{Fe}{xvi} & 335.5 & 45 & 0.441 & -0.698 & 16.4\\
\ion{Fe}{xvi} & 360.8 & 40 & 0.234 & -0.205 & 17.8 \\
\ion{Mg}{ix} & 368.0 & 40 & 6.16 & 0.0562 & 0.361 \\
\ion{Fe}{xv} & 417.3 & 40 & 0.0050 & 1.11 & 9.95 \\
\ion{Ne}{vii} & 465.0 & 46 & 2.23 & 0.177 & 0.00963 * \\
\ion{Si}{xii} & 499.0 & 40 & 0.486 & 1.32 & 7.33 \\
\ion{Si}{xii} & 520.7 & 40 & 0.111 & 2.56 & 15.2 \\
\ion{O}{iv} & 554.5 & 40 & 4.49 & 0.105 & 0.0611 \\
\ion{He}{i} & 584.0 & 40 & 10.3 & 0.166 & 0.260 \\
\ion{O}{iii} & 599.6 & 46 & 1.03 & 0.197 & 0.0408 * \\ 
\ion{Mg}{x} & 610.0 & 40 & 4.02 & 0.267 & 0.708 \\
\ion{Mg}{x} & 625.0 & 40 & 1.78 & 0.182 & 0.897 \\
\ion{O}{v} & 630.0 & 46 & 14.4 & 0.108 & 0.0802 * \\
\ion{N}{iv} & 765.2 & 46 & 2.45 & 0.128 & 0.00225 * \\
\ion{Ne}{viii} & 771.0 & 40 & 3.24 & 0.135 & 0.0342 \\
\ion{Ne}{viii} & 780.0 & 40 & 1.74 & 0.0754 & 0.0799 \\
\ion{S}{v} & 786.5 & 40 & 1.17 & 0.0307 & 0.0785 \\
\ion{O}{iv} & 787.7 & 40 & 2.16 & 0.0920 & 0.0352 \\
\ion{O}{iv} & 790.1 & 40 & 4.64 & 0.0807 & 0.0408 \\
\ion{C}{iii} & 977.1 & 40 & 54.7 & 0.129 & 0.147 \\
\ion{H}{i} & 1025.7 & 40 & 32.8 & 0.0486 & 0.351 \\
\ion{O}{vi} & 1031.9 & 40 & 23.3 & 0.0914 & 0.266\\
\ion{C}{ii} & 1036.6 & 40 & 4.13 & 0.119 & 0.309 \\
\ion{O}{vi} & 1037.6 & 40 & 10.5 & 0.173 & 0.298 \\

\hline
\end{tabular}
\end{center}
\end{table}

\begin{table}
\caption{Correlations of a selection of CDS irradiances with the F30 cm radio flux. P$_{\rm min}$ is the minimum value of the F30 observed, I$_{\rm min}$ is the minimum intensity observed for that line and a and b are the renormalised coefficients based on the linear regression fitting.}
\label{tab:coefficients_CDS}
\begin{center}
    \centering
    \begin{tabular}{@{}l c | c c c c @{}}
    \hline\hline
    Ion & $\lambda$ (\AA)  
    & \multicolumn{4}{c}{F30 cm} \\ 
    \hline 
   & & P$_{\rm min}$ & I$_{\rm min}$ & a & b  \\
    \hline
\ion{Fe}{xvi} & 335.5 & 44 & 0.952 & -0.315 & 9.66\\
\ion{Fe}{xvi} & 360.8 & 44 & 0.289 & -0.0889 & 16.6 \\
\ion{Mg}{ix} & 368.0 & 44 & 3.50 & 0.169 & 0.556 \\
\ion{O}{iv} & 554.5 & 44 & 3.07 & 0.156 & 0.0171 \\
\ion{He}{i} & 584.0 & 44 & 9.55 & 0.0586 & 0.458 \\
\ion{O}{iii} & 599.6 & 44 & 1.19 & 0.0744 & 0.0608 \\ 
\ion{Mg}{x} & 610.0 & 44 & 3.52 & 0.180 & 0.800 \\
\ion{Mg}{x} & 625.0 & 44 & 1.21 & 0.266 & 1.01 \\
\ion{O}{v} & 630.0 & 44 & 7.97 & 0.150 & 0.131 \\
\hline
\end{tabular}
\end{center}
\end{table}

\section{Conclusions}

We have expanded previous analyses found in the literature on EUV irradiances using more up to date CDS and EVE data and the latest calibration versions, as well as updated atomic data. 

With the version 8 of the EVE calibration, we now have general agreement between the CDS and EVE irradiances. Some differences are still present, as there are differences among
historical records, but the progress has been significant, if one recalls that even
the stronger line in the EUV, the resonance from He II,  was incorrectly measured by 
about a factor of two until we had the CDS and EVE measurements. 

The correlations with the proxies have turned out to be a very useful diagnostic of 
potential problems in the  measurements. Indeed we discovered some technical 
issues with EVE in this way.

We found that strong and hotter lines from the solar corona show a good correlation with the proxies and the solar cycle, although each has a different linear coefficient. We found that the correlation of lines forming at TR temperatures with the solar cycles and proxies is weaker compared to coronal lines, but still significant, with typical correlation coefficients around 0.9. This suggests that TR irradiances are less influenced by active regions than coronal irradiances but are not entirely unaffected. The He and H lines forming between the chromosphere and the TR vary more compared to the rest of the TR lines. 
These results are not new, but have now been better quantified 
and are based on long-term observations from 1998 (CDS) till 2024 (EVE).

We have found good agreement with many previous historical records, with the exception of a few  lines and longer wavelength lines, where almost all 
historical records indicate lower irradiances than those measured by EVE. 

The strong correlations across all proxies suggest that high-temperature plasma emissions are robust indicators of solar activity, as previously found in the literature. This aligns with the expectation that coronal emissions originate in regions dominated by strong magnetic fields, which correlate well with all activity measures. 
It is clear that the F30 flux and the Mg II index of activity are best at 
predicting the coronal EUV signal. 
In a follow-up paper, we will explore in more detail the reasons why 
the correlations vary for different lines, taking into account variations
in density and chemical abundances for example. We will then build a model
to predict the variations of 
EUV irradiances of the Sun based on proxies.

\section*{Acknowledgements}

ED acknowledges support from STFC (UK) via a studentship
(2882461 related to ST/Y509127/1). GDZ acknowledges support
from STFC (UK) via the consolidated grants to the atomic astro-
physics group at DAMTP, University of Cambridge (ST/P000665/1.
and ST/T000481/1).
CDS was built and operated by a consortium led by the 
Rutherford Appleton Laboratory (RAL).
SOHO is a mission of international cooperation between 
ESA and NASA.

We acknowledge the Laboratory for Atmospheric and Space Physics. (2005). LASP Interactive Solar Irradiance Datacenter. Laboratory for Atmospheric and Space Physics. https://doi.org/10.25980/L27Z-XD34

\section*{Data Availability}

We provide ascii tables with the linear coefficients between the irradiances of the main EUV lines and the four proxies, as well as  programs to read the tables and provide estimated irradiances and their uncertainties when a proxy value is 
given as input. See details on ZENODO.



\bibliographystyle{mnras}
\bibliography{paper} 

\begin{thebibliography}{}
\makeatletter
\relax
\def\mn@urlcharsother{\let\do\@makeother \do\$\do\&\do\#\do\^\do\_\do\%\do\~}
\def\mn@doi{\begingroup\mn@urlcharsother \@ifnextchar [ {\mn@doi@}
  {\mn@doi@[]}}
\def\mn@doi@[#1]#2{\def\@tempa{#1}\ifx\@tempa\@empty \href
  {http://dx.doi.org/#2} {doi:#2}\else \href {http://dx.doi.org/#2} {#1}\fi
  \endgroup}
\def\mn@eprint#1#2{\mn@eprint@#1:#2::\@nil}
\def\mn@eprint@arXiv#1{\href {http://arxiv.org/abs/#1} {{\tt arXiv:#1}}}
\def\mn@eprint@dblp#1{\href {http://dblp.uni-trier.de/rec/bibtex/#1.xml}
  {dblp:#1}}
\def\mn@eprint@#1:#2:#3:#4\@nil{\def\@tempa {#1}\def\@tempb {#2}\def\@tempc
  {#3}\ifx \@tempc \@empty \let \@tempc \@tempb \let \@tempb \@tempa \fi \ifx
  \@tempb \@empty \def\@tempb {arXiv}\fi \@ifundefined
  {mn@eprint@\@tempb}{\@tempb:\@tempc}{\expandafter \expandafter \csname
  mn@eprint@\@tempb\endcsname \expandafter{\@tempc}}}

\bibitem[\protect\citeauthoryear{{Andretta} \& {Del Zanna}}{{Andretta} \& {Del
  Zanna}}{2014}]{andretta_delzanna:2014}
{Andretta} V.,  {Del Zanna} G.,  2014, \mn@doi [\aap]
  {10.1051/0004-6361/201322841}, \href
  {http://adsabs.harvard.edu/abs/2014A%26A...563A..26A} {563, A26}

\bibitem[\protect\citeauthoryear{Andretta, Zanna  \& Jordan}{Andretta
  et~al.}{2003}]{andretta_euv_2003}
Andretta V.,  Zanna G.~D.,   Jordan S.~D.,  2003, \mn@doi [Astronomy \&
  Astrophysics] {10.1051/0004-6361:20021893}, 400, 737

\bibitem[\protect\citeauthoryear{Asplund, Amarsi  \& Grevesse}{Asplund
  et~al.}{2021}]{asplund_chemical_2021}
Asplund M.,  Amarsi A.~M.,   Grevesse N.,  2021, \mn@doi [Astronomy and
  Astrophysics] {10.1051/0004-6361/202140445}, 653, A141

\bibitem[\protect\citeauthoryear{{Brekke}, {Thompson}, {Woods}  \&
  {Eparvier}}{{Brekke} et~al.}{2000}]{brekke_etal:00}
{Brekke} P.,  {Thompson} W.~T.,  {Woods} T.~N.,   {Eparvier} F.~G.,  2000,
  \apj, 536, 959

\bibitem[\protect\citeauthoryear{{Chamberlin}, {Woods}  \&
  {Eparvier}}{{Chamberlin} et~al.}{2004}]{2004SPIE.5538...31C}
{Chamberlin} P.~C.,  {Woods} T.~N.,   {Eparvier} F.~G.,  2004, in {Soufli} R.,
  {Seely} J.~F.,  eds,  Society of Photo-Optical Instrumentation Engineers
  (SPIE) Conference Series Vol. 5538, Optical Constants of Materials for UV to
  X-Ray Wavelengths. pp 31--42, \mn@doi{10.1117/12.559874}

\bibitem[\protect\citeauthoryear{{Chamberlin}, {Hock}, {Crotser}, {Eparvier},
  {Furst}, {Triplett}, {Woodraska}  \& {Woods}}{{Chamberlin}
  et~al.}{2007}]{chamberlin_etal_07}
{Chamberlin} P.~C.,  {Hock} R.~A.,  {Crotser} D.~A.,  {Eparvier} F.~G.,
  {Furst} M.,  {Triplett} M.~A.,  {Woodraska} D.~L.,   {Woods} T.~N.,  2007, in
  Society of Photo-Optical Instrumentation Engineers (SPIE) Conference Series.
  , \mn@doi{10.1117/12.734116}

\bibitem[\protect\citeauthoryear{{Chamberlin} et~al.,}{{Chamberlin}
  et~al.}{2020}]{chamberlin_etal:2020}
{Chamberlin} P.~C.,  et~al., 2020, \mn@doi [Space Weather]
  {10.1029/2020SW00258810.1002/essoar.10503732.2}, \href
  {https://ui.adsabs.harvard.edu/abs/2020SpWea..1802588C} {18, e02588}

\bibitem[\protect\citeauthoryear{{Clette}, {Svalgaard}, {Vaquero}  \&
  {Cliver}}{{Clette} et~al.}{2015}]{clette_etal:2015}
{Clette} F.,  {Svalgaard} L.,  {Vaquero} J.~M.,   {Cliver} E.~W.,  2015, in
  {Balogh} A.,  {Hudson} H.,  {Petrovay} K.,   {von Steiger} R.,  eds, ,
  Vol.~53, The Solar Activity Cycle.
p.~35, \mn@doi{10.1007/978-1-4939-2584-1_3}

\bibitem[\protect\citeauthoryear{{Dammasch}, {Wilhelm}, {Curdt}  \&
  {Sch{\"u}hle}}{{Dammasch} et~al.}{1999}]{dammasch_etal:1999}
{Dammasch} I.~E.,  {Wilhelm} K.,  {Curdt} W.,   {Sch{\"u}hle} U.,  1999, in
  {A.~Wilson \& et al.} ed.,  ESA Special Publication Vol. 448, Magnetic Fields
  and Solar Processes. pp 1165--+

\bibitem[\protect\citeauthoryear{{Del Zanna}}{{Del
  Zanna}}{2002}]{delzanna02_issi-ADS}
{Del Zanna} G.,  2002, The Radiometric Calibration of SOHO (ESA SR-002).~Edited
  by A.~Pauluhn, M.C.E.~Huber and R.~von Steiger, p.283, \href
  {http://cdsads.u-strasbg.fr/abs/2002ISSIR...2..283D} {2, 283}

\bibitem[\protect\citeauthoryear{{Del Zanna}}{{Del
  Zanna}}{2019}]{delzanna:2019}
{Del Zanna} G.,  2019, \mn@doi [\aap] {10.1051/0004-6361/201834842}, \href
  {https://ui.adsabs.harvard.edu/abs/2019A&A...624A..36D} {624, A36}

\bibitem[\protect\citeauthoryear{{Del Zanna} \& {Andretta}}{{Del Zanna} \&
  {Andretta}}{2011}]{delzanna_andretta:2011}
{Del Zanna} G.,  {Andretta} V.,  2011, \mn@doi [\aap]
  {10.1051/0004-6361/201016106}, \href
  {http://adsabs.harvard.edu/abs/2011A%26A...528A.139D} {528, A139+}

\bibitem[\protect\citeauthoryear{Del~Zanna \& Andretta}{Del~Zanna \&
  Andretta}{2015b}]{del_zanna_euv_2015}
Del~Zanna G.,  Andretta V.,  2015b, \mn@doi [Astronomy \& Astrophysics]
  {10.1051/0004-6361/201526804}, 584, A29

\bibitem[\protect\citeauthoryear{{Del Zanna} \& {Andretta}}{{Del Zanna} \&
  {Andretta}}{2015a}]{delzanna_andretta:2015}
{Del Zanna} G.,  {Andretta} V.,  2015a, \mn@doi [\aap]
  {10.1051/0004-6361/201526804}, \href
  {http://adsabs.harvard.edu/abs/2015A%26A...584A..29D} {584, A29}

\bibitem[\protect\citeauthoryear{{Del Zanna}, {Bromage}, {Landi}  \&
  {Landini}}{{Del Zanna} et~al.}{2001}]{delzanna01_cdscal}
{Del Zanna} G.,  {Bromage} B.~J.~I.,  {Landi} E.,   {Landini} M.,  2001, \aap,
  379, 708

\bibitem[\protect\citeauthoryear{Del~Zanna, Andretta, Chamberlin, Woods  \&
  Thompson}{Del~Zanna et~al.}{2010a}]{del_zanna_euv_2010}
Del~Zanna G.,  Andretta V.,  Chamberlin P.~C.,  Woods T.~N.,   Thompson W.~T.,
  2010a, \mn@doi [Astronomy and Astrophysics, Volume 518, id.A49,
  {\textless}NUMPAGES{\textgreater}11{\textless}/NUMPAGES{\textgreater} pp.]
  {10.1051/0004-6361/200912904}, 518, A49

\bibitem[\protect\citeauthoryear{{Del Zanna}, {Andretta}, {Chamberlin}, {Woods}
   \& {Thompson}}{{Del Zanna} et~al.}{2010b}]{delzanna_etal:10_cdscal}
{Del Zanna} G.,  {Andretta} V.,  {Chamberlin} P.~C.,  {Woods} T.~N.,
  {Thompson} W.~T.,  2010b, \mn@doi [\aap] {10.1051/0004-6361/200912904}, \href
  {http://adsabs.harvard.edu/abs/2010A%26A...518A..49D} {518, A49+}

\bibitem[\protect\citeauthoryear{{Deliporanidou} \& {Del
  Zanna}}{{Deliporanidou} \& {Del Zanna}}{2024}]{Deliporanidou_2024}
{Deliporanidou} E.,  {Del Zanna} G.,  2024, \mn@doi [\mnras]
  {10.1093/mnras/stae2299}, \href
  {https://ui.adsabs.harvard.edu/abs/2024MNRAS.534.3989D} {534, 3989}

\bibitem[\protect\citeauthoryear{{Dudok de Wit}, {Kretzschmar}, {Aboudarham},
  {Amblard}, {Auch{\`e}re}  \& {Lilensten}}{{Dudok de Wit}
  et~al.}{2008}]{2008AdSpR..42..903D}
{Dudok de Wit} T.,  {Kretzschmar} M.,  {Aboudarham} J.,  {Amblard} P.~O.,
  {Auch{\`e}re} F.,   {Lilensten} J.,  2008, \mn@doi [Advances in Space
  Research] {10.1016/j.asr.2007.04.019}, \href
  {https://ui.adsabs.harvard.edu/abs/2008AdSpR..42..903D} {42, 903}

\bibitem[\protect\citeauthoryear{{Dudok de Wit}, {Kopp}, {Fr{\"o}hlich}  \&
  {Sch{\"o}ll}}{{Dudok de Wit} et~al.}{2017}]{2017GeoRL..44.1196D}
{Dudok de Wit} T.,  {Kopp} G.,  {Fr{\"o}hlich} C.,   {Sch{\"o}ll} M.,  2017,
  \mn@doi [\grl] {10.1002/2016GL071866}, \href
  {https://ui.adsabs.harvard.edu/abs/2017GeoRL..44.1196D} {44, 1196}

\bibitem[\protect\citeauthoryear{{Dufresne}, {Del Zanna}  \&
  {Mason}}{{Dufresne} et~al.}{2023}]{dufresne_etal:2023}
{Dufresne} R.~P.,  {Del Zanna} G.,   {Mason} H.~E.,  2023, \mn@doi [\mnras]
  {10.1093/mnras/stad794}, \href
  {https://ui.adsabs.harvard.edu/abs/2023MNRAS.521.4696D} {521, 4696}

\bibitem[\protect\citeauthoryear{{Dufresne}, {Del Zanna}, {Young}, {Dere},
  {Deliporanidou}, {Barnes}  \& {Landi}}{{Dufresne}
  et~al.}{2024}]{dufresne_chianti_2024}
{Dufresne} R.~P.,  {Del Zanna} G.,  {Young} P.~R.,  {Dere} K.~P.,
  {Deliporanidou} E.,  {Barnes} W.~T.,   {Landi} E.,  2024, \mn@doi [\apj]
  {10.3847/1538-4357/ad6765}, \href
  {https://ui.adsabs.harvard.edu/abs/2024ApJ...974...71D} {974, 71}

\bibitem[\protect\citeauthoryear{{Dupree} \& {Reeves}}{{Dupree} \&
  {Reeves}}{1971}]{dupree_reeves:1971}
{Dupree} A.~K.,  {Reeves} E.~M.,  1971, \mn@doi [\apj] {10.1086/150924}, \href
  {https://ui.adsabs.harvard.edu/abs/1971ApJ...165..599D} {165, 599}

\bibitem[\protect\citeauthoryear{{Fontenla}, {White}, {Fox}, {Avrett}  \&
  {Kurucz}}{{Fontenla} et~al.}{1999}]{Fontenla_1999}
{Fontenla} J.,  {White} O.~R.,  {Fox} P.~A.,  {Avrett} E.~H.,   {Kurucz} R.~L.,
   1999, \mn@doi [\apj] {10.1086/307258}, \href
  {https://ui.adsabs.harvard.edu/abs/1999ApJ...518..480F} {518, 480}

\bibitem[\protect\citeauthoryear{{Harrison} \& {et}}{{Harrison} \&
  {et}}{1995}]{harrison95}
{Harrison} R.~A.,  {et} a.,  1995, \solphys, 162, 233

\bibitem[\protect\citeauthoryear{Harrison et~al.}{Harrison
  et~al.}{1995}]{Harrison-etal:95}
Harrison R.~A.,  et~al., 1995, Sol.~Phys., 162, 233

\bibitem[\protect\citeauthoryear{{Heroux} \& {Higgins}}{{Heroux} \&
  {Higgins}}{1977}]{heroux_higgins:1977}
{Heroux} L.,  {Higgins} J.~E.,  1977, \mn@doi [\jgr] {10.1029/JA082i022p03307},
  \href {http://adsabs.harvard.edu/abs/1977JGR....82.3307H} {82, 3307}

\bibitem[\protect\citeauthoryear{{Heroux} \& {Hinteregger}}{{Heroux} \&
  {Hinteregger}}{1978}]{heroux_hinteregger:1978}
{Heroux} L.,  {Hinteregger} H.~E.,  1978, \mn@doi [\jgr]
  {10.1029/JA083iA11p05305}, \href
  {https://ui.adsabs.harvard.edu/abs/1978JGR....83.5305H} {83, 5305}

\bibitem[\protect\citeauthoryear{{Heroux}, {Cohen}  \& {Higgins}}{{Heroux}
  et~al.}{1974}]{heroux_etal:1974}
{Heroux} L.,  {Cohen} M.,   {Higgins} J.~E.,  1974, \mn@doi [\jgr]
  {10.1029/JA079i034p05237}, \href
  {http://adsabs.harvard.edu/abs/1974JGR....79.5237H} {79, 5237}

\bibitem[\protect\citeauthoryear{Higgins}{Higgins}{1976}]{higgins_solar_1976}
Higgins J.~E.,  1976, \mn@doi [Journal of Geophysical Research]
  {10.1029/JA081i007p01301}, 81, 1301

\bibitem[\protect\citeauthoryear{{Hinteregger}, {Fukui}  \&
  {Gilson}}{{Hinteregger} et~al.}{1981}]{Hinteregger_1981}
{Hinteregger} H.~E.,  {Fukui} K.,   {Gilson} B.~R.,  1981, \mn@doi [\grl]
  {10.1029/GL008i011p01147}, \href
  {https://ui.adsabs.harvard.edu/abs/1981GeoRL...8.1147H} {8, 1147}

\bibitem[\protect\citeauthoryear{{Hock}, {Chamberlin}, {Woods}, {Crotser},
  {Eparvier}, {Woodraska}  \& {Woods}}{{Hock} et~al.}{2010}]{hock_etal_09}
{Hock} R.~A.,  {Chamberlin} P.~C.,  {Woods} T.~N.,  {Crotser} D.,  {Eparvier}
  F.~G.,  {Woodraska} D.~L.,   {Woods} E.~C.,  2010, \mn@doi [\solphys]
  {10.1007/s11207-010-9520-9}, \href
  {http://adsabs.harvard.edu/abs/2010SoPh..tmp...37H} {pp 37--+}

\bibitem[\protect\citeauthoryear{{Hock}, {Chamberlin}, {Woods}, {Crotser},
  {Eparvier}, {Woodraska}  \& {Woods}}{{Hock} et~al.}{2012}]{hock_etal:2012}
{Hock} R.~A.,  {Chamberlin} P.~C.,  {Woods} T.~N.,  {Crotser} D.,  {Eparvier}
  F.~G.,  {Woodraska} D.~L.,   {Woods} E.~C.,  2012, \mn@doi [\solphys]
  {10.1007/s11207-010-9520-9}, \href
  {http://adsabs.harvard.edu/abs/2012SoPh..275..145H} {275, 145}

\bibitem[\protect\citeauthoryear{{Lean}, {White}, {Livingston}  \&
  {Picone}}{{Lean} et~al.}{2001}]{Lean_2001}
{Lean} J.~L.,  {White} O.~R.,  {Livingston} W.~C.,   {Picone} J.~M.,  2001,
  \mn@doi [\jgr] {10.1029/2000JA000340}, \href
  {https://ui.adsabs.harvard.edu/abs/2001JGR...10610645L} {106, 10645}

\bibitem[\protect\citeauthoryear{Lean, Woods, Eparvier, Meier, Strickland,
  Correira  \& Evans}{Lean et~al.}{2011a}]{lean_solar_2011}
Lean J.~L.,  Woods T.~N.,  Eparvier F.~G.,  Meier R.~R.,  Strickland D.~J.,
  Correira J.~T.,   Evans J.~S.,  2011a, \mn@doi [Journal of Geophysical
  Research: Space Physics] {10.1029/2010JA015901}, 116

\bibitem[\protect\citeauthoryear{{Lean}, {Woods}, {Eparvier}, {Meier},
  {Strickland}, {Correira}  \& {Evans}}{{Lean} et~al.}{2011b}]{lean_etal:2011}
{Lean} J.~L.,  {Woods} T.~N.,  {Eparvier} F.~G.,  {Meier} R.~R.,  {Strickland}
  D.~J.,  {Correira} J.~T.,   {Evans} J.~S.,  2011b, \mn@doi [Journal of
  Geophysical Research (Space Physics)] {10.1029/2010JA015901}, \href
  {https://ui.adsabs.harvard.edu/abs/2011JGRA..116.1102L} {116, A01102}

\bibitem[\protect\citeauthoryear{{Lezzi}, {Andretta}, {Murabito}  \& {Del
  Zanna}}{{Lezzi} et~al.}{2023}]{Lezzi_2023}
{Lezzi} S.~M.,  {Andretta} V.,  {Murabito} M.,   {Del Zanna} G.,  2023, \mn@doi
  [\aap] {10.1051/0004-6361/202347414}, \href
  {https://ui.adsabs.harvard.edu/abs/2023A&A...680A..61L} {680, A61}

\bibitem[\protect\citeauthoryear{Linsky}{Linsky}{2019}]{linsky_host_2019}
Linsky J.,  2019, \mn@doi [Lecture Notes in Physics, Berlin Springer Verlag]
  {10.1007/978-3-030-11452-7}, 955

\bibitem[\protect\citeauthoryear{{Livingston}, {Wallace}, {White}  \&
  {Giampapa}}{{Livingston} et~al.}{2007}]{livingston_etal:07}
{Livingston} W.,  {Wallace} L.,  {White} O.~R.,   {Giampapa} M.~S.,  2007,
  \mn@doi [\apj] {10.1086/511127}, \href
  {http://adsabs.harvard.edu/abs/2007ApJ...657.1137L} {657, 1137}

\bibitem[\protect\citeauthoryear{{Livingston}, {White}, {Wallace}  \&
  {Harvey}}{{Livingston} et~al.}{2010}]{livingston_etal:2010}
{Livingston} W.,  {White} O.~R.,  {Wallace} L.,   {Harvey} J.,  2010, \memsai,
  \href {https://ui.adsabs.harvard.edu/abs/2010MmSAI..81..643L} {81, 643}

\bibitem[\protect\citeauthoryear{Malinovsky \& Heroux}{Malinovsky \&
  Heroux}{1973}]{malinovsky_heroux:1973}
Malinovsky L.,  Heroux M.,  1973, \apj, \href
  {http://adsabs.harvard.edu/abs/1973ApJ...181.1009M} {181, 1009}

\bibitem[\protect\citeauthoryear{Malinovsky, Heroux, Malinovsky  \&
  Heroux}{Malinovsky et~al.}{1973}]{malinovsky_analysis_1973}
Malinovsky L.,  Heroux M.,  Malinovsky L.,   Heroux M.,  1973, \mn@doi [ApJ]
  {10.1086/152108}, 181, 1009

\bibitem[\protect\citeauthoryear{Namekata, Toriumi, Airapetian, Shoda, Watanabe
   \& Notsu}{Namekata et~al.}{2023}]{namekata_reconstructing_2023}
Namekata K.,  Toriumi S.,  Airapetian V.~S.,  Shoda M.,  Watanabe K.,   Notsu
  Y.,  2023, \mn@doi [The Astrophysical Journal] {10.3847/1538-4357/acbe38},
  945, 147

\bibitem[\protect\citeauthoryear{{Oster}}{{Oster}}{1983}]{oster:1983}
{Oster} L.,  1983, \mn@doi [\jgr] {10.1029/JA088iA11p09037}, \href
  {https://ui.adsabs.harvard.edu/abs/1983JGR....88.9037O} {88, 9037}

\bibitem[\protect\citeauthoryear{{Reeves} \& {Parkinson}}{{Reeves} \&
  {Parkinson}}{1970a}]{reeves_parkinson:1970a}
{Reeves} E.~M.,  {Parkinson} W.~H.,  1970a, \mn@doi [\ao]
  {10.1364/AO.9.001201}, \href
  {https://ui.adsabs.harvard.edu/abs/1970ApOpt...9.1201R} {9, 1201}

\bibitem[\protect\citeauthoryear{{Reeves} \& {Parkinson}}{{Reeves} \&
  {Parkinson}}{1970b}]{reeves_parkinson:1970b}
{Reeves} E.~M.,  {Parkinson} W.~H.,  1970b, \mn@doi [\apjs] {10.1086/190217},
  \href {https://ui.adsabs.harvard.edu/abs/1970ApJS...21....1R} {21, 1}

\bibitem[\protect\citeauthoryear{Reeves, Timothy, Withbroe  \& Huber}{Reeves
  et~al.}{1977}]{reeves_etal:1977b}
Reeves E.~M.,  Timothy J.~G.,  Withbroe G.~L.,   Huber M. C.~E.,  1977, \ao,
  \href {http://adsabs.harvard.edu/abs/1977ApOpt..16..849R} {16, 849}

\bibitem[\protect\citeauthoryear{{Richards}, {Fennelly}  \& {Torr}}{{Richards}
  et~al.}{1994}]{Richards_1994}
{Richards} P.~G.,  {Fennelly} J.~A.,   {Torr} D.~G.,  1994, \mn@doi [\jgr]
  {10.1029/94JA00518}, \href
  {https://ui.adsabs.harvard.edu/abs/1994JGR....99.8981R} {99, 8981}

\bibitem[\protect\citeauthoryear{Sanz-Forcada, Micela, Ribas, Pollock, Eiroa,
  Velasco, Solano  \& García-Álvarez}{Sanz-Forcada
  et~al.}{2011}]{sanz_forcada_estimation_2011}
Sanz-Forcada J.,  Micela G.,  Ribas I.,  Pollock A. M.~T.,  Eiroa C.,  Velasco
  A.,  Solano E.,   García-Álvarez D.,  2011, \mn@doi [Astronomy and
  Astrophysics] {10.1051/0004-6361/201116594}, 532, A6

\bibitem[\protect\citeauthoryear{{Schonfeld}, {White}, {Henney}, {Arge}  \&
  {McAteer}}{{Schonfeld} et~al.}{2015}]{2015ApJ...808...29S}
{Schonfeld} S.~J.,  {White} S.~M.,  {Henney} C.~J.,  {Arge} C.~N.,   {McAteer}
  R.~T.~J.,  2015, \mn@doi [\apj] {10.1088/0004-637X/808/1/29}, \href
  {https://ui.adsabs.harvard.edu/abs/2015ApJ...808...29S} {808, 29}

\bibitem[\protect\citeauthoryear{{Solomon} \& {Qian}}{{Solomon} \&
  {Qian}}{2005}]{Solomon_Qian_2005}
{Solomon} S.~C.,  {Qian} L.,  2005, \mn@doi [Journal of Geophysical Research
  (Space Physics)] {10.1029/2005JA011160}, \href
  {https://ui.adsabs.harvard.edu/abs/2005JGRA..11010306S} {110, A10306}

\bibitem[\protect\citeauthoryear{{Svalgaard} \& {Sun}}{{Svalgaard} \&
  {Sun}}{2016}]{2016usc..confE..28S}
{Svalgaard} L.,  {Sun} X.,  2016, in {Pesnell} W.~D.,  {Thompson} B.,  eds, SDO
  2016: Unraveling the Sun's Complexity. p.~28

\bibitem[\protect\citeauthoryear{Tapping}{Tapping}{2013}]{tapping2013}
Tapping K.~F.,  2013, \mn@doi [Space Weather] {10.1002/swe.20064}, 11, 394

\bibitem[\protect\citeauthoryear{{Timothy} \& {Timothy}}{{Timothy} \&
  {Timothy}}{1970}]{timothy_timothy:1970}
{Timothy} A.~F.,  {Timothy} J.~G.,  1970, \mn@doi [\jgr]
  {10.1029/JA075i034p06950}, \href
  {http://adsabs.harvard.edu/abs/1970JGR....75.6950T} {75, 6950}

\bibitem[\protect\citeauthoryear{{Timothy}, {Chambers}, {Dentremont}, {Lanham}
  \& {Reeves}}{{Timothy} et~al.}{1975}]{timothy_etal:1975}
{Timothy} J.~G.,  {Chambers} R.~M.,  {Dentremont} A.~M.,  {Lanham} N.~W.,
  {Reeves} E.~M.,  1975, Space Science Instrumentation, \href
  {https://ui.adsabs.harvard.edu/abs/1975SSI.....1...23T} {1, 23}

\bibitem[\protect\citeauthoryear{{Tobiska}, {Woods}, {Eparvier}, {Viereck},
  {Floyd}, {Bouwer}, {Rottman}  \& {White}}{{Tobiska}
  et~al.}{2000}]{Tobiska_2000}
{Tobiska} W.~K.,  {Woods} T.,  {Eparvier} F.,  {Viereck} R.,  {Floyd} L.,
  {Bouwer} D.,  {Rottman} G.,   {White} O.~R.,  2000, \mn@doi [Journal of
  Atmospheric and Solar-Terrestrial Physics] {10.1016/S1364-6826(00)00070-5},
  \href {https://ui.adsabs.harvard.edu/abs/2000JASTP..62.1233T} {62, 1233}

\bibitem[\protect\citeauthoryear{Toriumi, Airapetian, Namekata  \&
  Notsu}{Toriumi et~al.}{2022}]{toriumi_universal_2022}
Toriumi S.,  Airapetian V.~S.,  Namekata K.,   Notsu Y.,  2022, \mn@doi [The
  Astrophysical Journal Supplement Series] {10.3847/1538-4365/ac8b15}, 262, 46

\bibitem[\protect\citeauthoryear{{Vernazza} \& {Reeves}}{{Vernazza} \&
  {Reeves}}{1978}]{vernazza_reeves:1978}
{Vernazza} J.~E.,  {Reeves} E.~M.,  1978, \mn@doi [\apjs] {10.1086/190539},
  \href {http://adsabs.harvard.edu/abs/1978ApJS...37..485V} {37, 485}

\bibitem[\protect\citeauthoryear{{Vourlidas} \& {Bruinsma}}{{Vourlidas} \&
  {Bruinsma}}{2018}]{vourlidas_2018}
{Vourlidas} A.,  {Bruinsma} S.,  2018, \mn@doi [Space Weather]
  {10.1002/2017SW001725}, \href
  {https://ui.adsabs.harvard.edu/abs/2018SpWea..16....5V} {16, 5}

\bibitem[\protect\citeauthoryear{{Warren}}{{Warren}}{2005}]{warren:2005}
{Warren} H.~P.,  2005, \mn@doi [\apjs] {10.1086/427171}, \href
  {http://ukads.nottingham.ac.uk/abs/2005ApJS..157..147W} {157, 147}

\bibitem[\protect\citeauthoryear{{Wilhelm} et~al.,}{{Wilhelm}
  et~al.}{1995}]{wilhelm95}
{Wilhelm} K.,  et~al., 1995, \solphys, 162, 189

\bibitem[\protect\citeauthoryear{{Wilhelm} et~al.,}{{Wilhelm}
  et~al.}{1998a}]{wilhelm_etal:98}
{Wilhelm} K.,  et~al., 1998a, \aap, \href
  {http://adsabs.harvard.edu/abs/1998A%26A...334..685W} {334, 685}

\bibitem[\protect\citeauthoryear{{Wilhelm} et~al.,}{{Wilhelm}
  et~al.}{1998b}]{wilhelm_etal:1998a}
{Wilhelm} K.,  et~al., 1998b, \aap, \href
  {http://adsabs.harvard.edu/abs/1998A%26A...334..685W} {334, 685}

\bibitem[\protect\citeauthoryear{{Woods} \& {Rottman}}{{Woods} \&
  {Rottman}}{1990}]{1990JGR....95.6227W}
{Woods} T.~N.,  {Rottman} G.~J.,  1990, \mn@doi [\jgr]
  {10.1029/JA095iA05p06227}, \href
  {https://ui.adsabs.harvard.edu/abs/1990JGR....95.6227W} {95, 6227}

\bibitem[\protect\citeauthoryear{{Woods}, {Rottman}, {Bailey}, {Solomon}  \&
  {Worden}}{{Woods} et~al.}{1998}]{woods_etal:1998}
{Woods} T.~N.,  {Rottman} G.~J.,  {Bailey} S.~M.,  {Solomon} S.~C.,   {Worden}
  J.~R.,  1998, \mn@doi [\solphys] {10.1023/A:1004912310883}, \href
  {https://ui.adsabs.harvard.edu/abs/1998SoPh..177..133W} {177, 133}

\bibitem[\protect\citeauthoryear{{Woods} et~al.,}{{Woods}
  et~al.}{2005}]{woods_etal:2005}
{Woods} T.~N.,  et~al., 2005, \mn@doi [Journal of Geophysical Research (Space
  Physics)] {10.1029/2004JA010765}, \href
  {http://adsabs.harvard.edu/abs/2005JGRA..11001312W} {110, 1312}

\bibitem[\protect\citeauthoryear{{Woods} et~al.,}{{Woods}
  et~al.}{2009}]{woods_etal:2009}
{Woods} T.~N.,  et~al., 2009, \mn@doi [\grl] {10.1029/2008GL036373}, \href
  {http://adsabs.harvard.edu/abs/2009GeoRL..3601101W} {36, 1101}

\bibitem[\protect\citeauthoryear{{Woods} et~al.,}{{Woods}
  et~al.}{2012a}]{woods_etal:2012}
{Woods} T.~N.,  et~al., 2012a, \mn@doi [\solphys] {10.1007/s11207-009-9487-6},
  \href {http://adsabs.harvard.edu/abs/2012SoPh..275..115W} {275, 115}

\bibitem[\protect\citeauthoryear{{Woods} et~al.,}{{Woods}
  et~al.}{2012b}]{Woods2012_EVE}
{Woods} T.~N.,  et~al., 2012b, \mn@doi [\solphys] {10.1007/s11207-009-9487-6},
  \href {https://ui.adsabs.harvard.edu/abs/2012SoPh..275..115W} {275, 115}

\makeatother
\end{thebibliography}




\appendix

\section{MEGS-A and MEGS-B Plots}
\label{append:extra plots}

\begin{figure*}
    \centerline{
        \includegraphics[width=9cm, keepaspectratio, trim=3cm 1.5cm 3cm 1.5cm, clip]{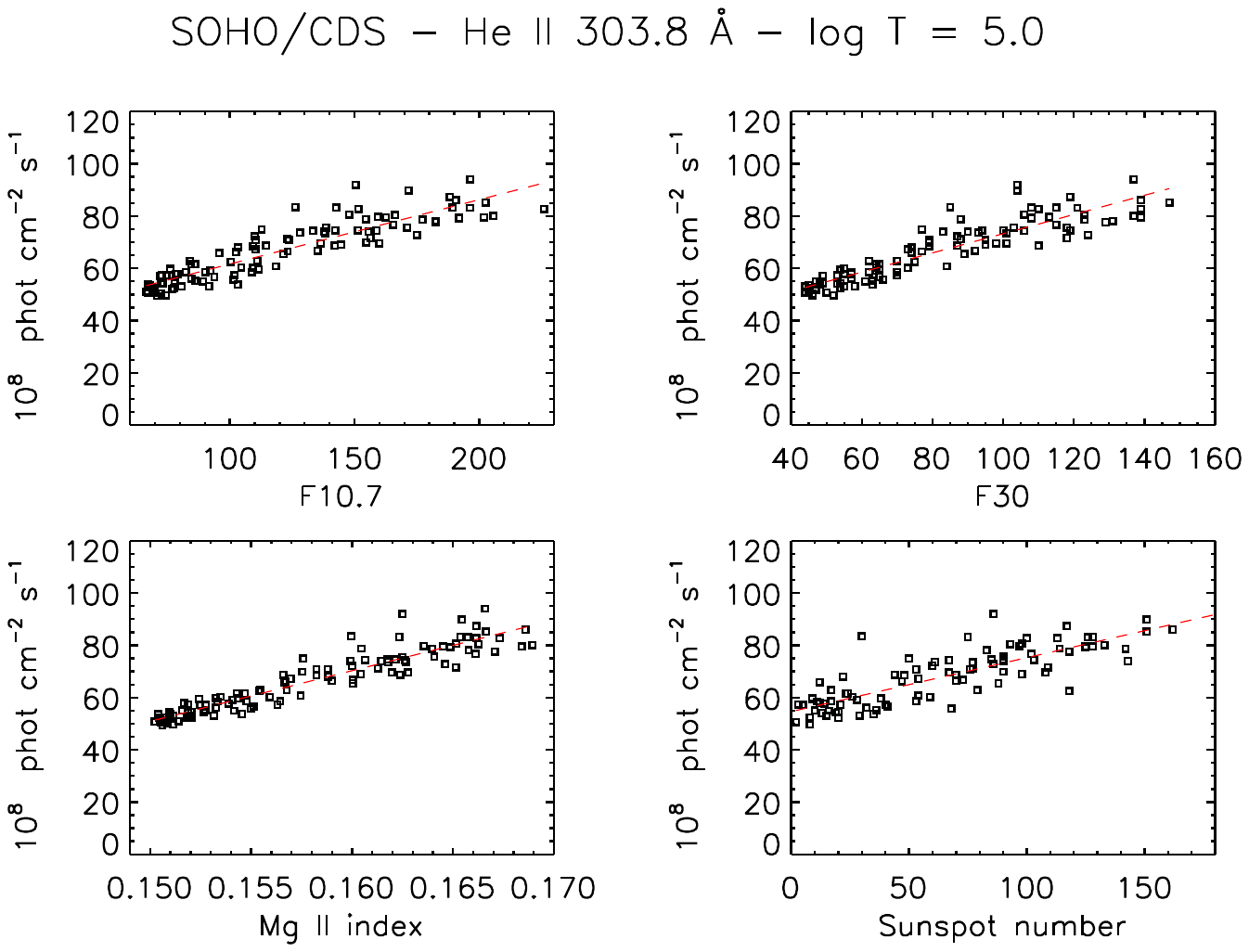}
        \includegraphics[width=9cm, keepaspectratio, trim=3cm 1.5cm 3cm 1.5cm, clip]{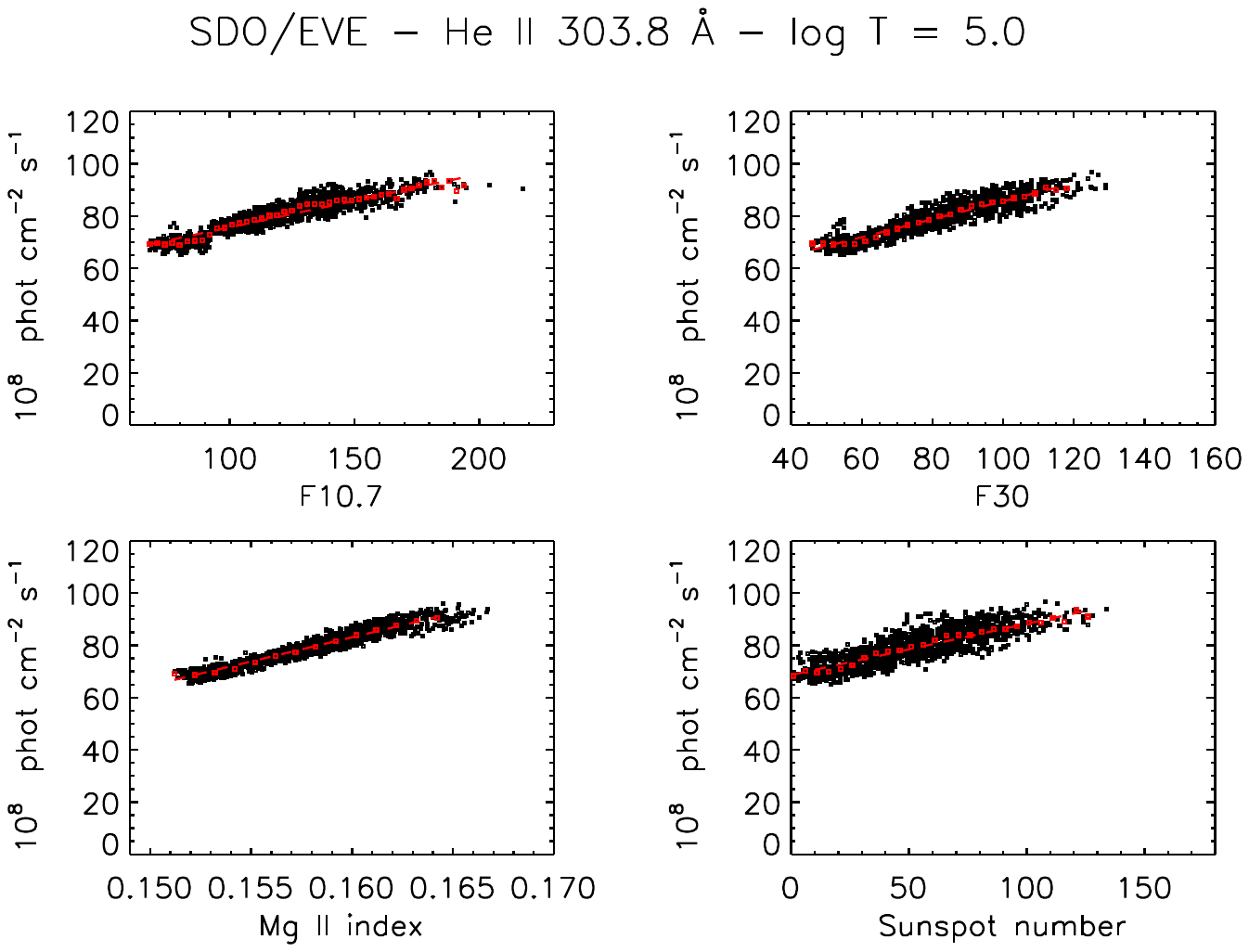}}        
    \caption{Correlation between the irradiances of SOHO/CDS and SDO/EVE for the 304 \,\AA\ \ion{He}{ii} with the daily F10.7 and F30 cm radio fluxes (\(10^{-22}\,\mathrm{W\,m^{-2}\,Hz^{-1}}\)), the Mg II index, and the Sunspot Number.}
    \label{fig:he_ii_304}
\end{figure*}

\begin{figure*}
  \centerline{
    \includegraphics[width=9cm, keepaspectratio, trim=3cm 1.5cm 3cm 1.5cm, clip]{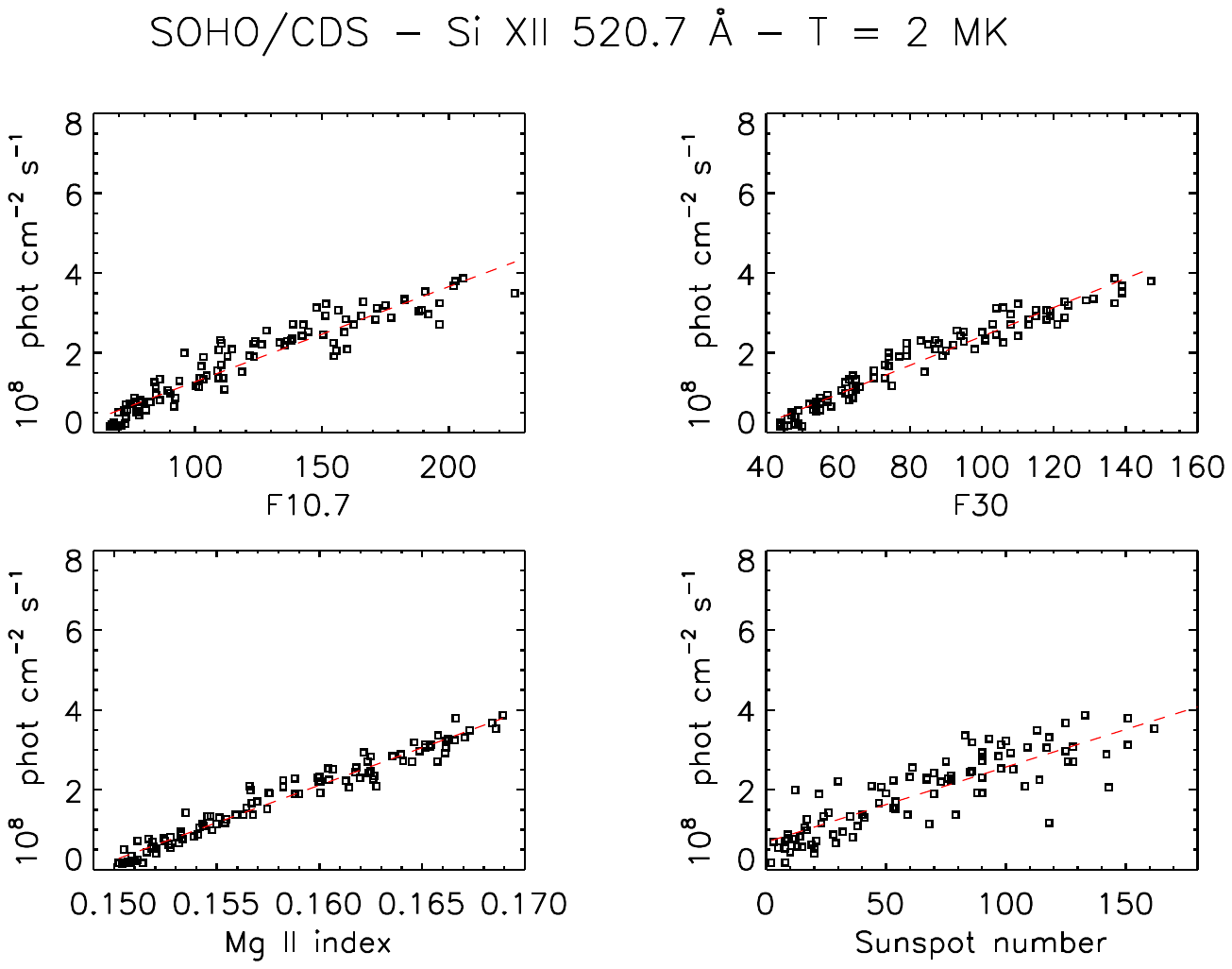}
    \includegraphics[width=9cm, keepaspectratio, trim=3cm 1.5cm 3cm 1.5cm, clip]{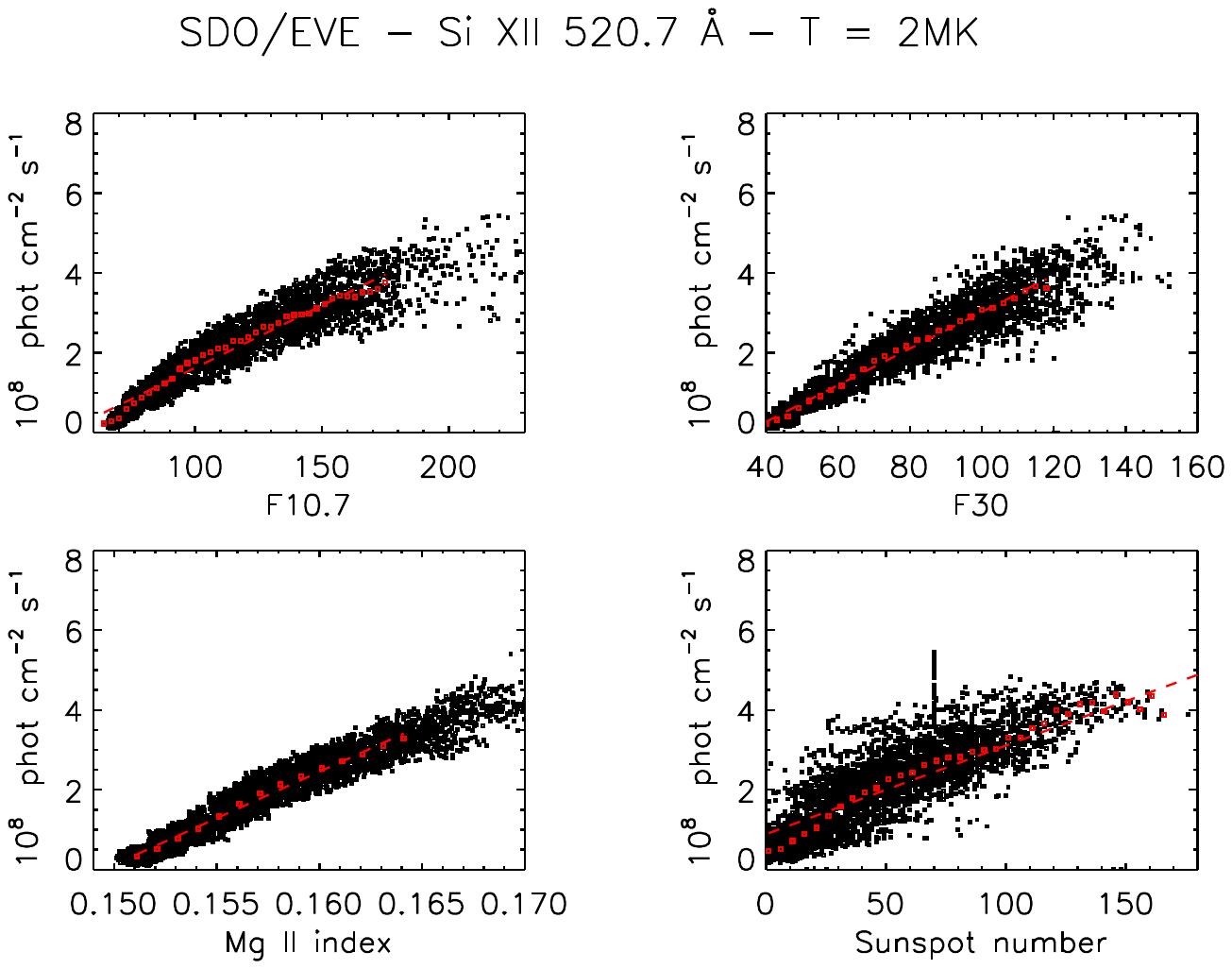}}
    
    \caption{Correlation between the irradiances of SOHO/CDS and SDO/EVE for the 520.7\,\AA\ \ion{Si}{xii} hot coronal line, with the daily F10.7 and F30 cm radio fluxes (\(10^{-22}\,\mathrm{W\,m^{-2}\,Hz^{-1}}\)), the Mg II index, and the Sunspot Number.}
    \label{fig:si_520}
\end{figure*}

This appendix presents further plots of irradiances as a function of the four activity proxies analysed in this study. These include spectral lines from both the MEGS-A and MEGS-B datasets from SDO EVE, as well as the respective plots for the same line from SOHO CDS. 

Fig.~\ref{fig:he_ii_304} shows the 304 \,\AA\ \ion{He}{ii} line using CDS data on the LHS and EVE MEGS-A data on the RHS. 
The difference in the absolute scale is due to the 
fact that CDS resolves the \ion{He}{ii} lines from the 
\ion{Si}{xi} line, whilst EVE does not. Otherwise the 
CDS and EVE agree, as shown previously by e.g. 
\cite{del_zanna_euv_2010}. 

As discussed earlier in the paper and in previous studies, the He and H lines are known to exhibit variability that is intermediate between TR and coronal lines, due to their formation in the chromosphere-corona transition layer, where the effects of active regions are significant but not as dominant as in coronal lines.
(e.g., \cite{Fontenla_1999}; \cite{Lean_2001}). This also emphasizes the \ion{He}{ii} line’s utility as a diagnostic for solar activity and its sensitivity to both short-term and long-term solar cycle changes.

Fig.~\ref{fig:si_520} shows the 520.7\,\AA\ \ion{Si}{xii} coronal line formed at around 2 MK. As expected, this line varies significantly with the proxies of activity and in addition it appears slightly stronger on the EVE data set comparing to the CDS. 

\begin{figure*}
      \centerline{
        \includegraphics[width=9cm, keepaspectratio, trim=3cm 1.5cm 3cm 1.5cm, clip]{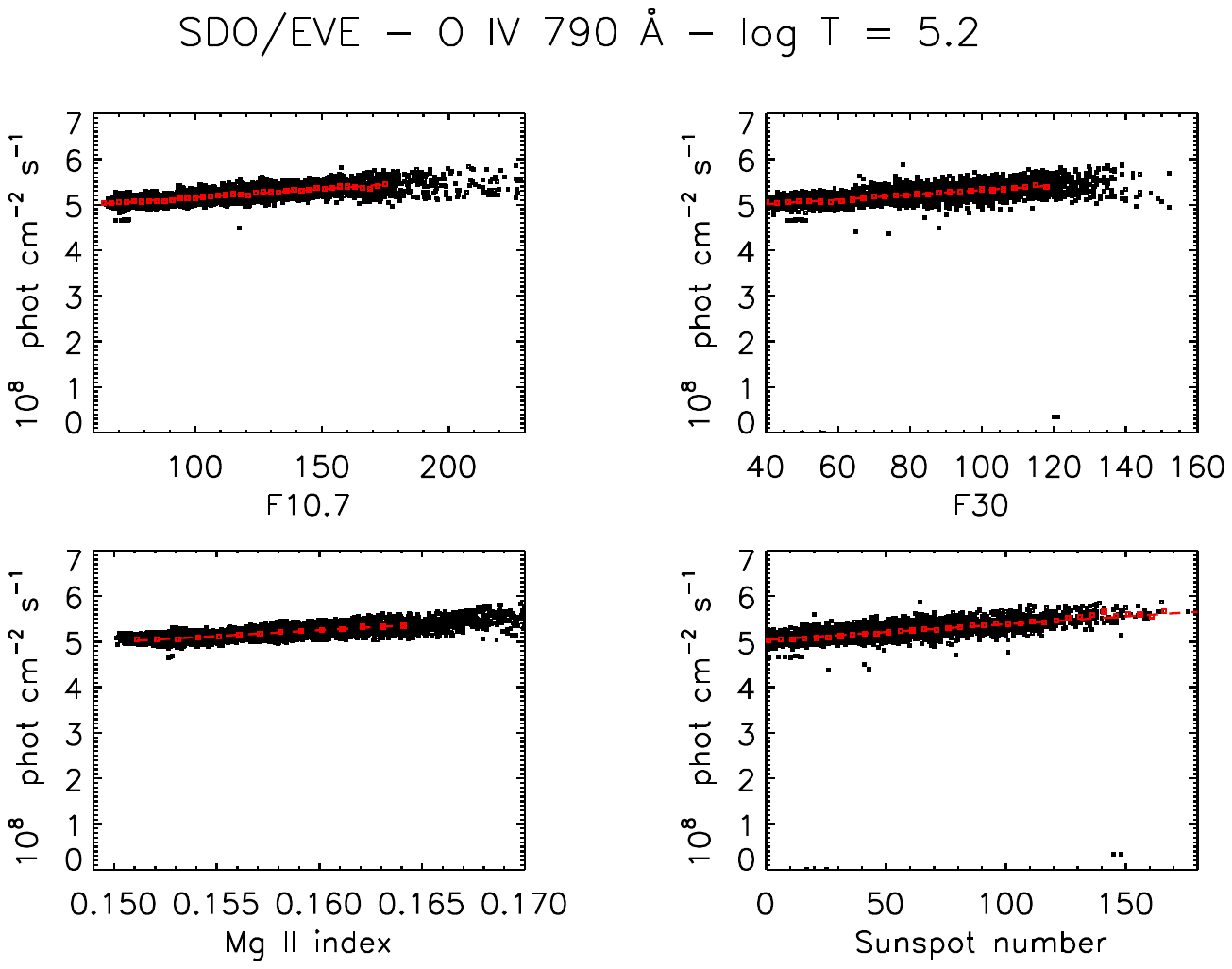}
        \includegraphics[width=9cm, keepaspectratio, trim=3cm 1.5cm 3cm 1.5cm, clip]{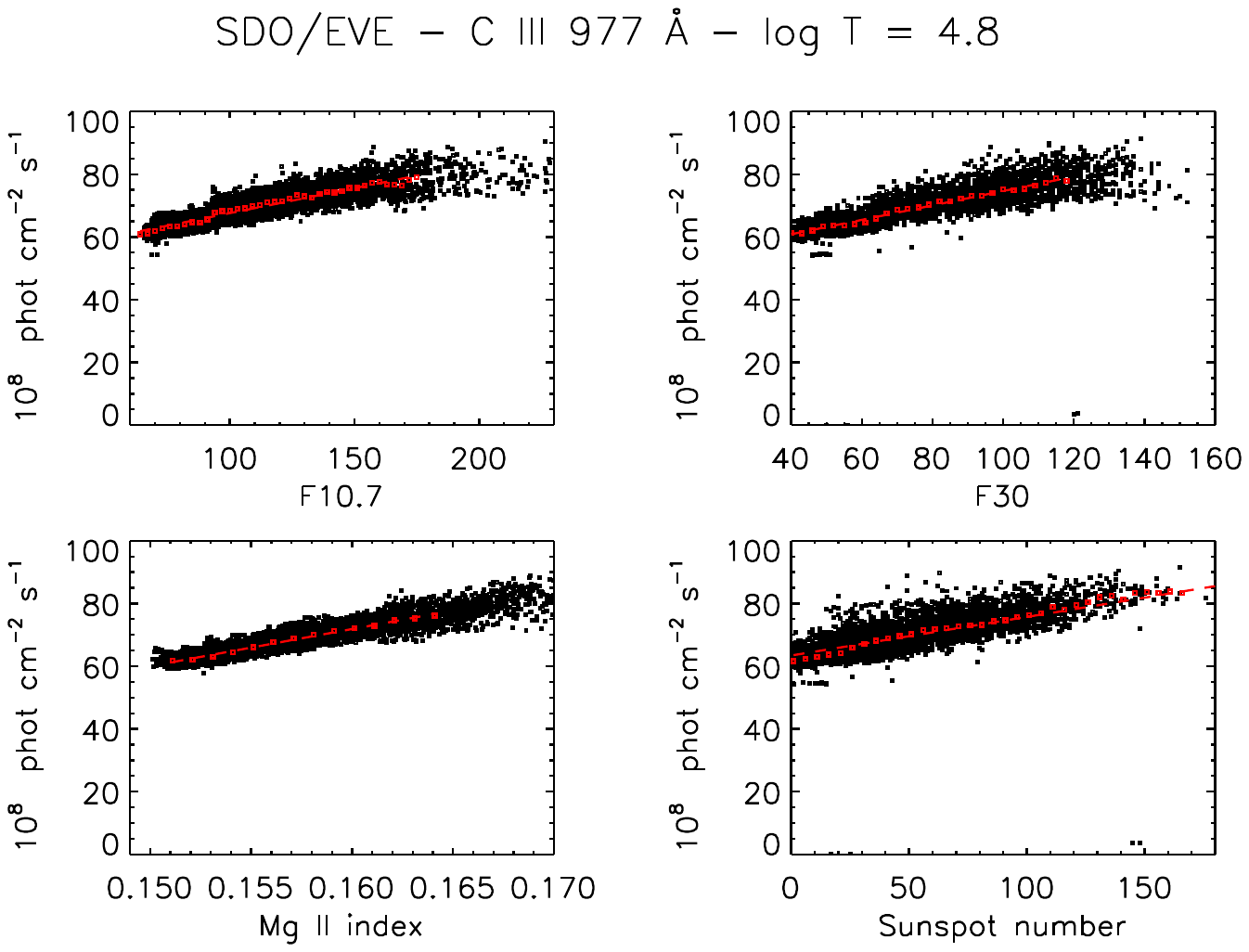}}       

         \centerline{
        \includegraphics[width=9cm, keepaspectratio, trim=3cm 1.5cm 3cm 1.5cm, clip]{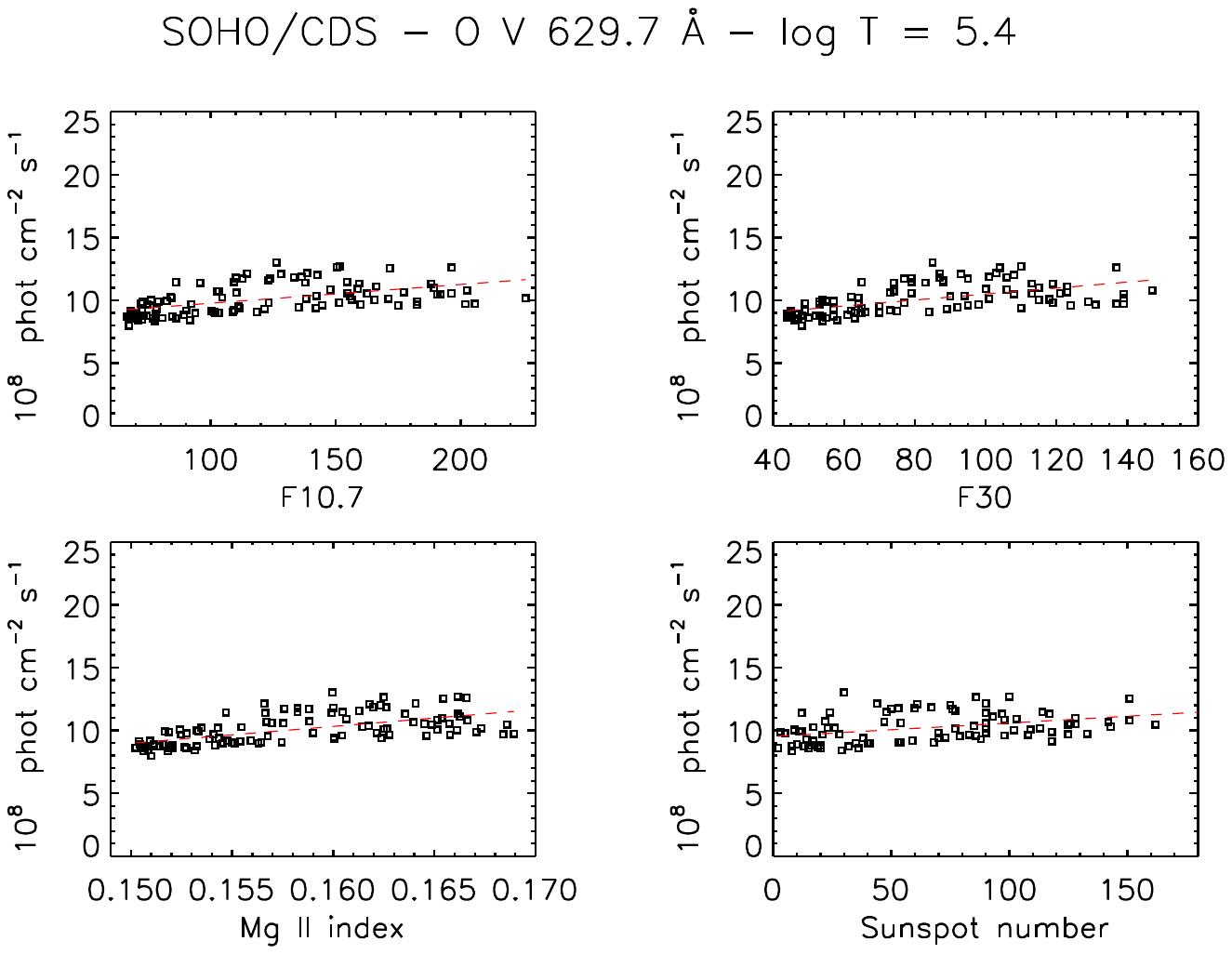}
        \includegraphics[width=9cm, keepaspectratio, trim=3cm 1.5cm 3cm 1.5cm, clip]{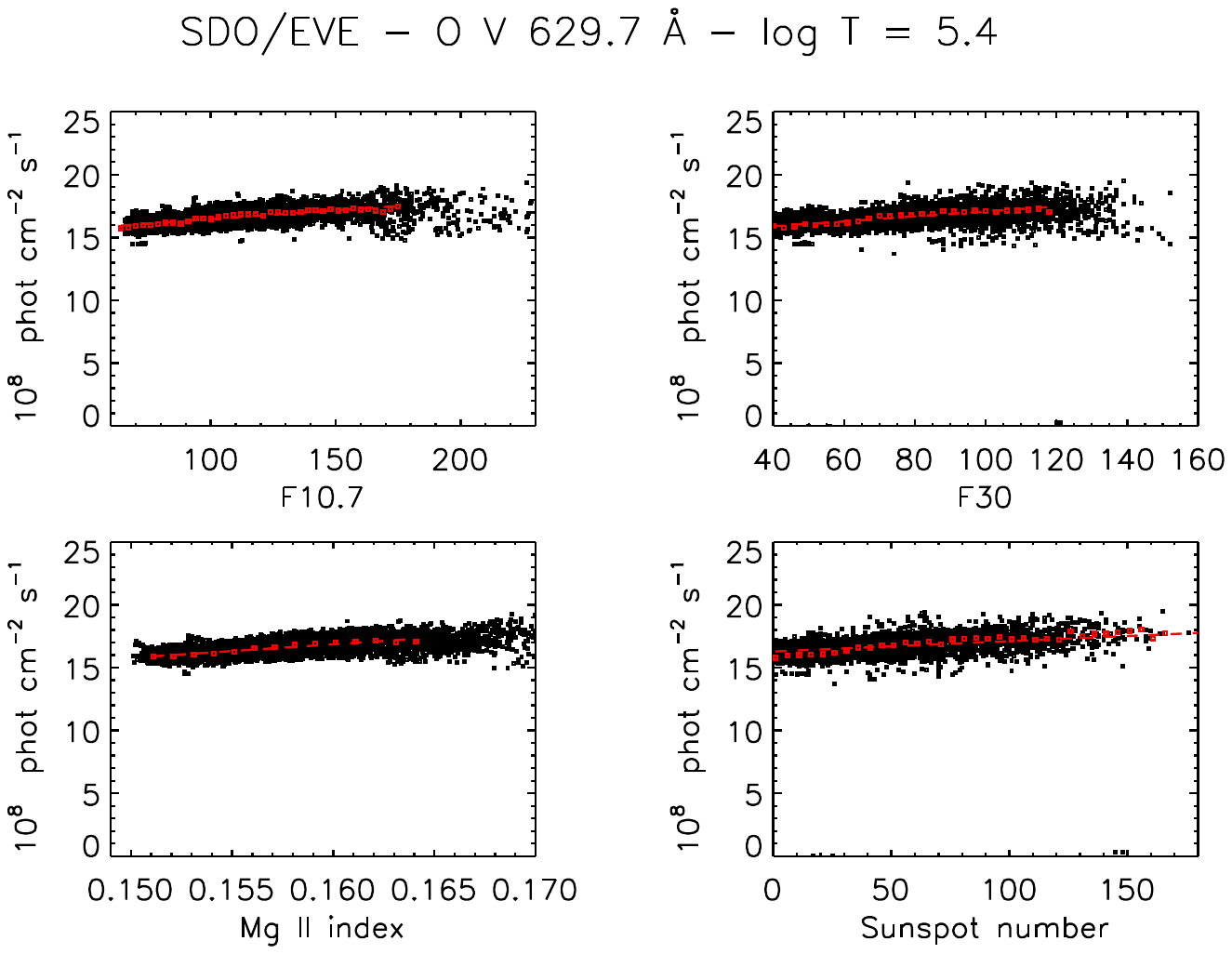}}       
    \caption{Correlation between the irradiances of SOHO/CDS for the 630\,\AA\ \ion{O}{v} and SDO/EVE for the 630\,\AA\ \ion{O}{v}, 790\, \AA\ \ion{O}{iv} and 977\, \AA\ \ion{C}{iii} transition region lines, with the daily F10.7 and F30 cm radio fluxes (\(10^{-22}\,\mathrm{W\,m^{-2}\,Hz^{-1}}\)), the Mg II index, and the Sunspot Number.}
    \label{fig:630_790_977}
\end{figure*}

Fig.~\ref{fig:630_790_977} shows three additional TR lines, which are the 630\,\AA\ \ion{O}{v}, the 790\, \AA\ \ion{O}{iv} and the 977\, \AA\ \ion{C}{iii}. The 977\, \AA\ \ion{C}{iii} on the upper RHS forms at approximately 80.000 K in the lower TR and it exhibits moderate variability with proxies of activity, compared with the 630\,\AA\ \ion{O}{v} and the 790\, \AA\ \ion{O}{iv} which form at higher temperatures but exhibit even less variability. 
It remains to be established if such variation in 
the \ion{C}{iii} is real, see below 
where we compare it with  historical records.

Also, it remains to be established if the CDS or the EVE calibration
for the 630\,\AA\ \ion{O}{v} line are correct. We recall that 
the CDS calibration is based on a  LASP sounding rocket 
\citep{brekke_etal:00} which was calibrated on the ground as the EVE 
and PEVE. The CDS irradiances agree with the LASP measurement but not
with EVE or PEVE. The historical records also show a large scatter of values. The Skylab quiet Sun observations are consistent with the 
LASP/CDS, as some other previous measurements, but not all.

On a side note,  lines formed at similar 
temperatures could in principle have a different behaviour 
with the proxies: proper modelling needs to take into account other factors such as photo-ionization and time-dependent ionization, just
to mention two.  

\section{Historical Records}
\label{append:Historical Records}

\begin{figure*}
\centerline{
        \includegraphics[angle=-90, width=6cm]{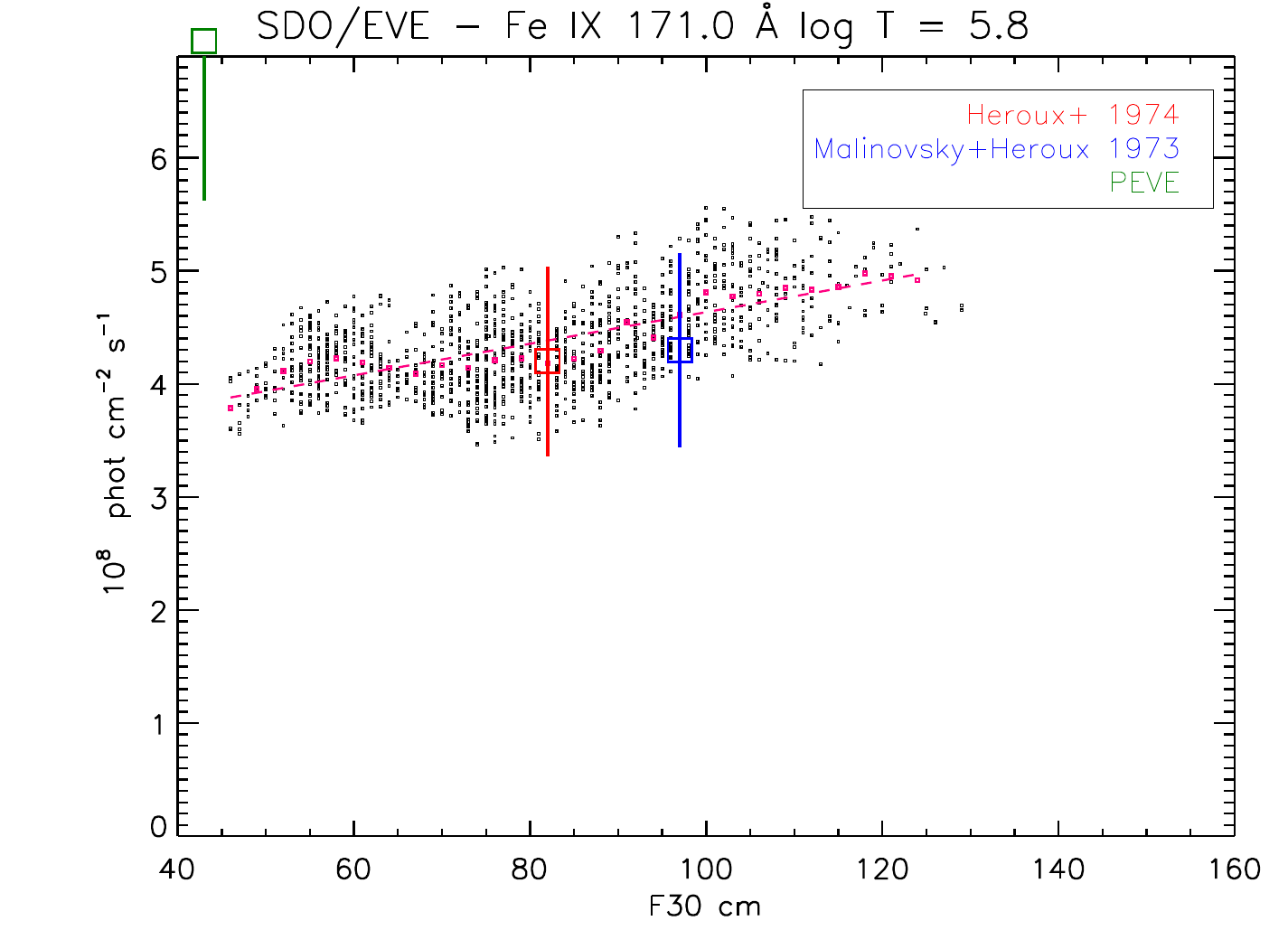}
        \includegraphics[angle=-90, width=6cm]{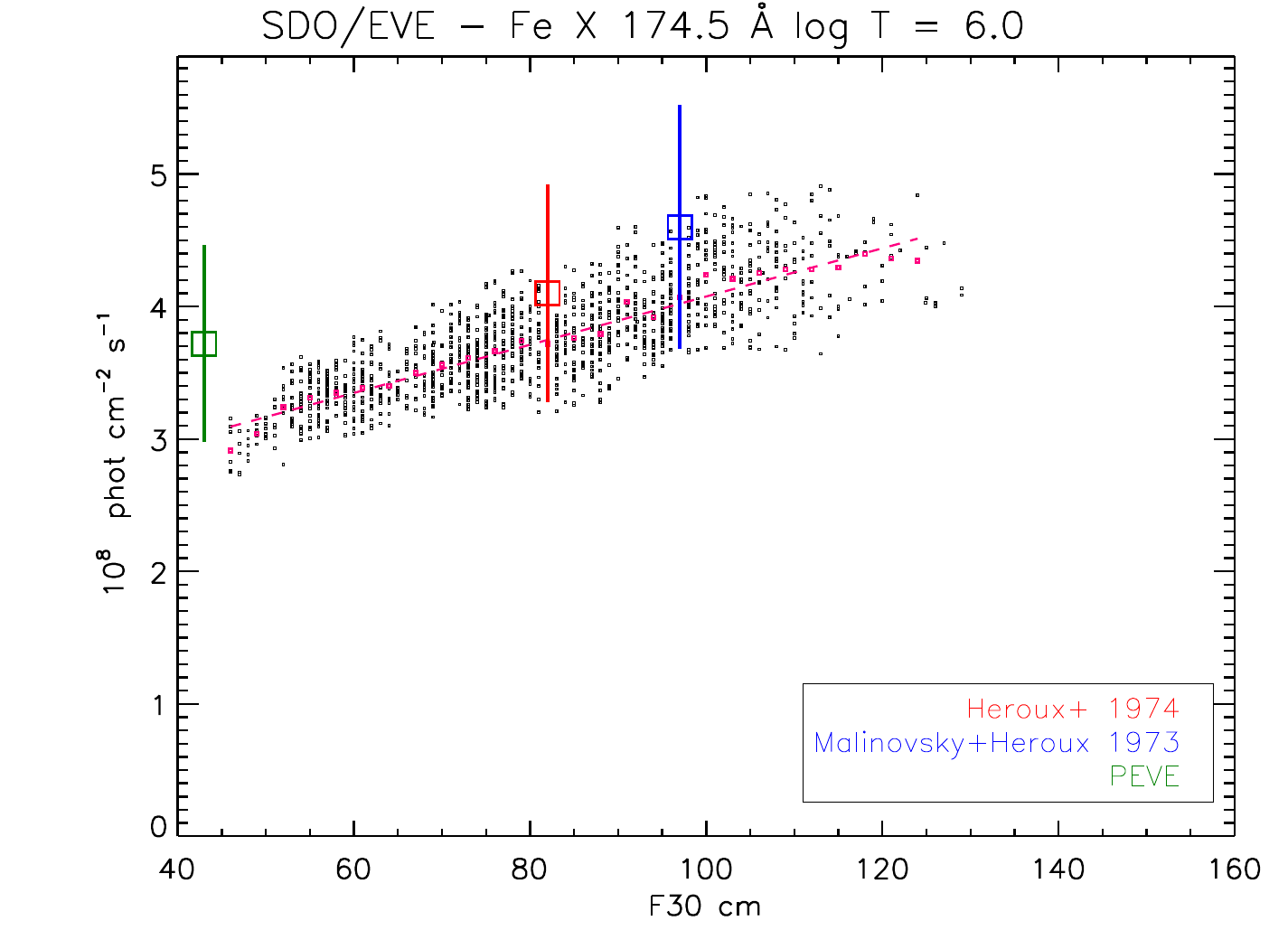}
        \includegraphics[angle=-90, width=6cm]{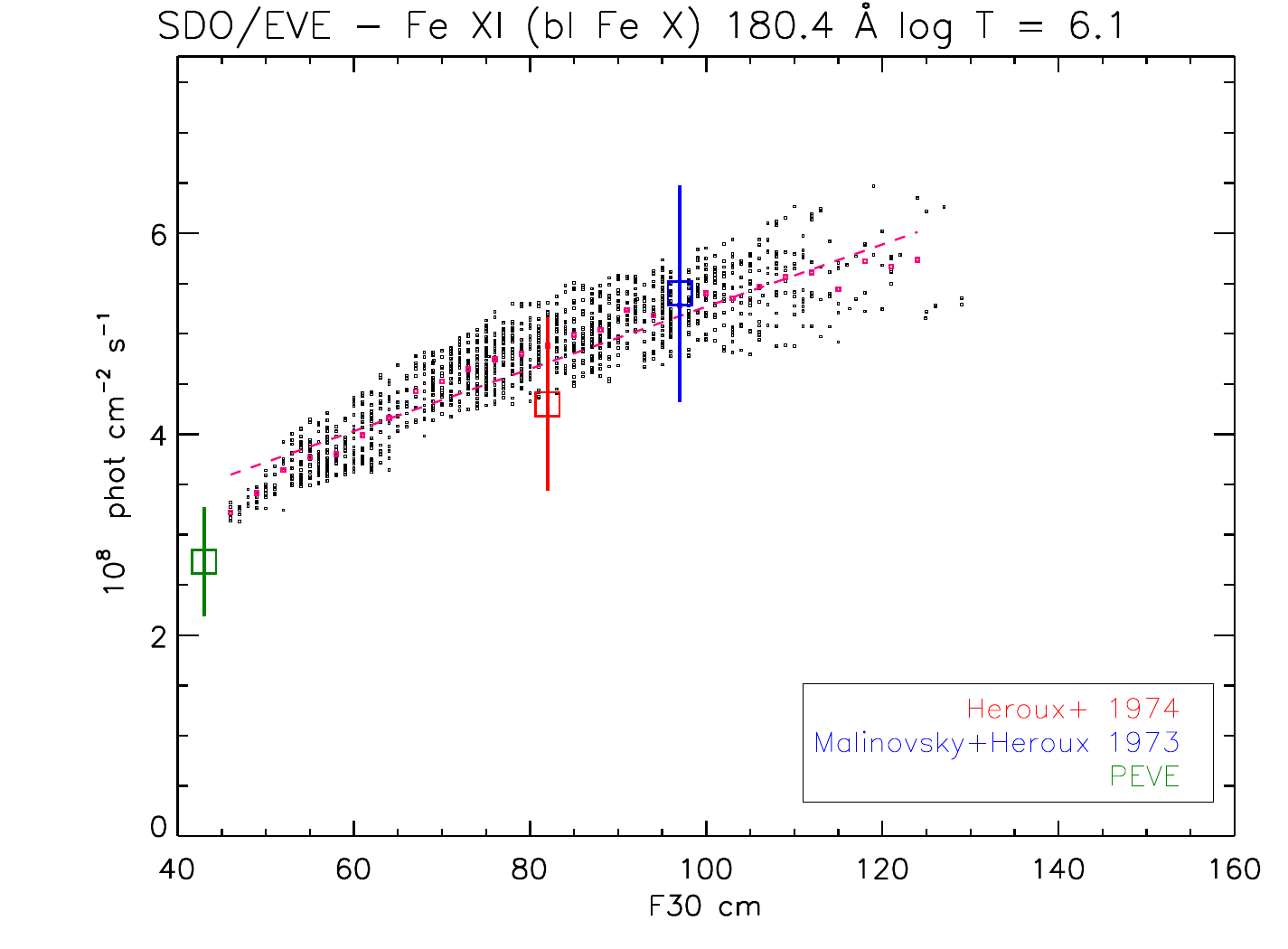}}
\centerline{
        \includegraphics[angle=-90, width=6cm]{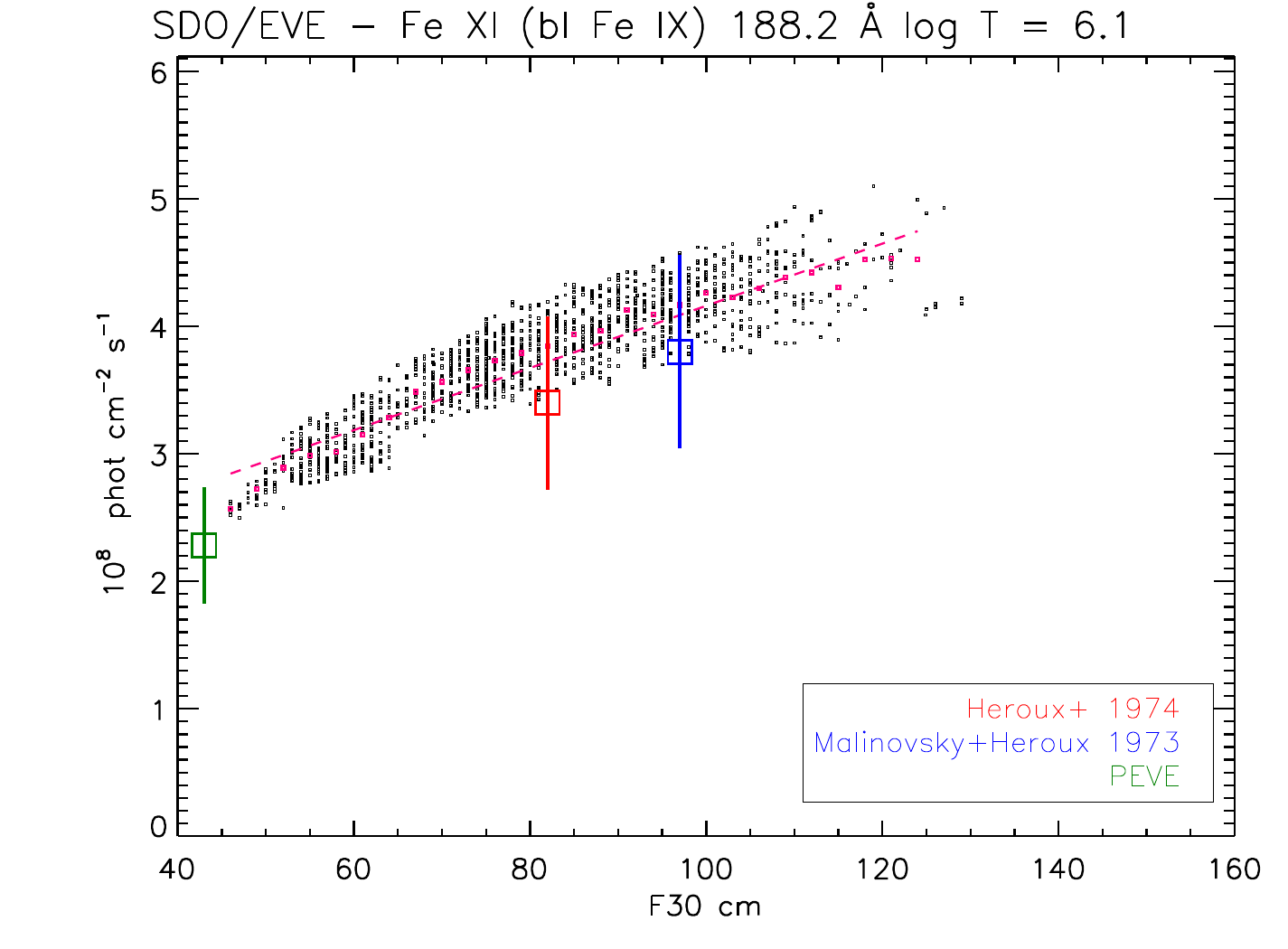}
        \includegraphics[angle=-90, width=6cm]{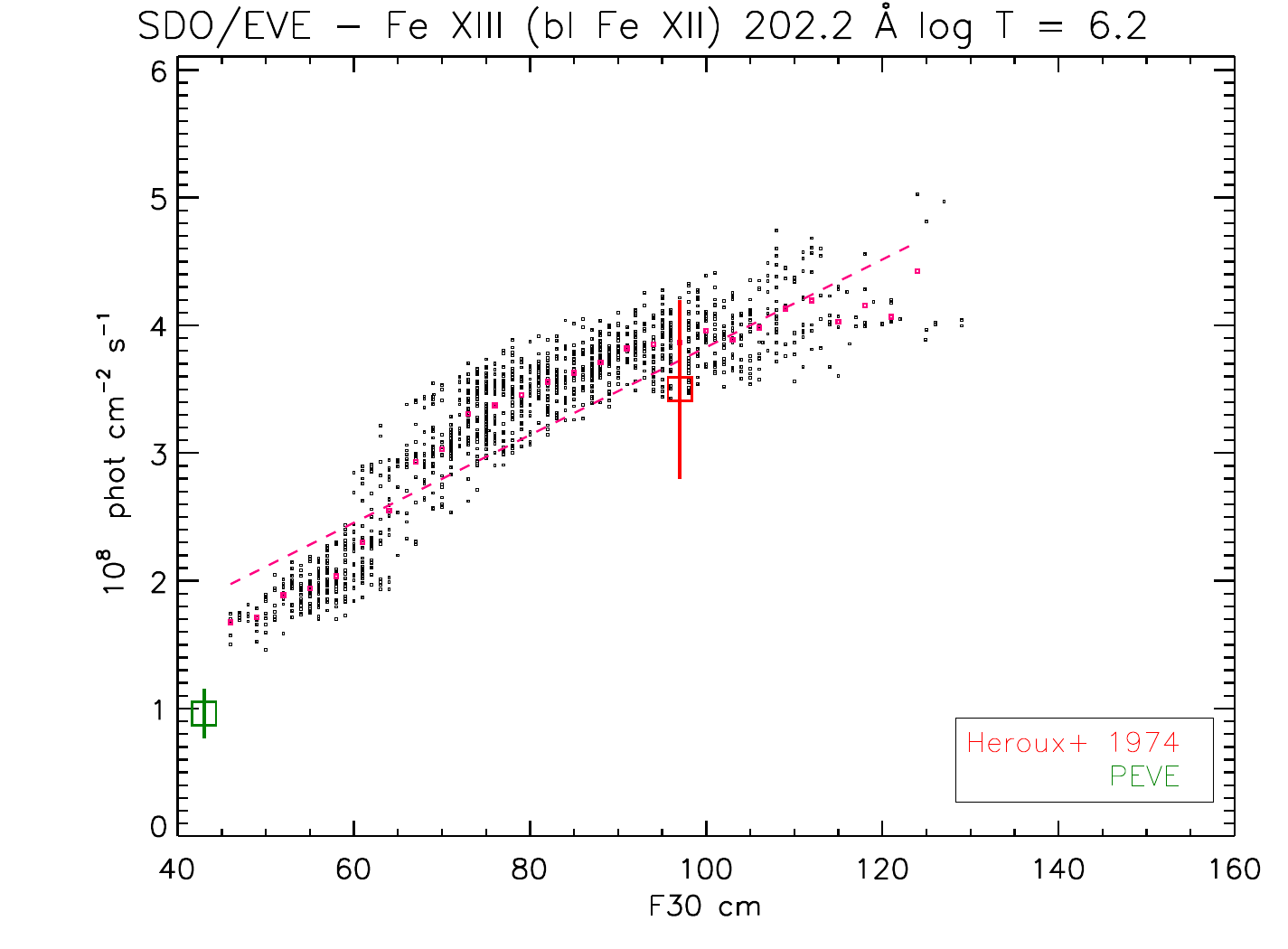}
        \includegraphics[angle=-90, width=6cm]{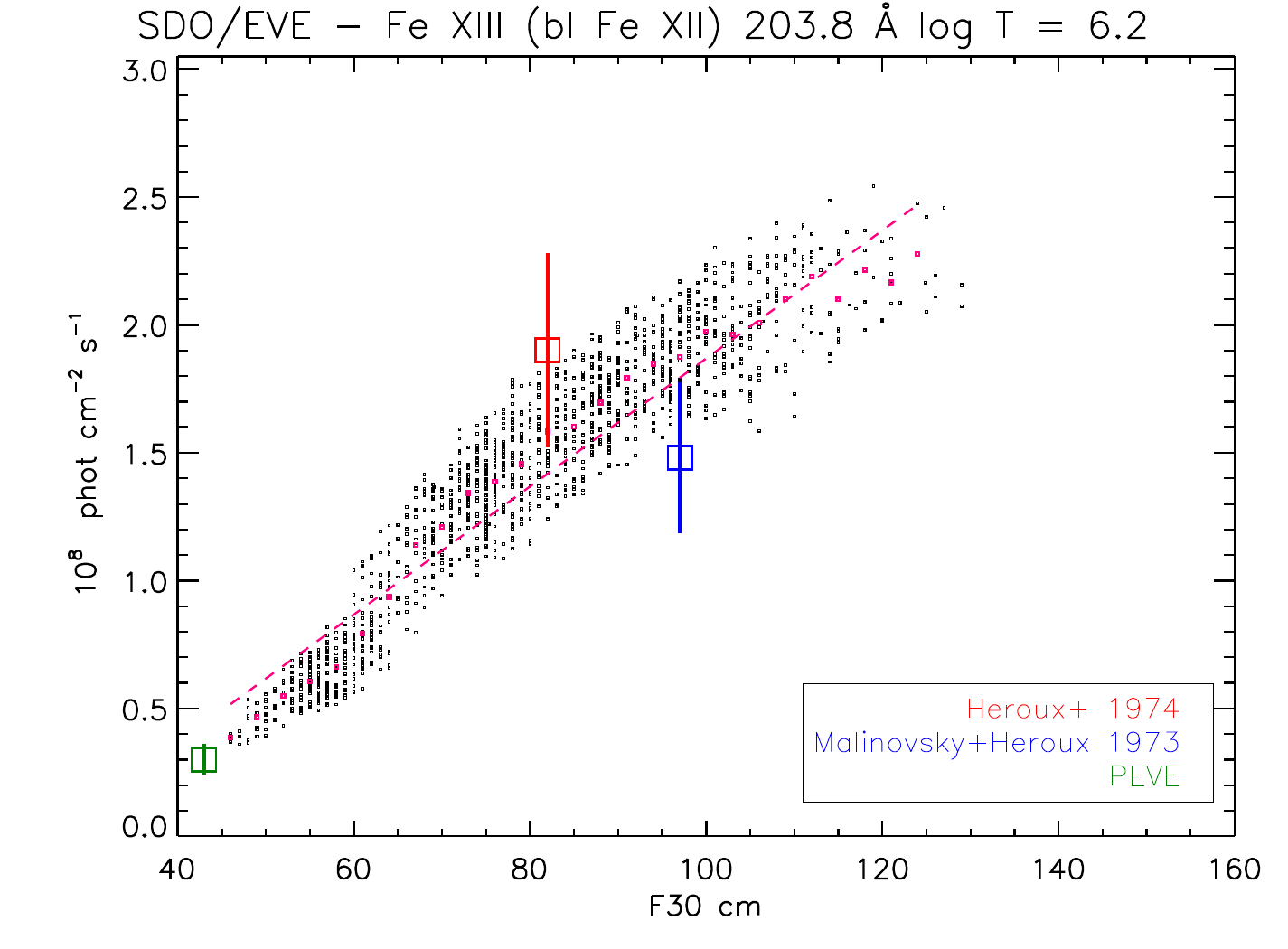}}
\centerline{
        \includegraphics[angle=-90, width=6cm]{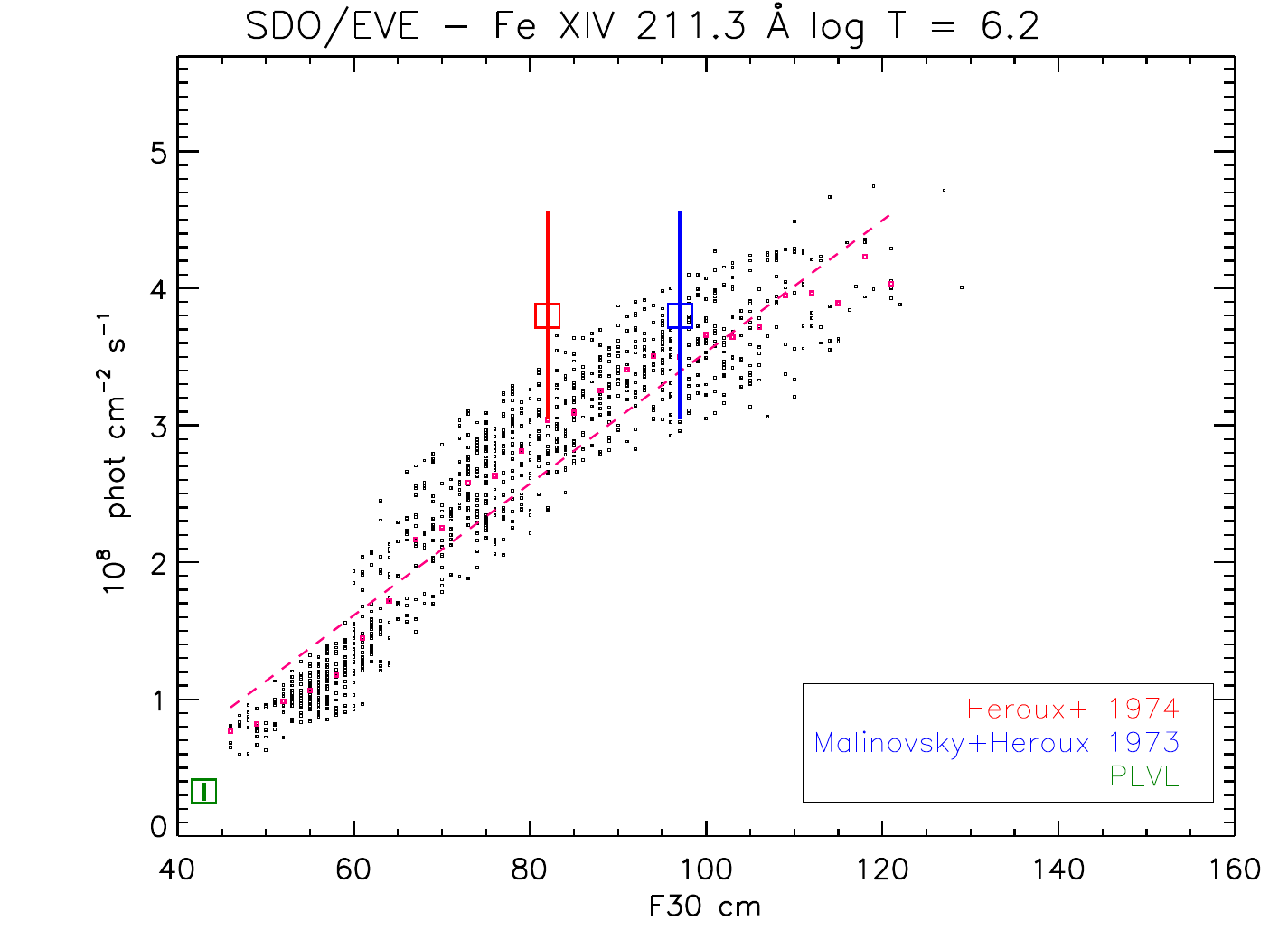}
        \includegraphics[angle=-90, width=6cm]{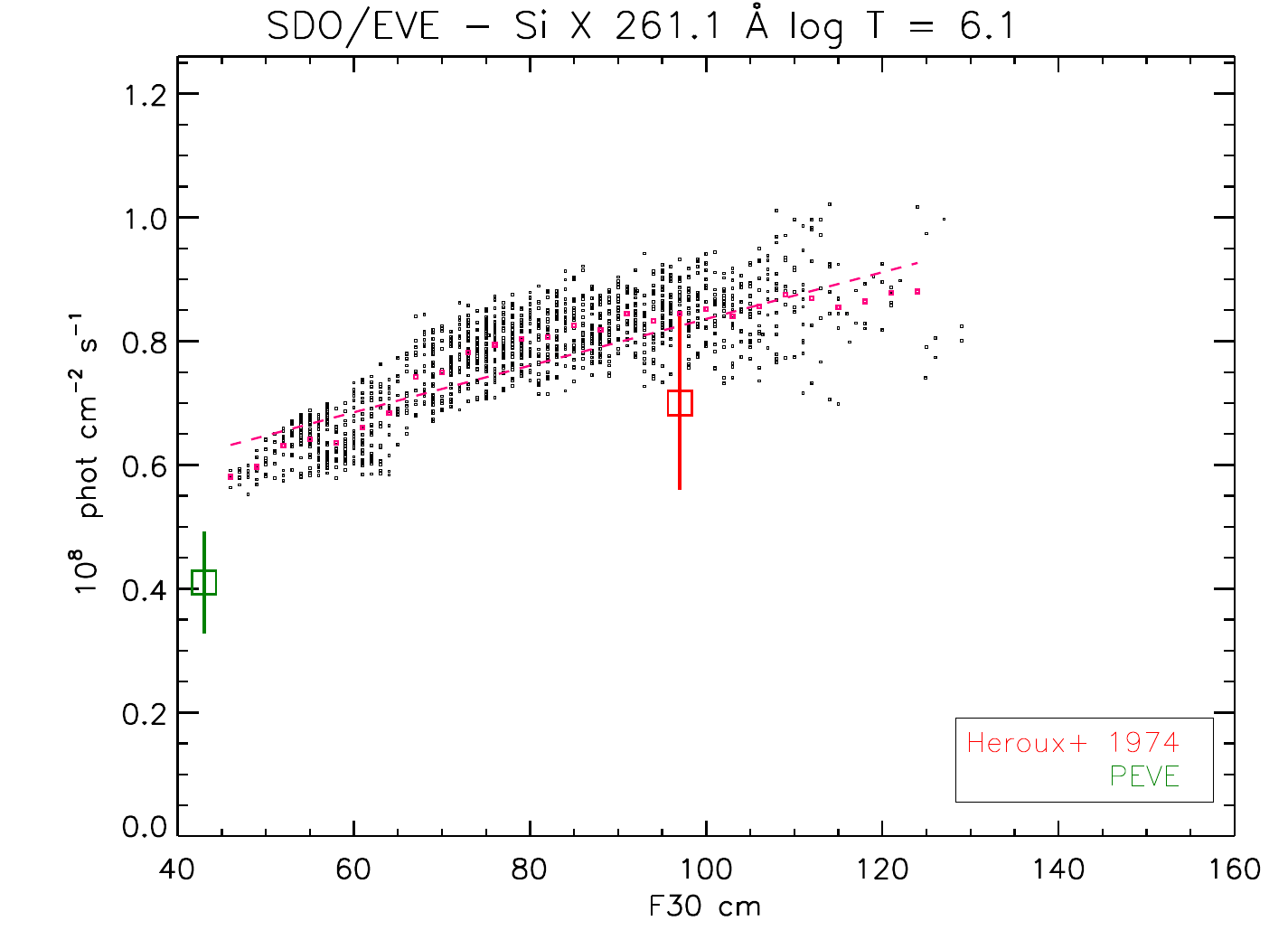}
        \includegraphics[angle=-90, width=6cm]{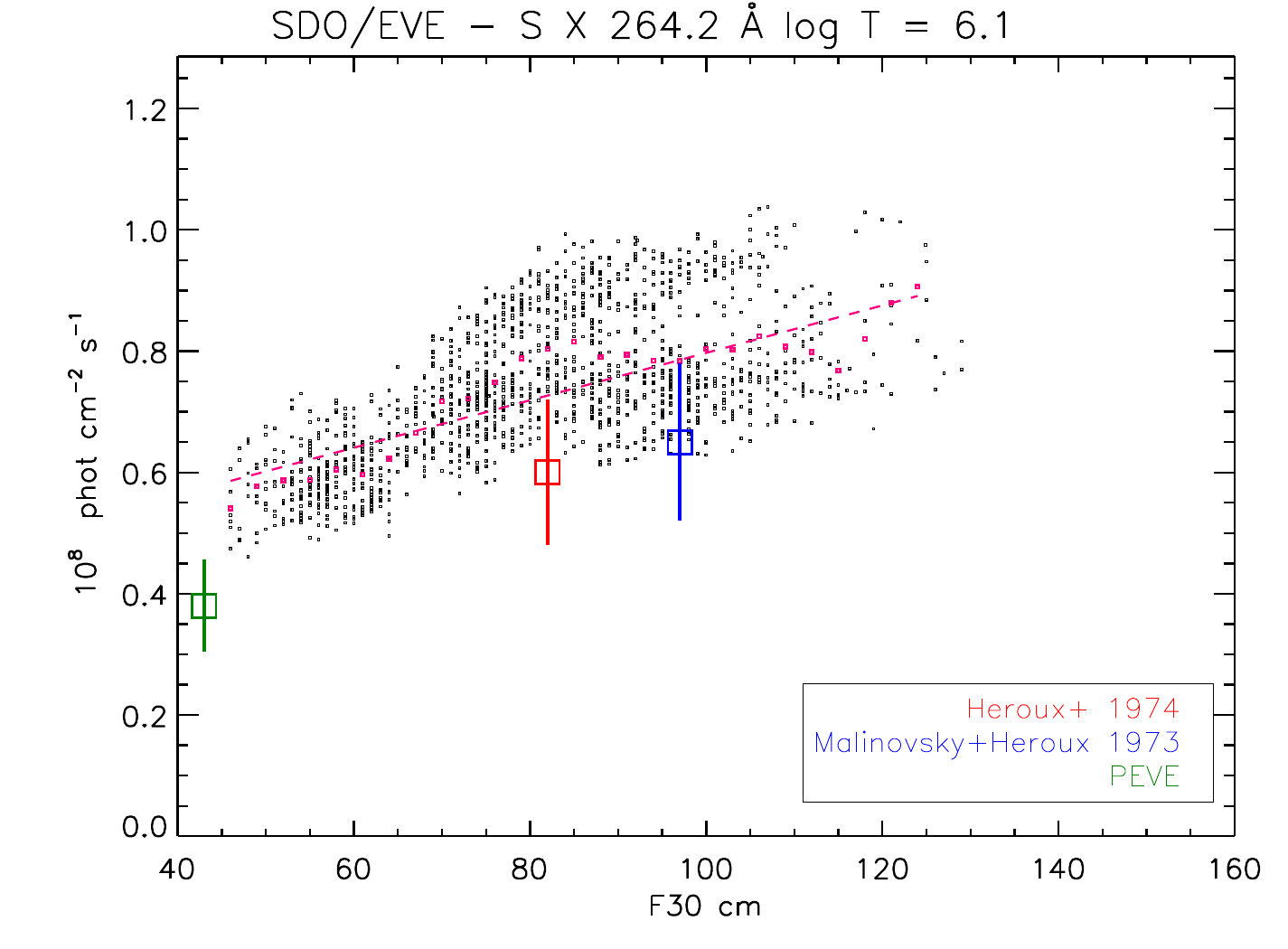}}
\centerline{
        \includegraphics[angle=-90, width=6cm]{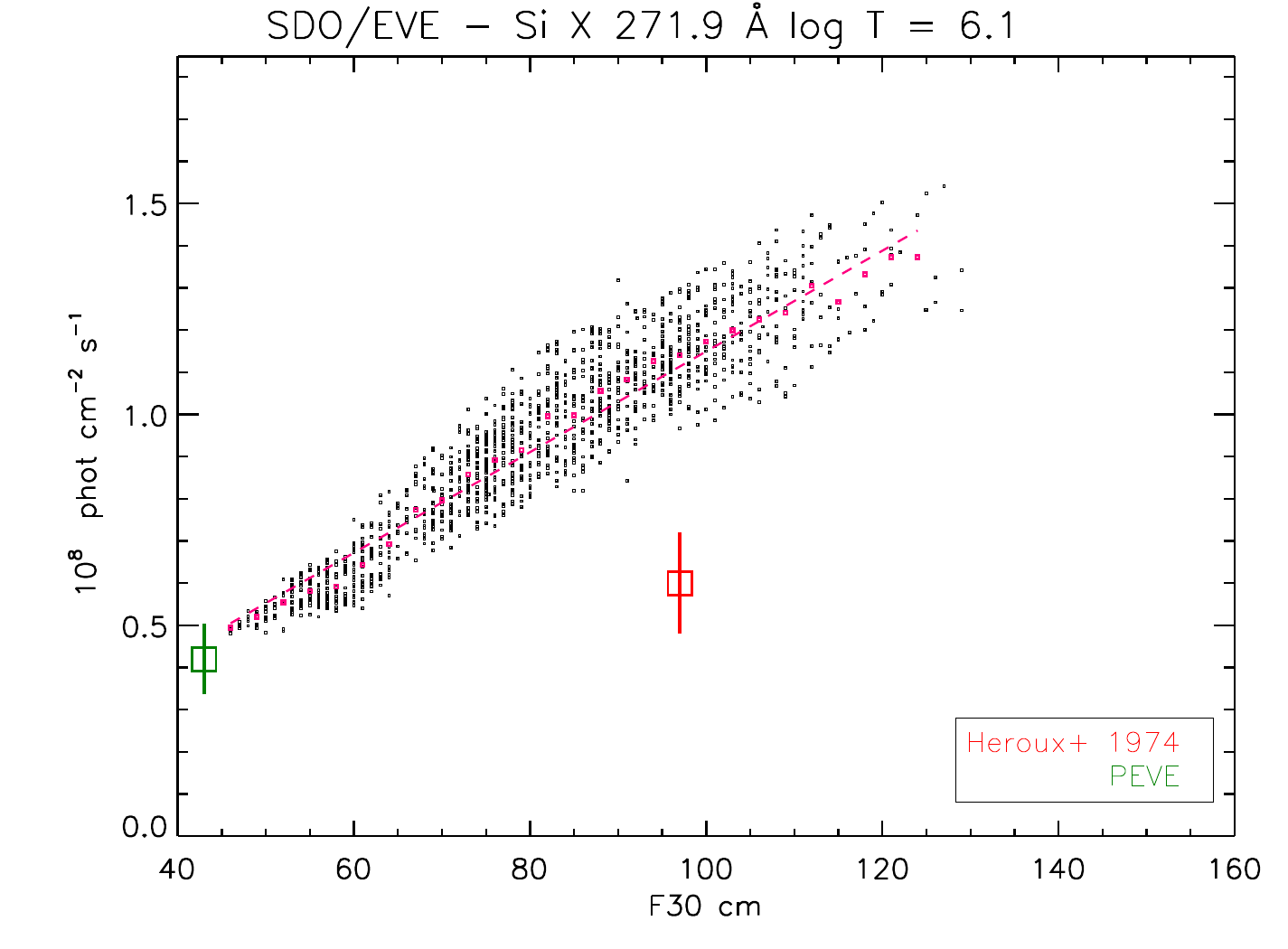}
        \includegraphics[angle=-90, width=6cm]{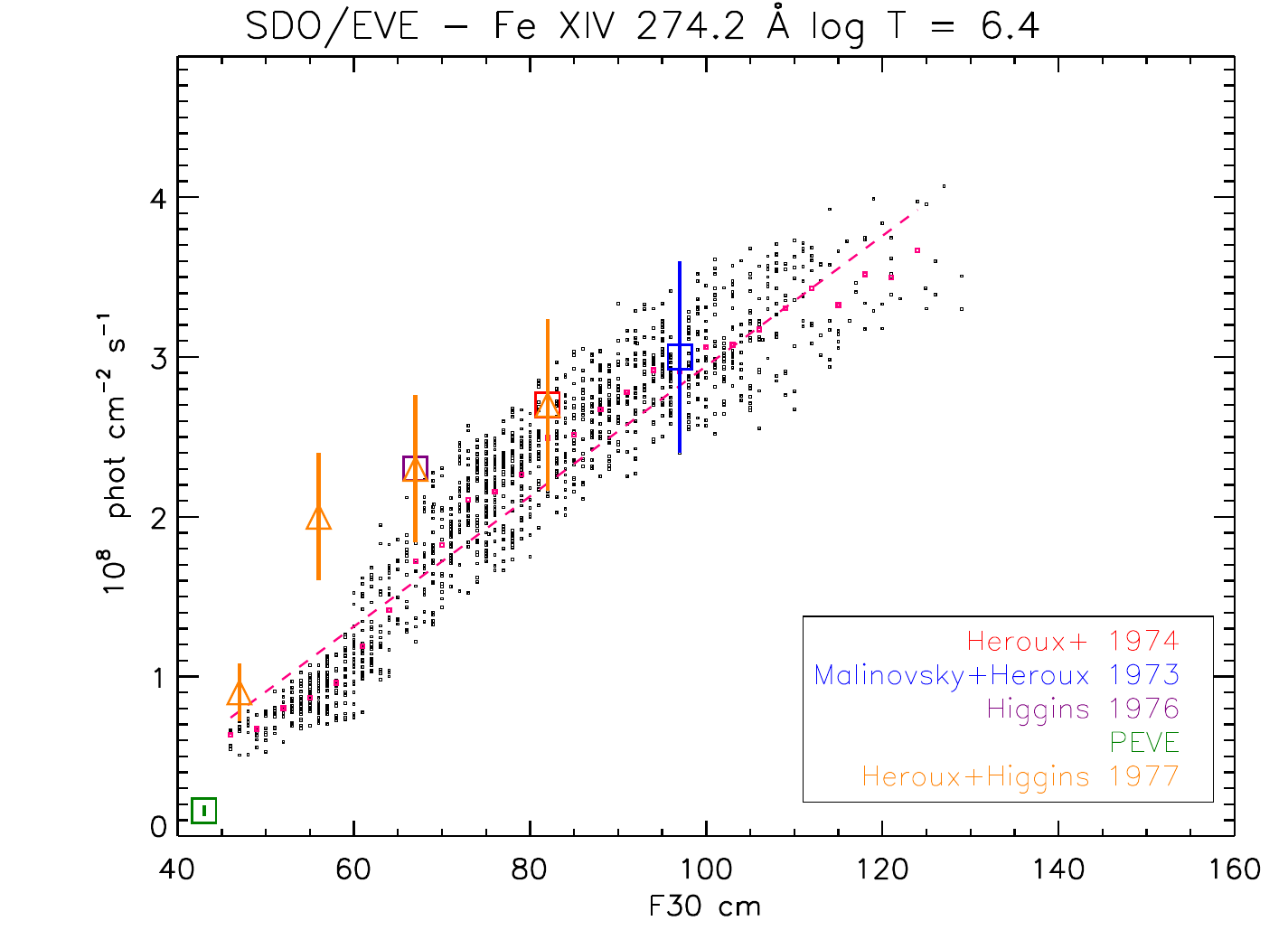}
        \includegraphics[angle=-90, width=6cm]{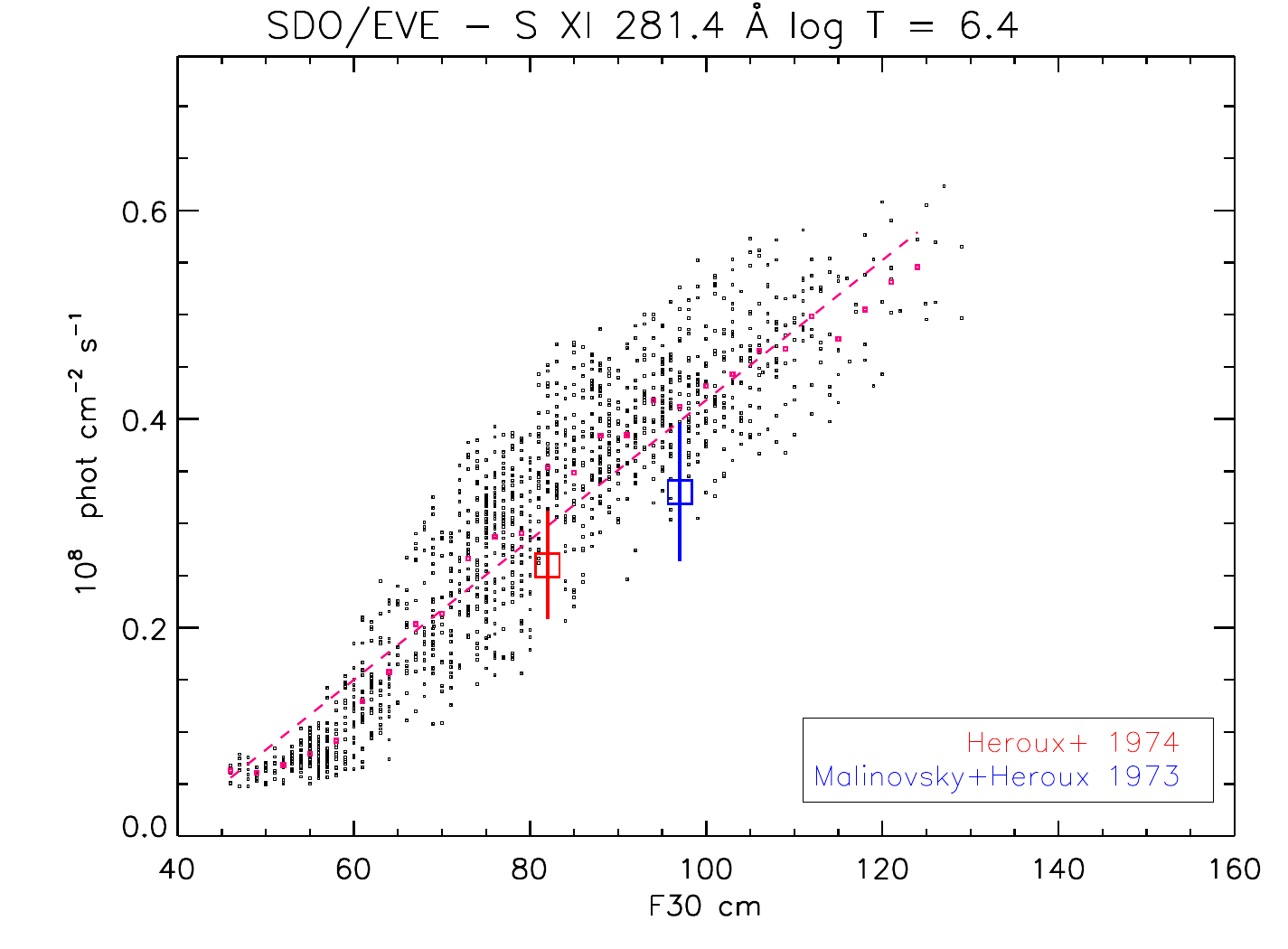}}
\centerline{
        \includegraphics[angle=-90, width=6cm]{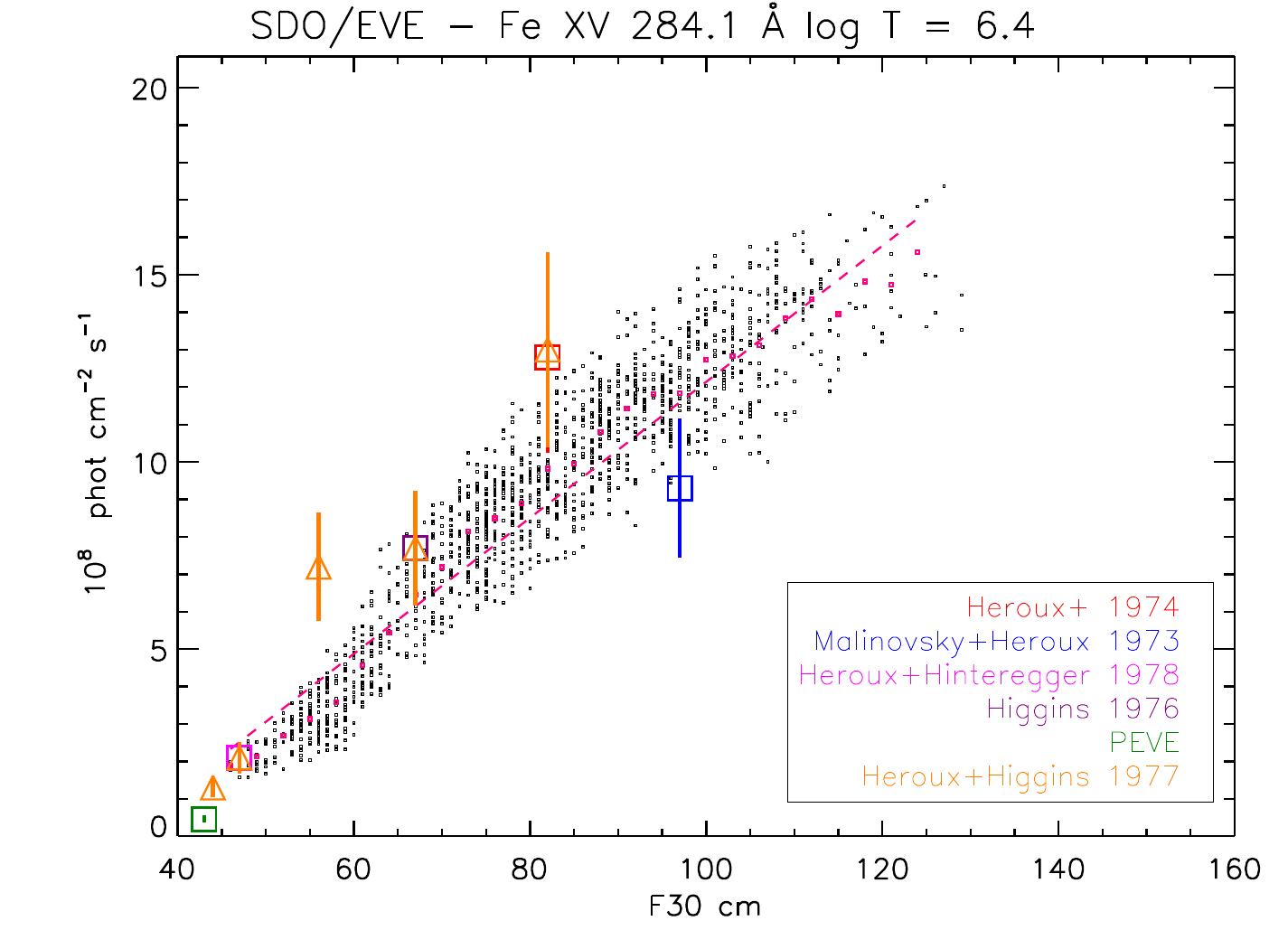}
        \includegraphics[angle=-90, width=6cm]{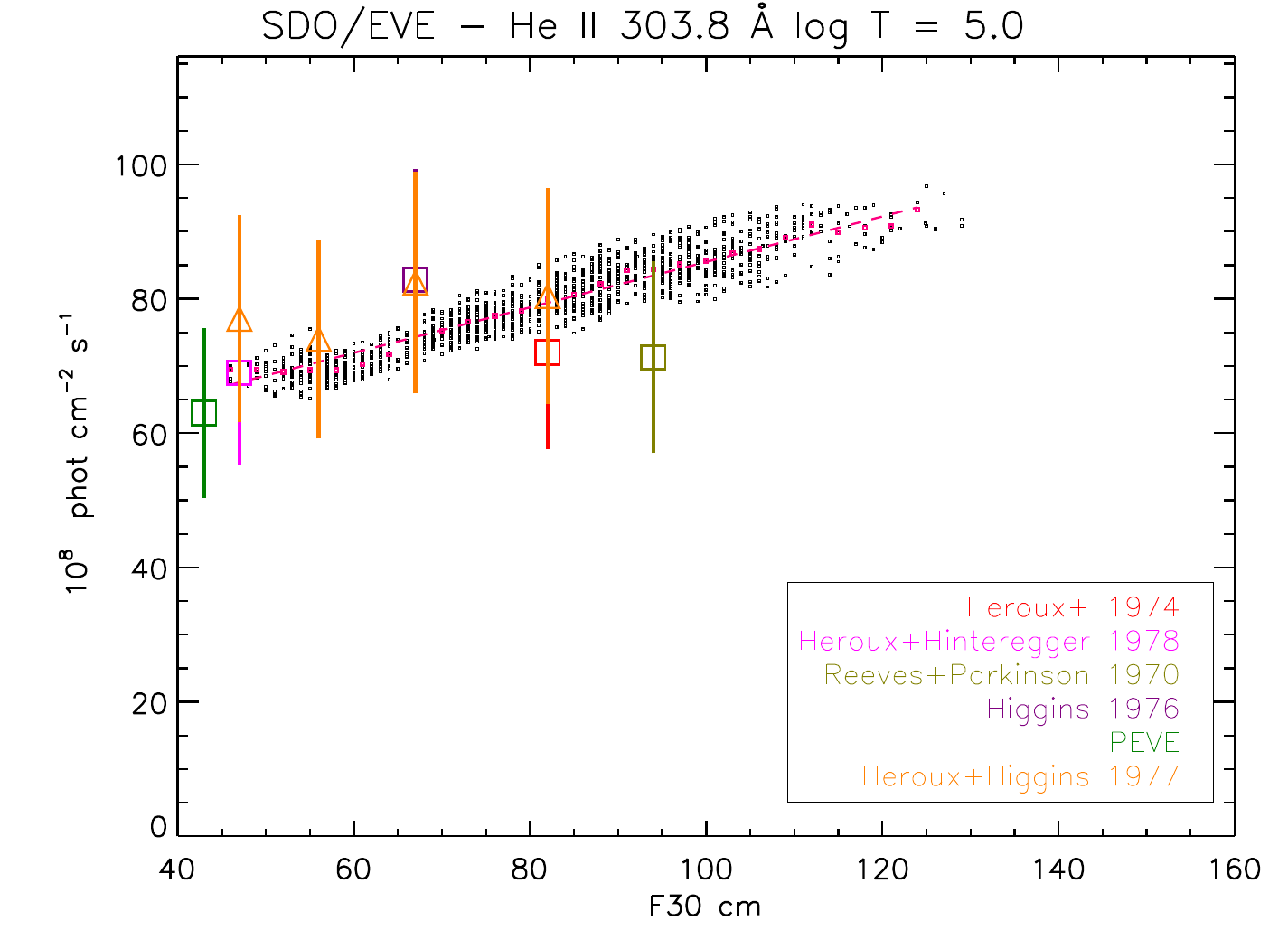}
        }
    \caption{Correlations of irradiances of various spectral lines measured by SDO/EVE MEGS-A and the F30 cm radio flux, with historical records from the literature and PEVE measurements from the solar minimum in 2008.}
    \label{fig:MEGSA_his_rec}
\end{figure*}

\begin{figure*}
 \centerline{
        \includegraphics[angle=-90, width=6cm, keepaspectratio]{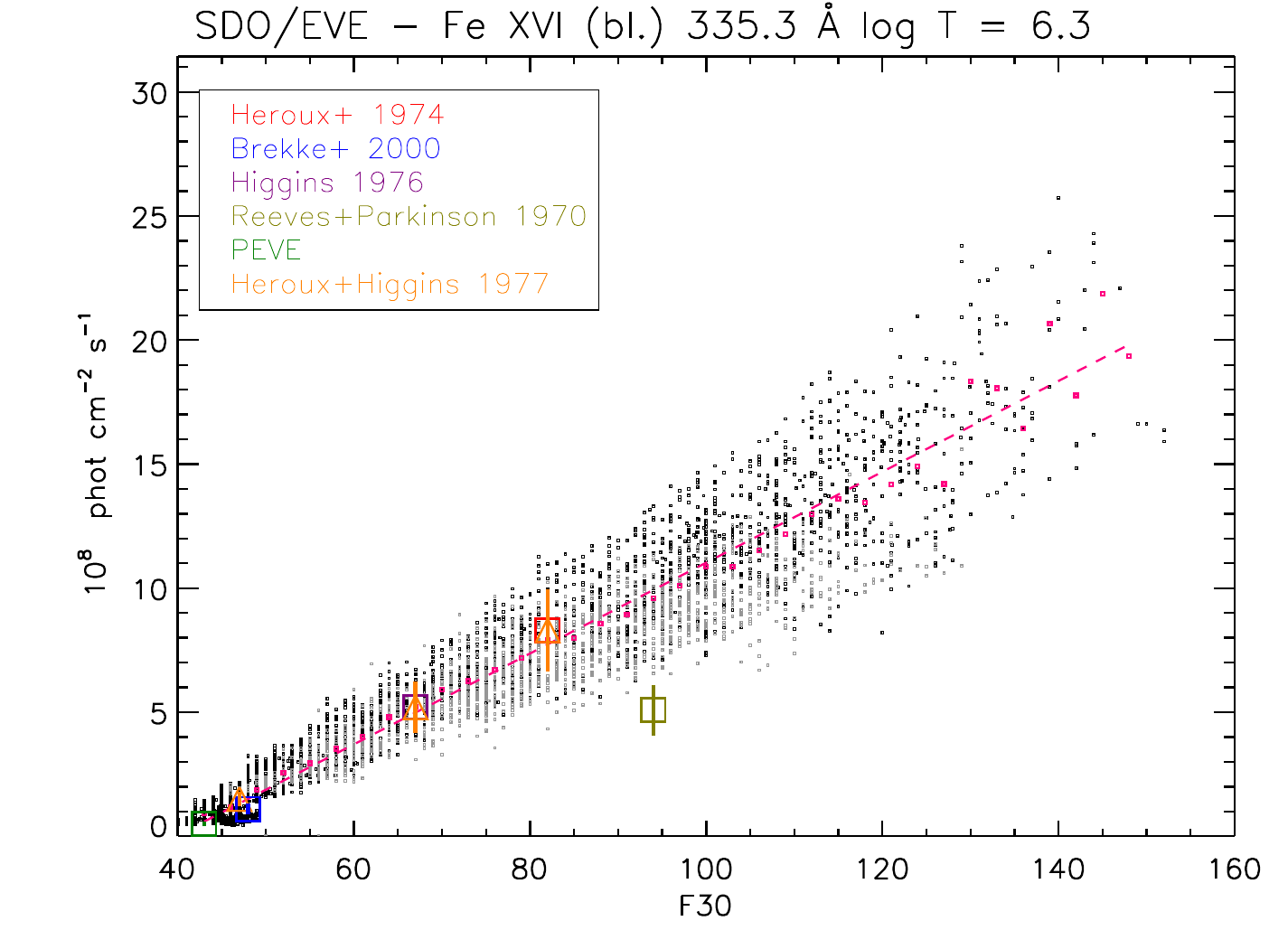}
        \includegraphics[angle=-90, width=6cm, keepaspectratio]{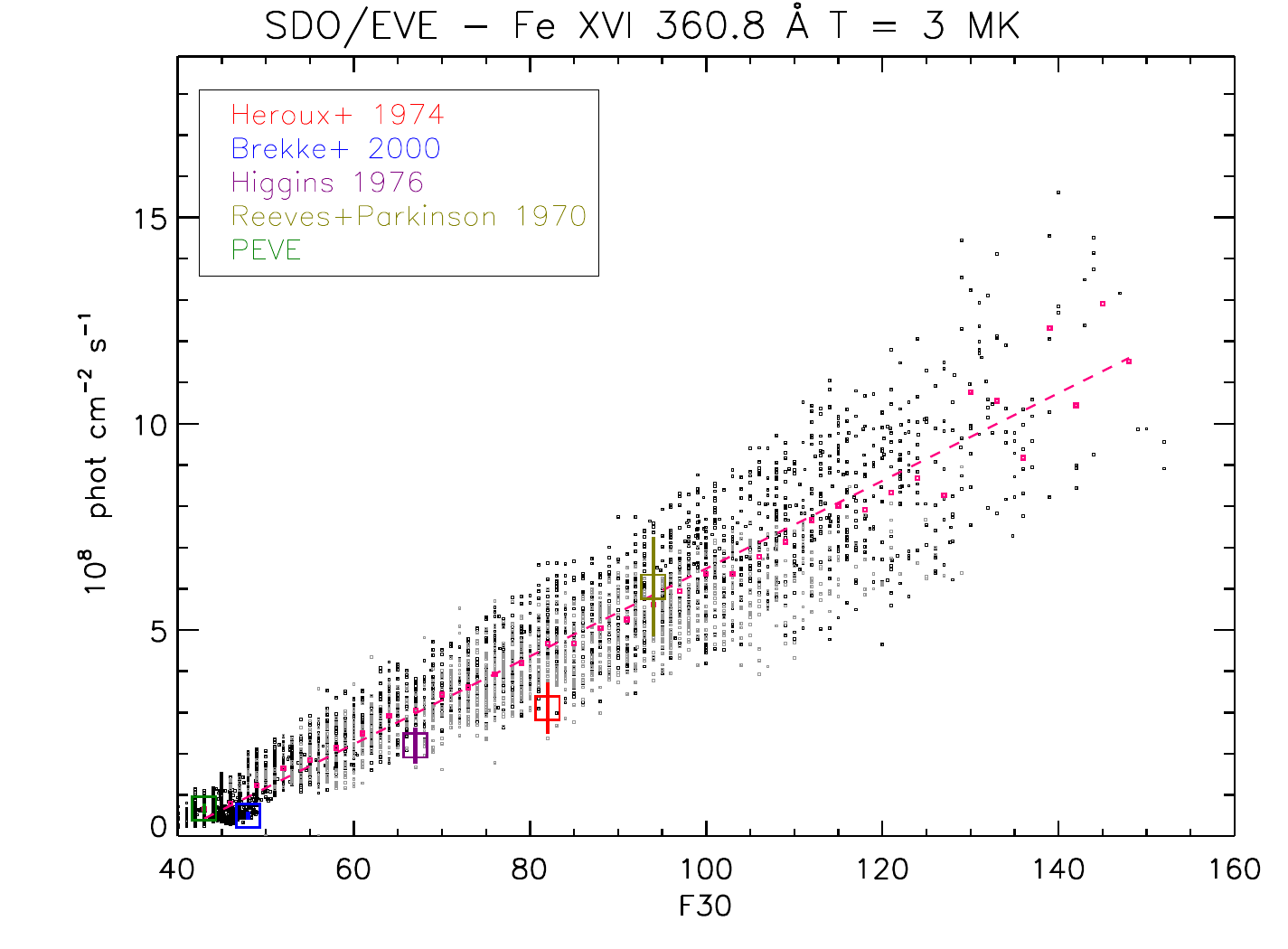}
        \includegraphics[angle=-90, width=6cm, keepaspectratio]{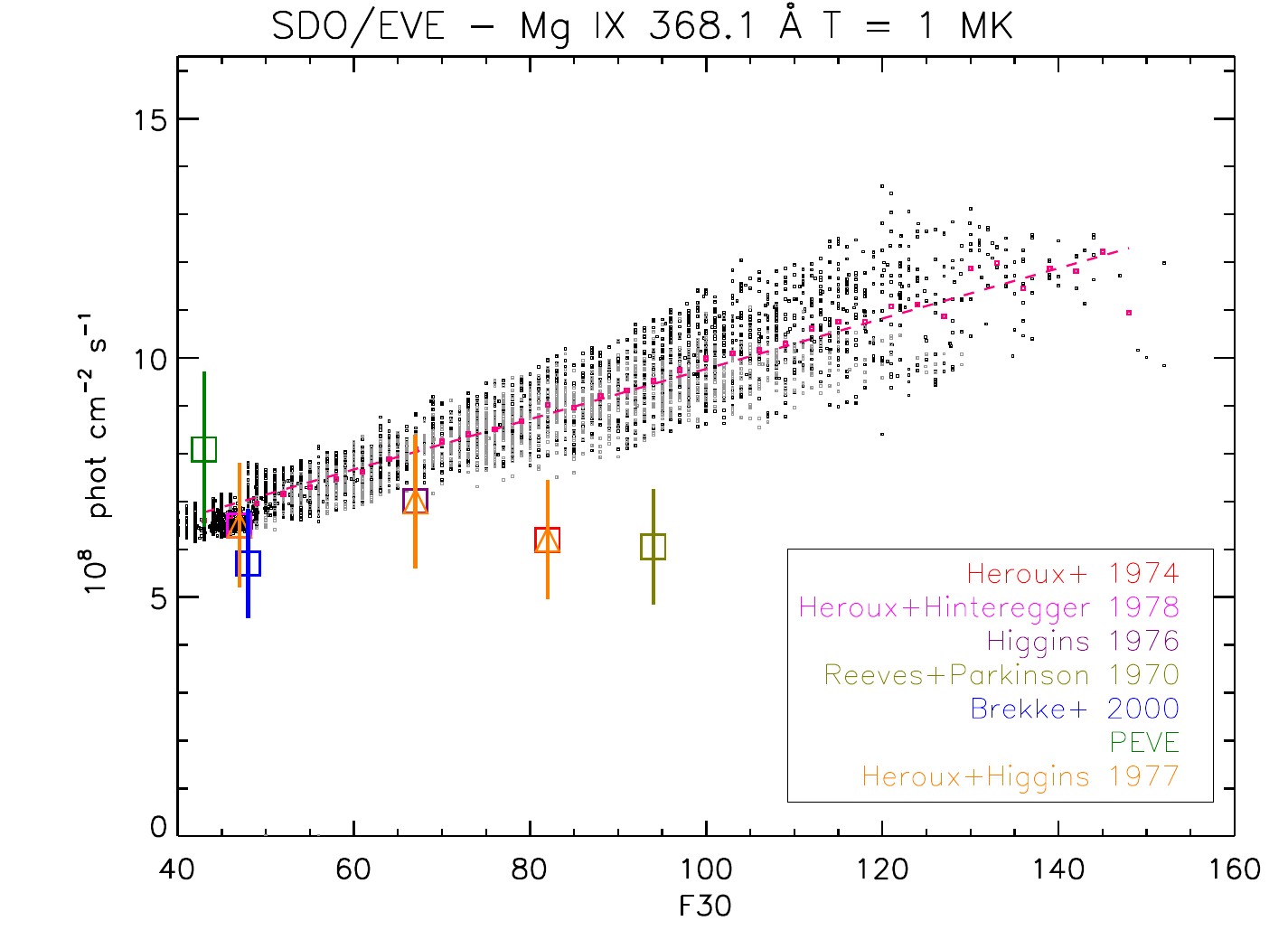}}
 \centerline{
        \includegraphics[angle=-90, width=6cm, keepaspectratio]{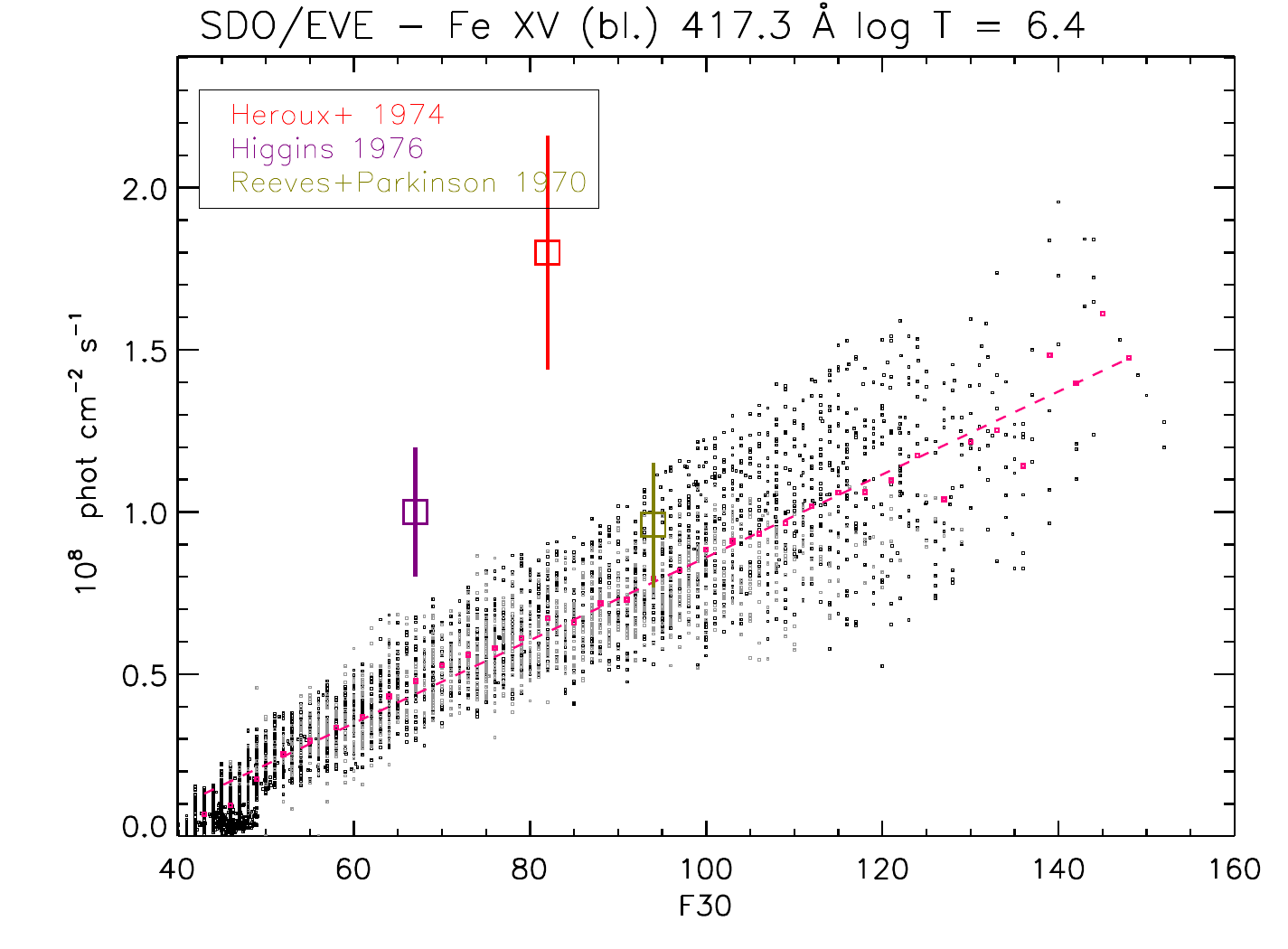}
        \includegraphics[angle=-90, width=6cm, keepaspectratio]{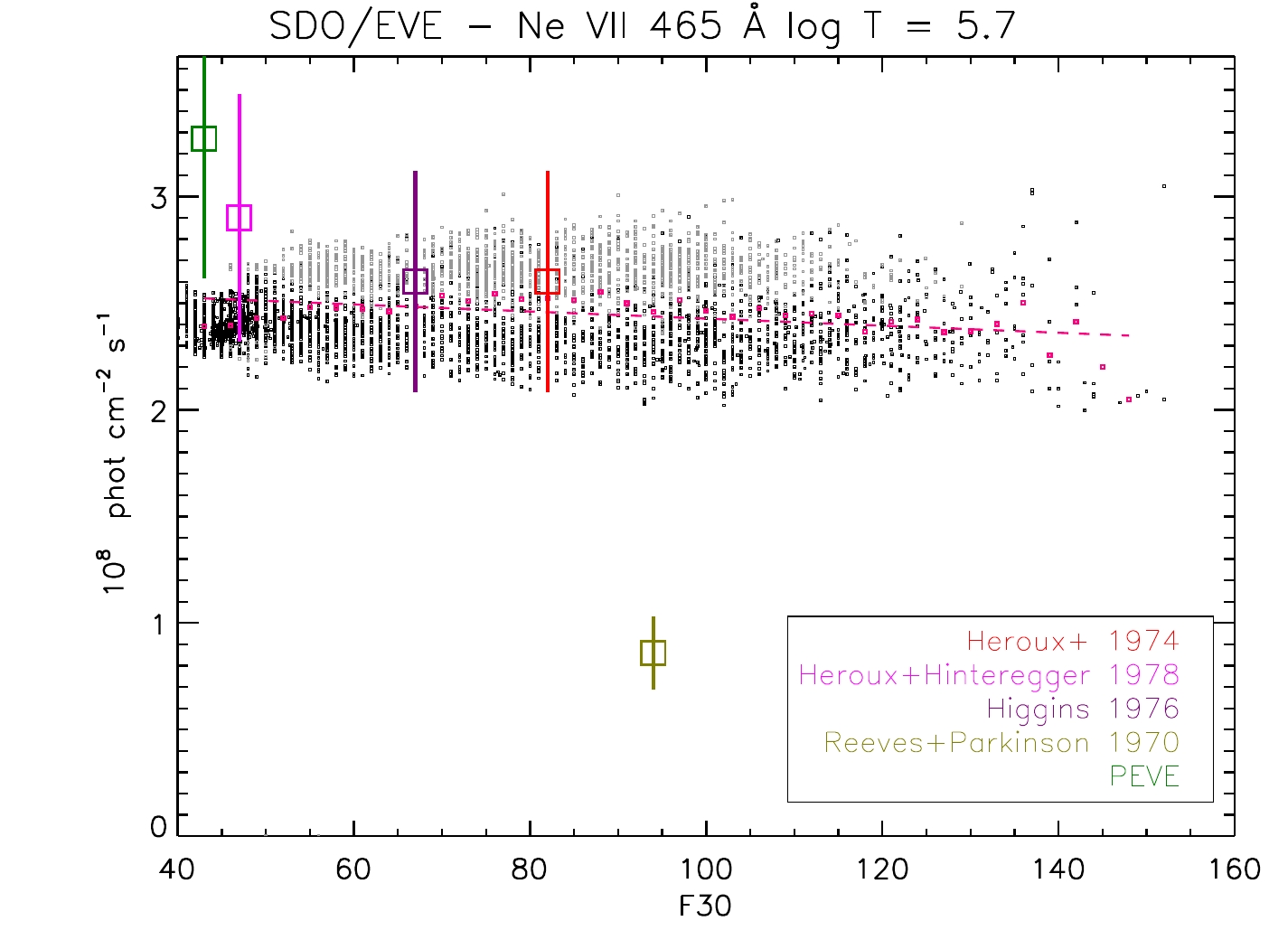}
        \includegraphics[angle=-90, width=6cm, keepaspectratio]{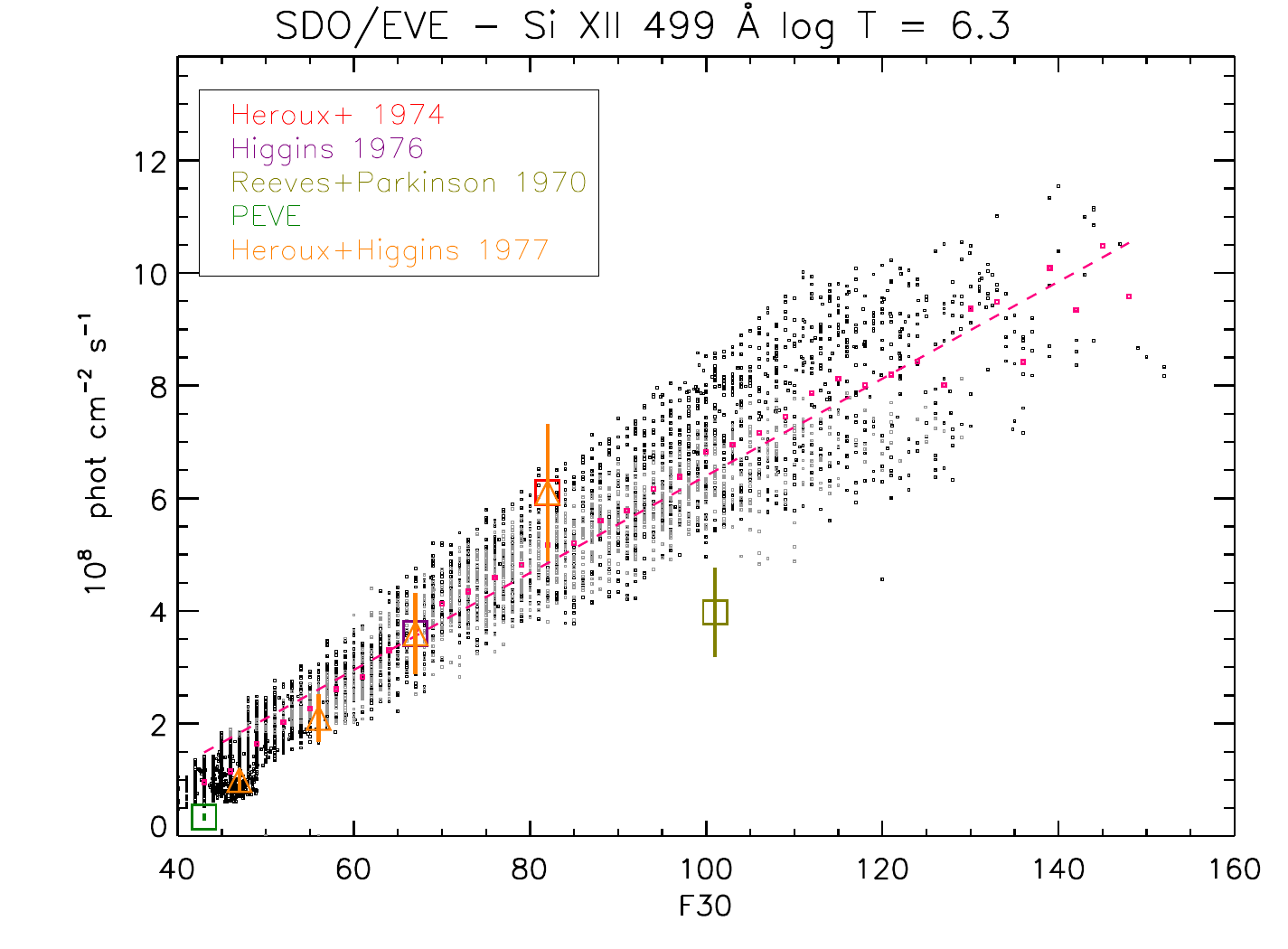}}
 \centerline{
        \includegraphics[angle=-90, width=6cm, keepaspectratio]{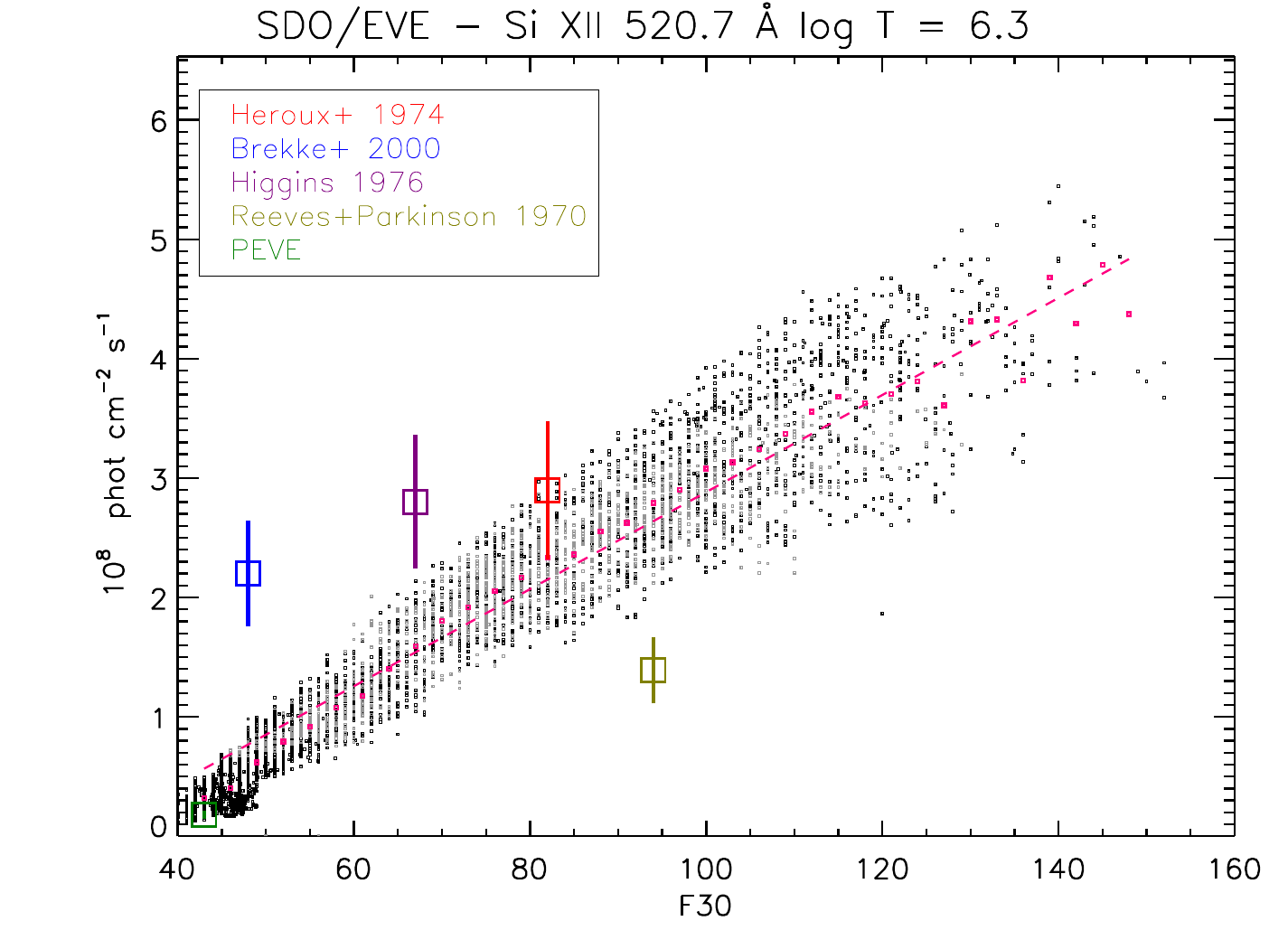}
        \includegraphics[angle=-90, width=6cm, keepaspectratio]{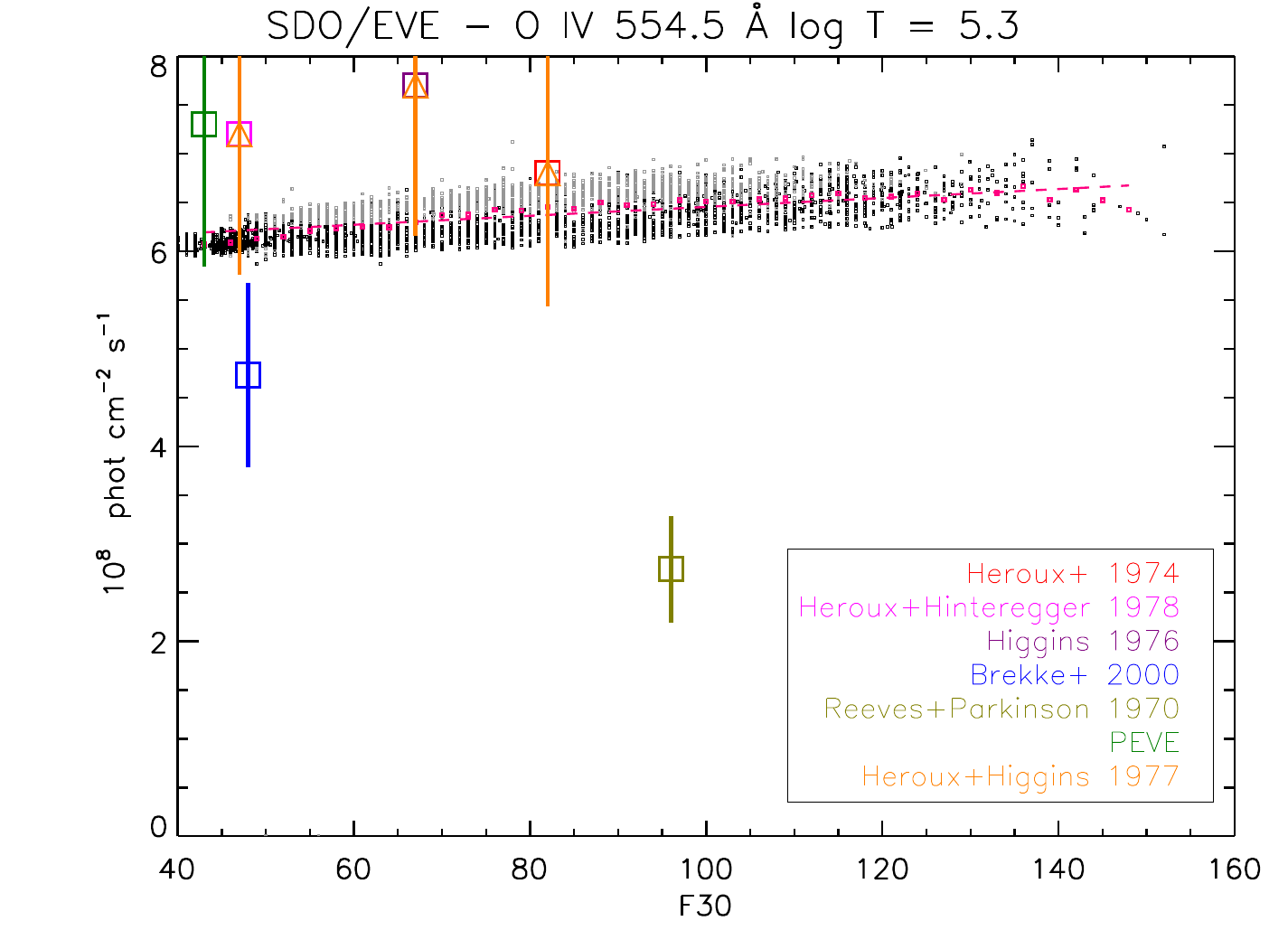}
        \includegraphics[angle=-90, width=6cm, keepaspectratio]{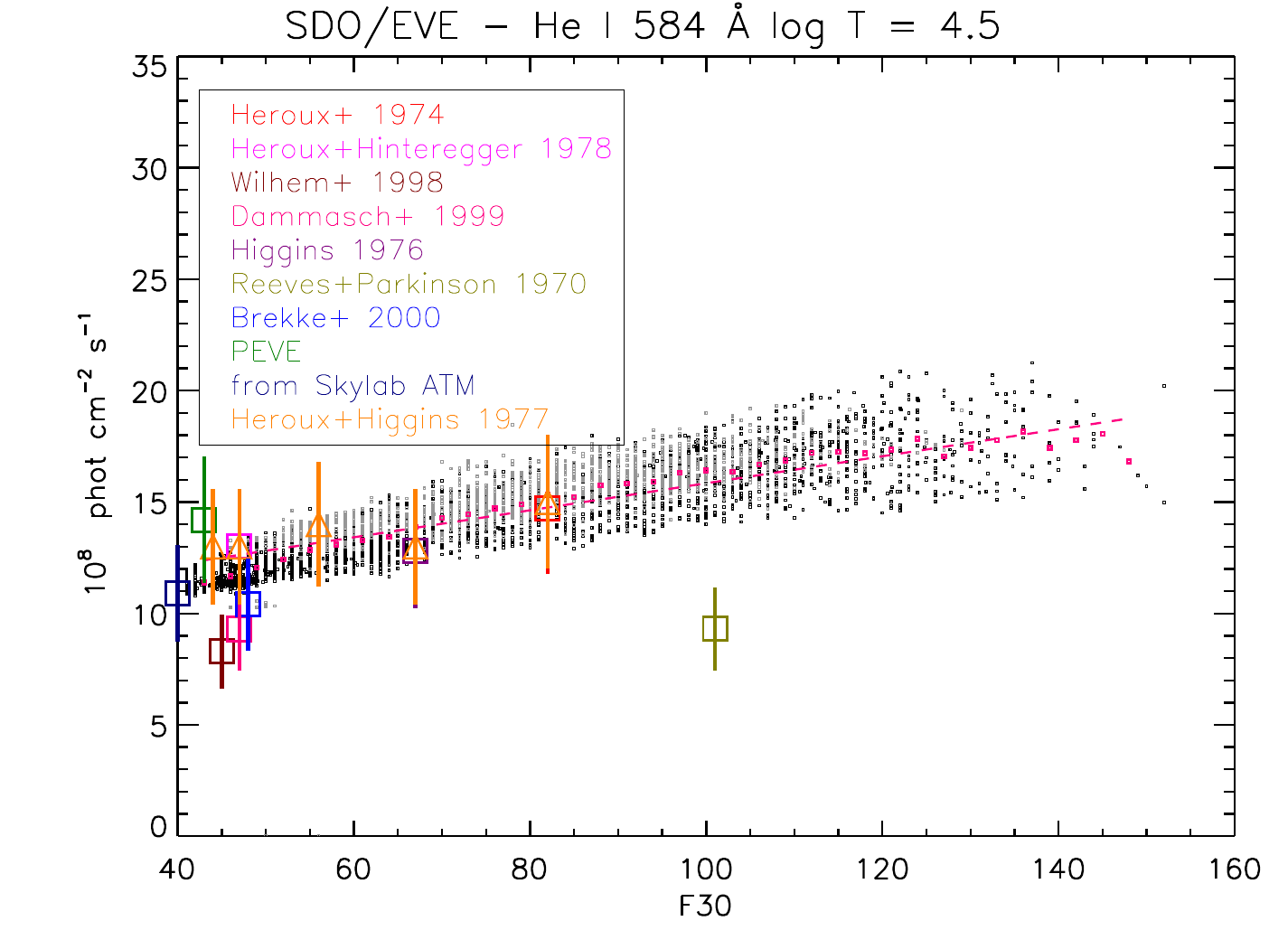}}
 \centerline{
        \includegraphics[angle=-90, width=6cm, keepaspectratio]{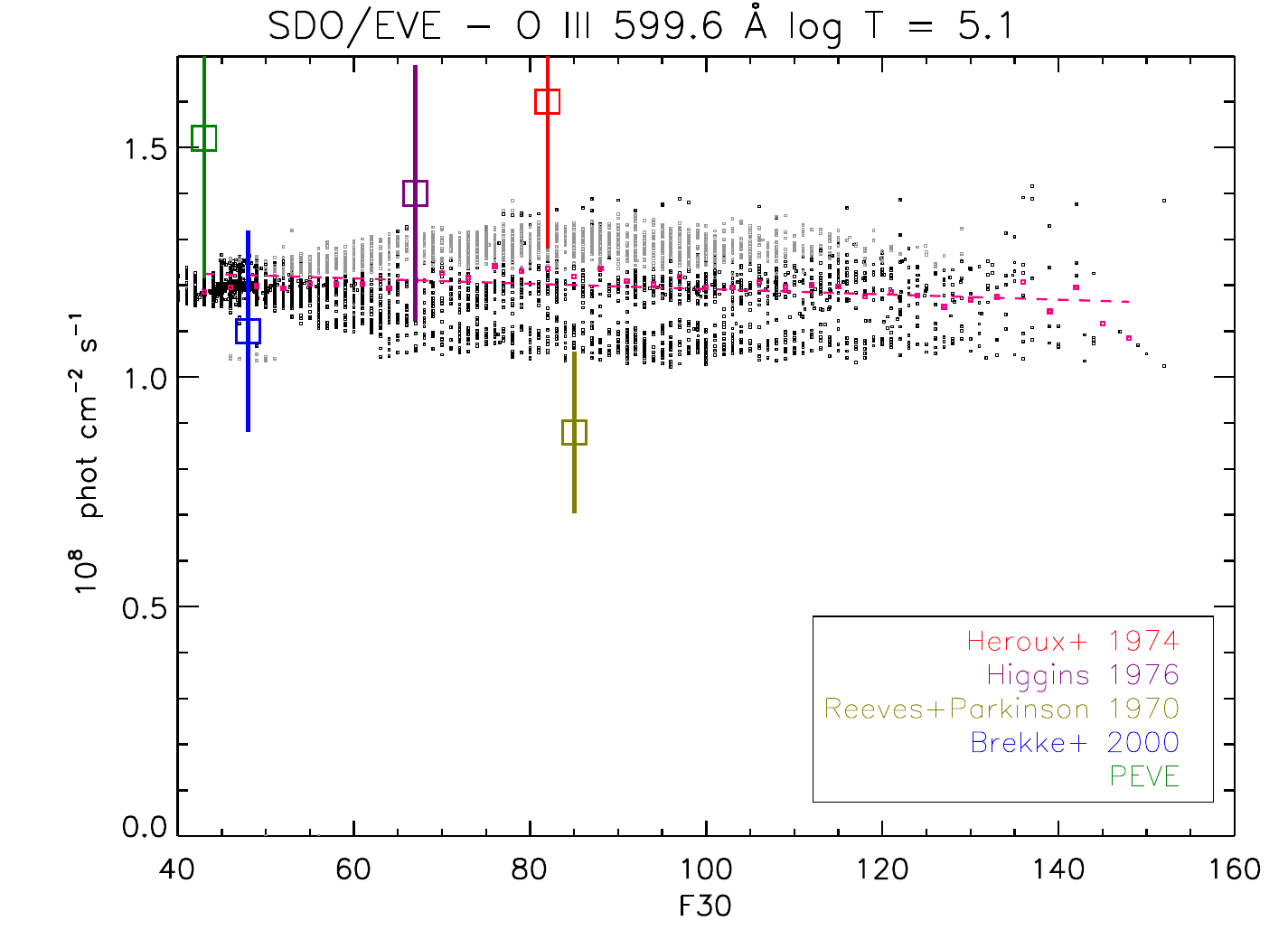}
        \includegraphics[angle=-90, width=6cm, keepaspectratio]{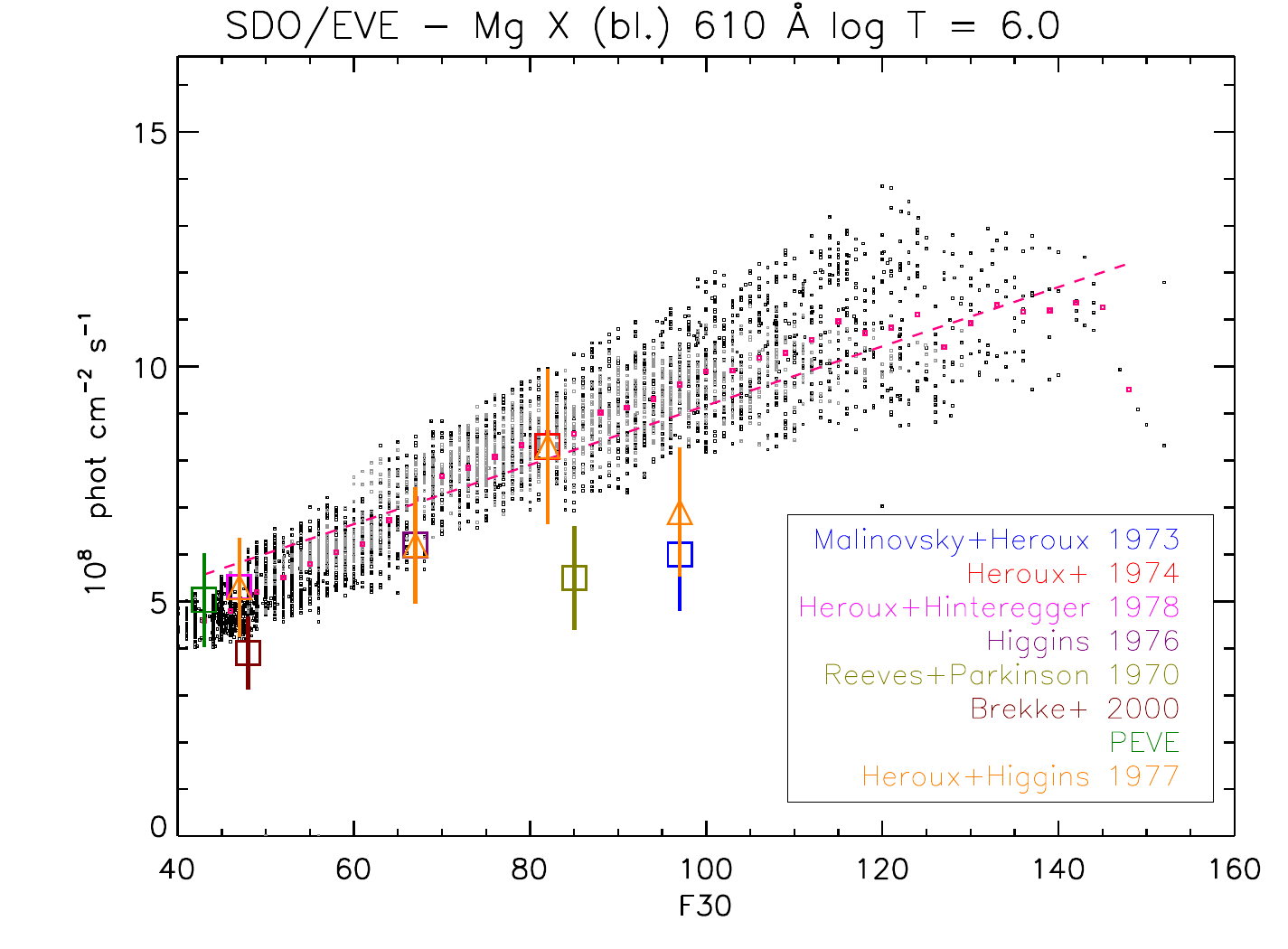}
        \includegraphics[angle=-90, width=6cm, keepaspectratio]{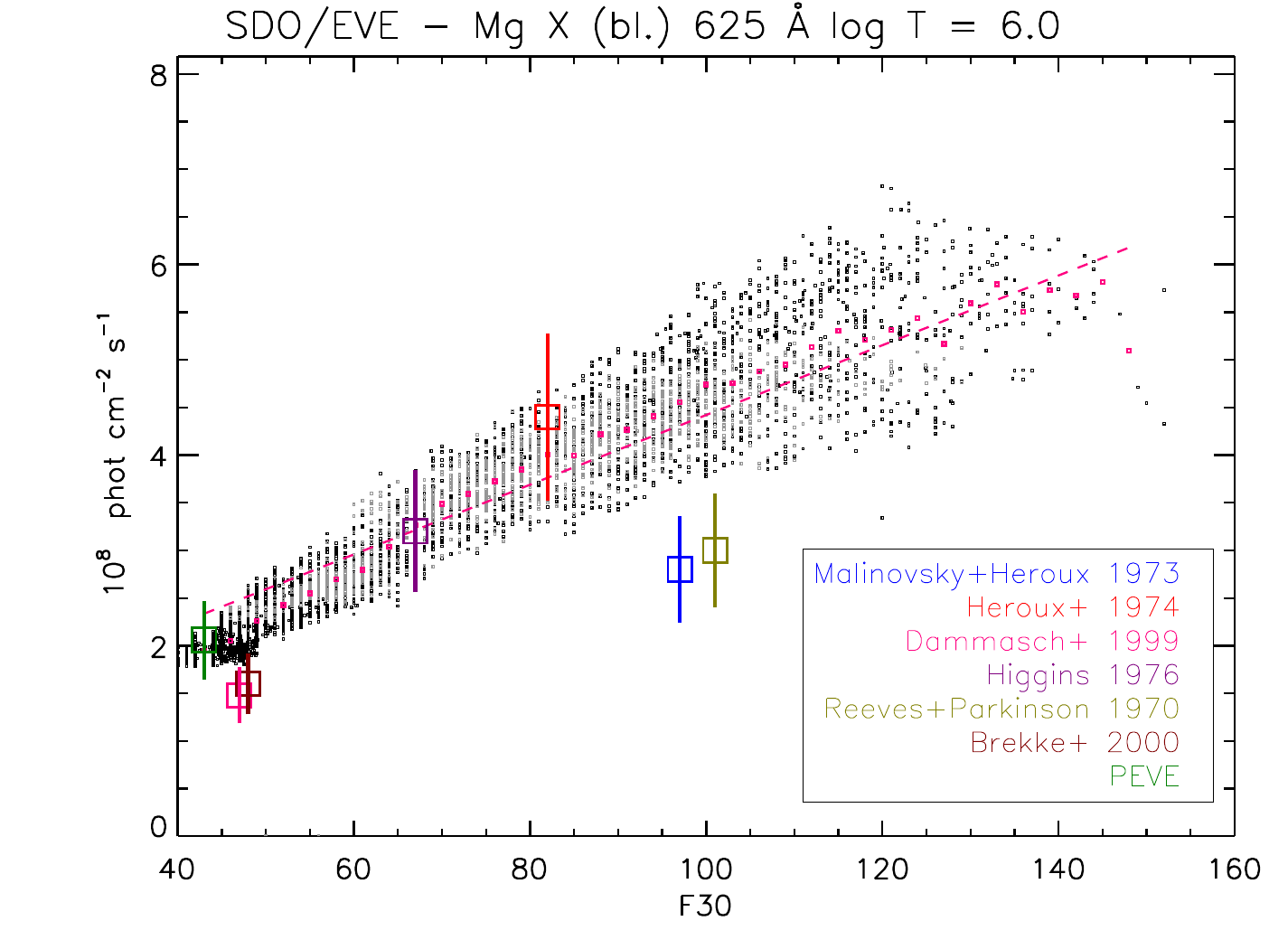}}
        \caption{Correlations of irradiances of various spectral lines measured by SDO/EVE MEGS-B and the F30 cm radio flux, with historical records from the literature and PEVE measurements from the solar minimum in 2008. Black scatter points indicate observations from 2010 until 2024 while the over-plotted grey points indicate observations from 2010 until 2014, after the failure of MEGS-A instrument.}
    \label{fig:MEGS_B_his_rec1}
\end{figure*}

\begin{figure*}
 \centerline{
        \includegraphics[angle=-90, width=6cm,      keepaspectratio]{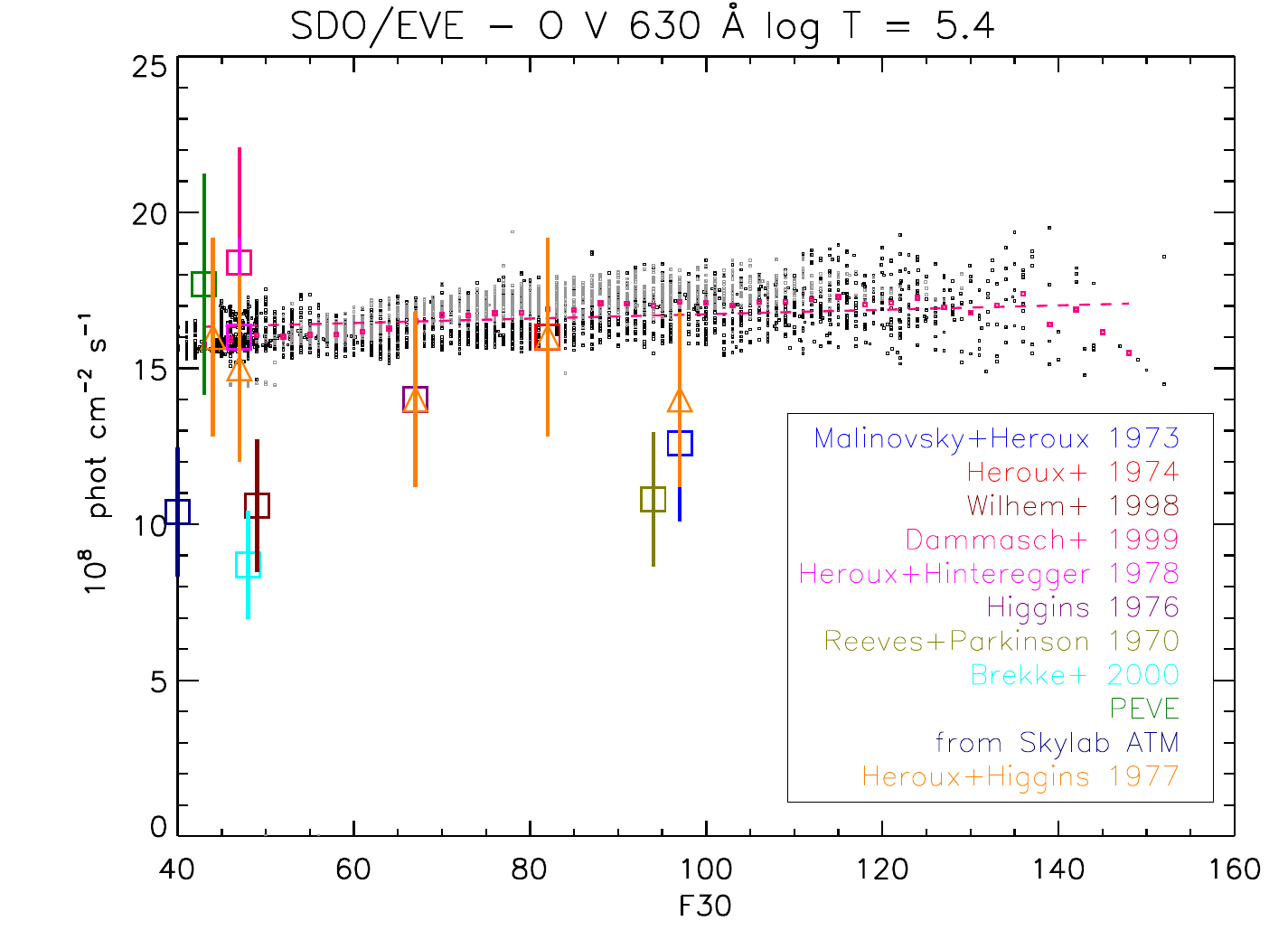}
        \includegraphics[angle=-90, width=6cm, keepaspectratio]{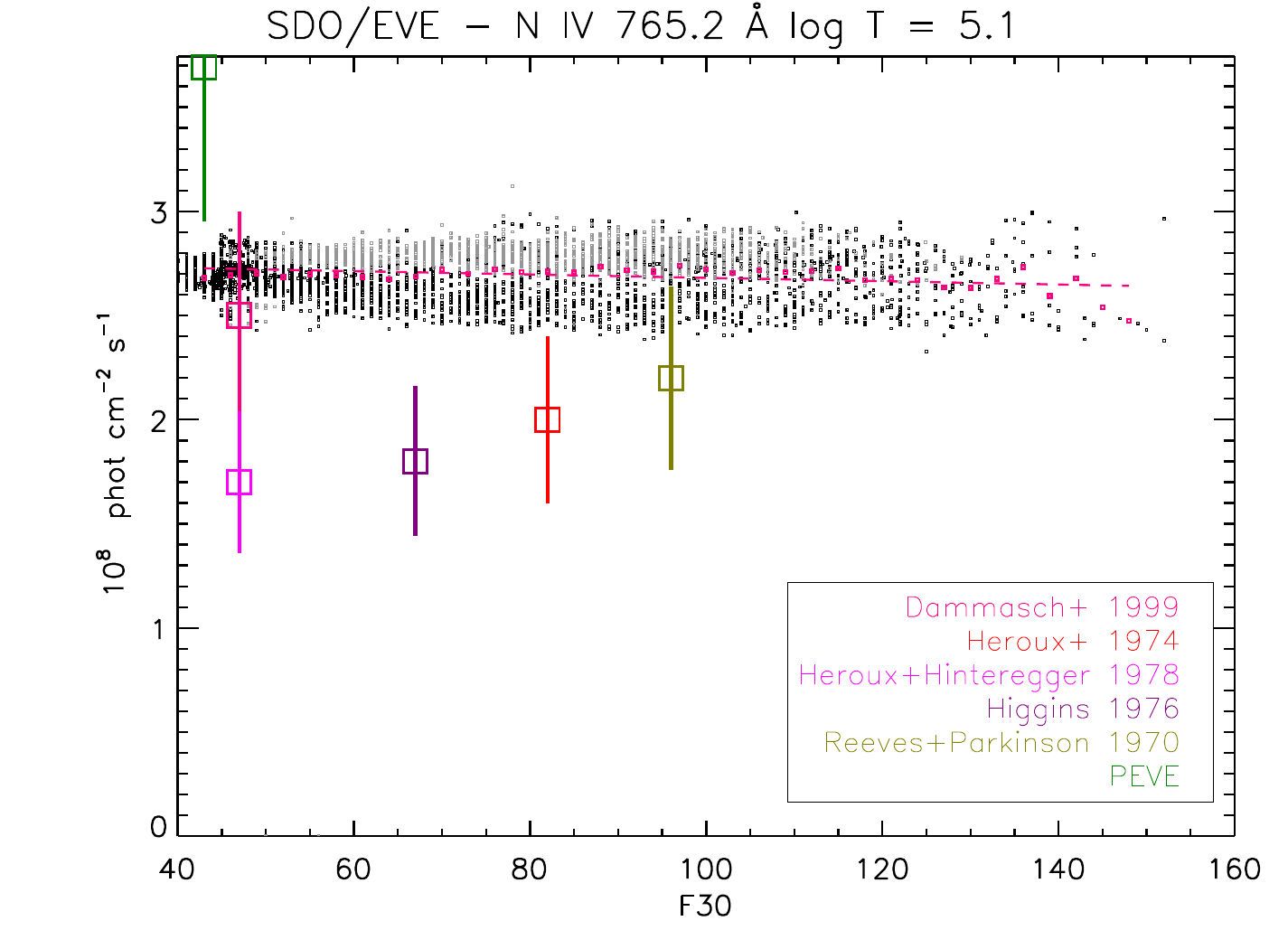}
        \includegraphics[angle=-90, width=6cm, keepaspectratio]{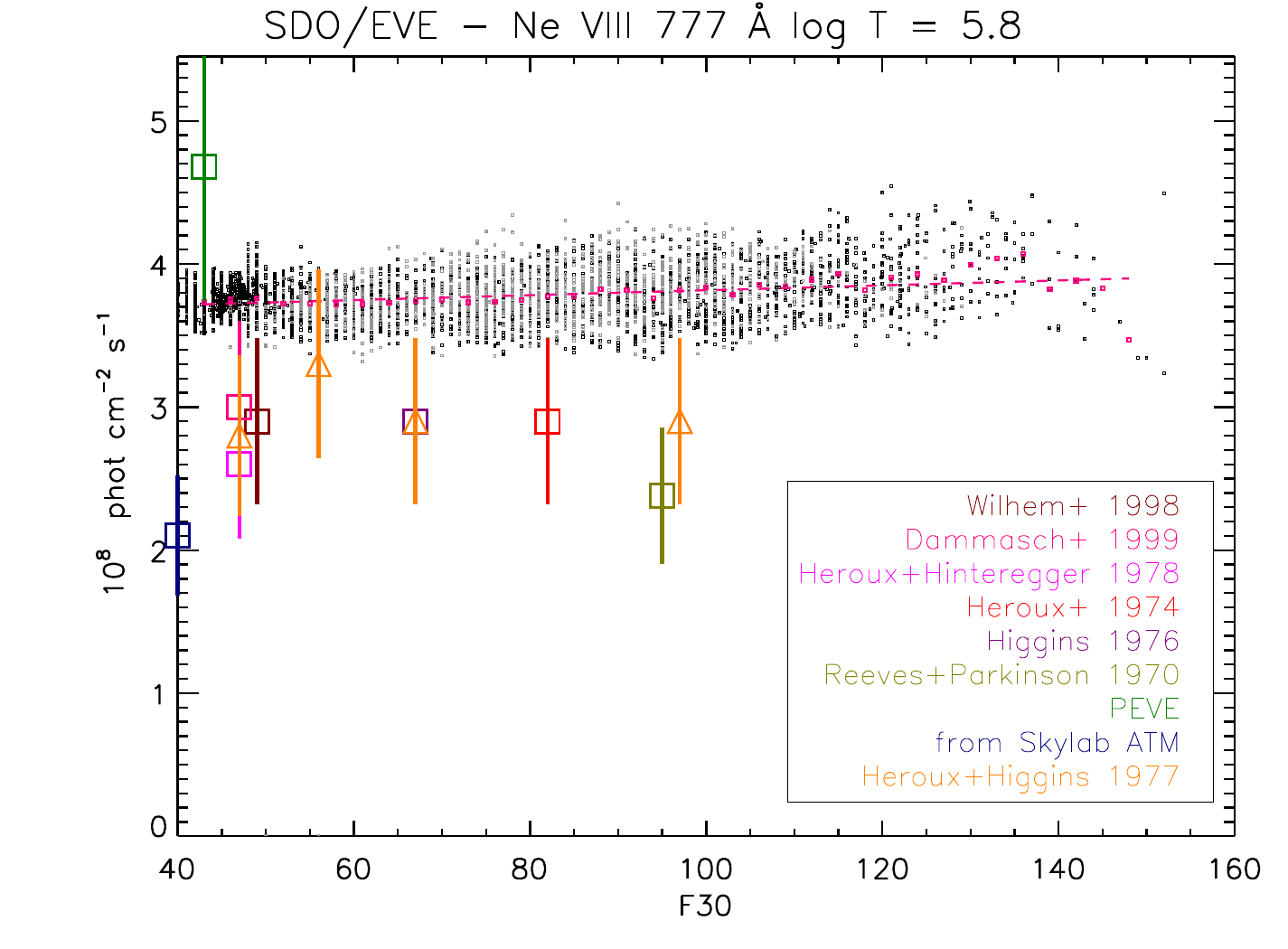}}
\centerline{
        \includegraphics[angle=-90, width=6cm, keepaspectratio]{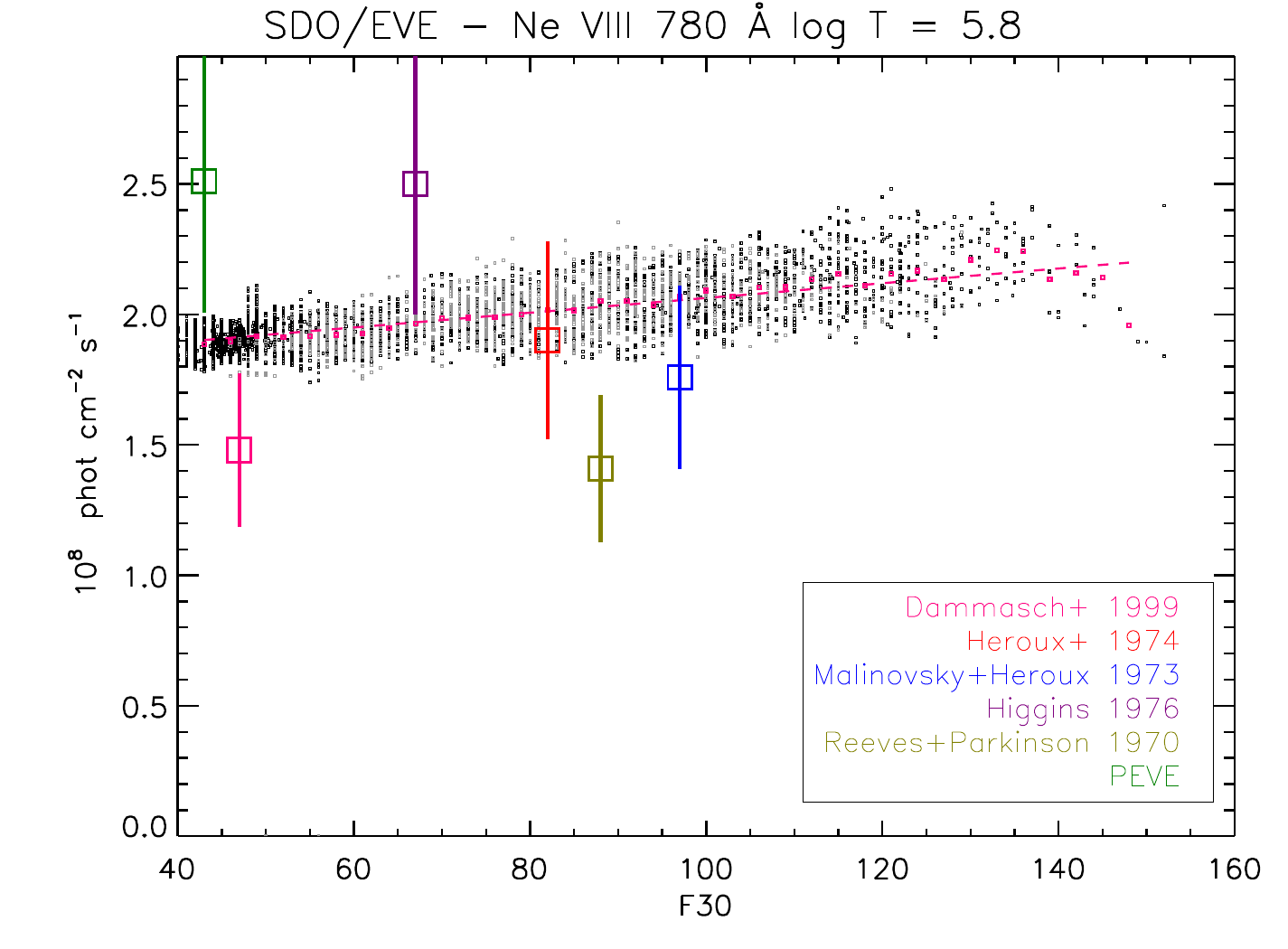}
        \includegraphics[angle=-90, width=6cm, keepaspectratio]{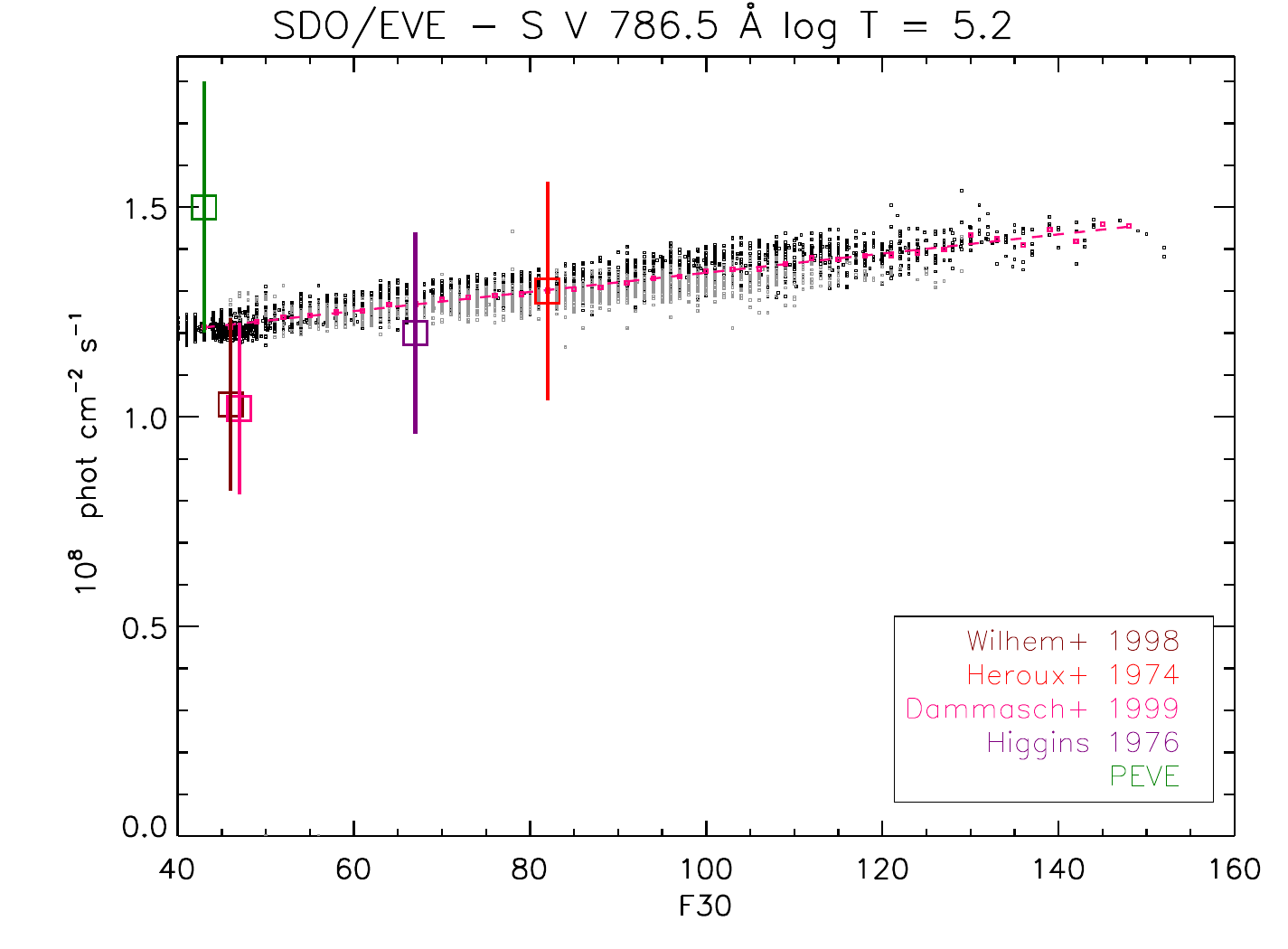}
        \includegraphics[angle=-90, width=6cm, keepaspectratio]{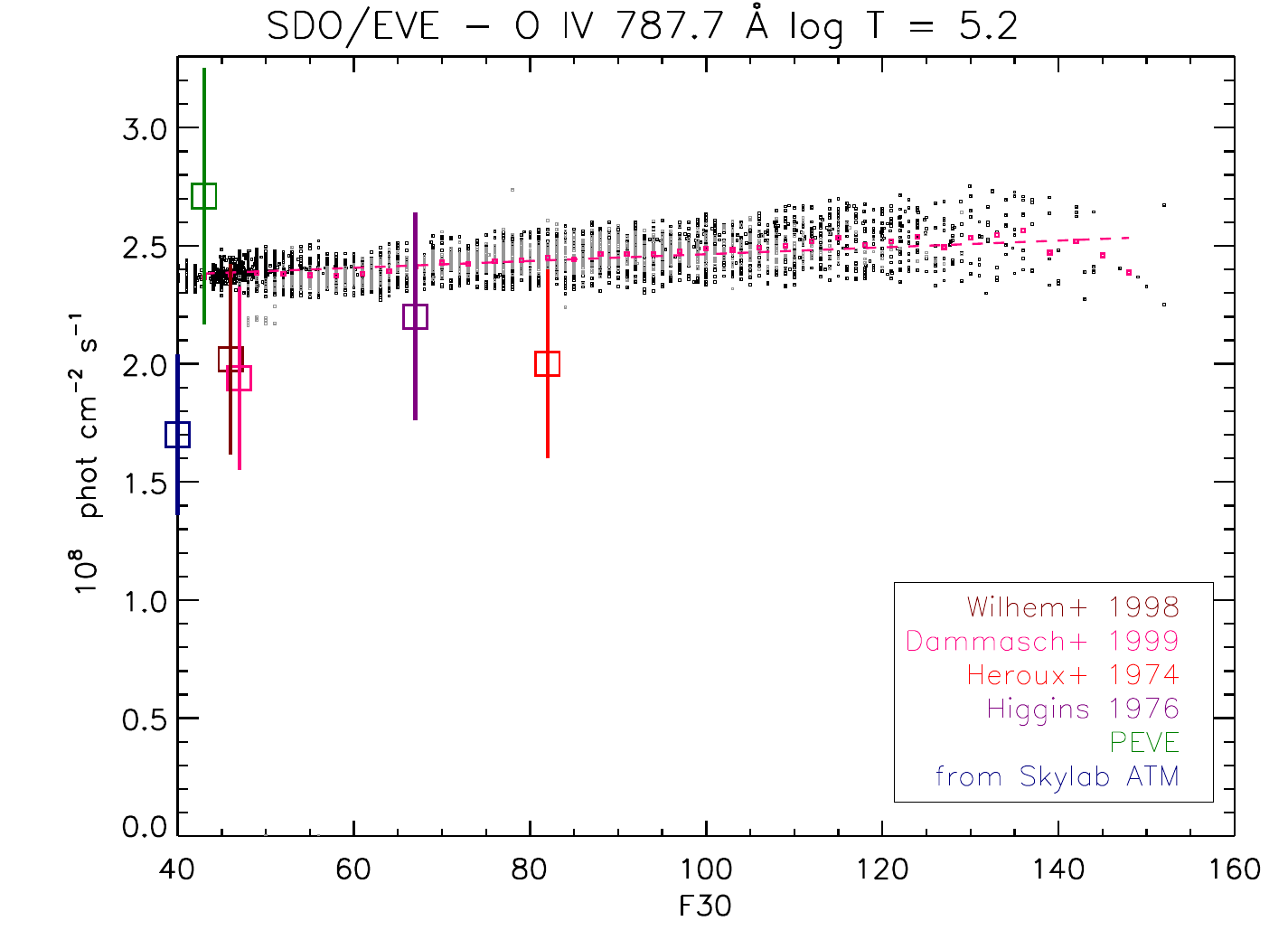}}
\centerline{
        \includegraphics[angle=-90, width=6cm, keepaspectratio]{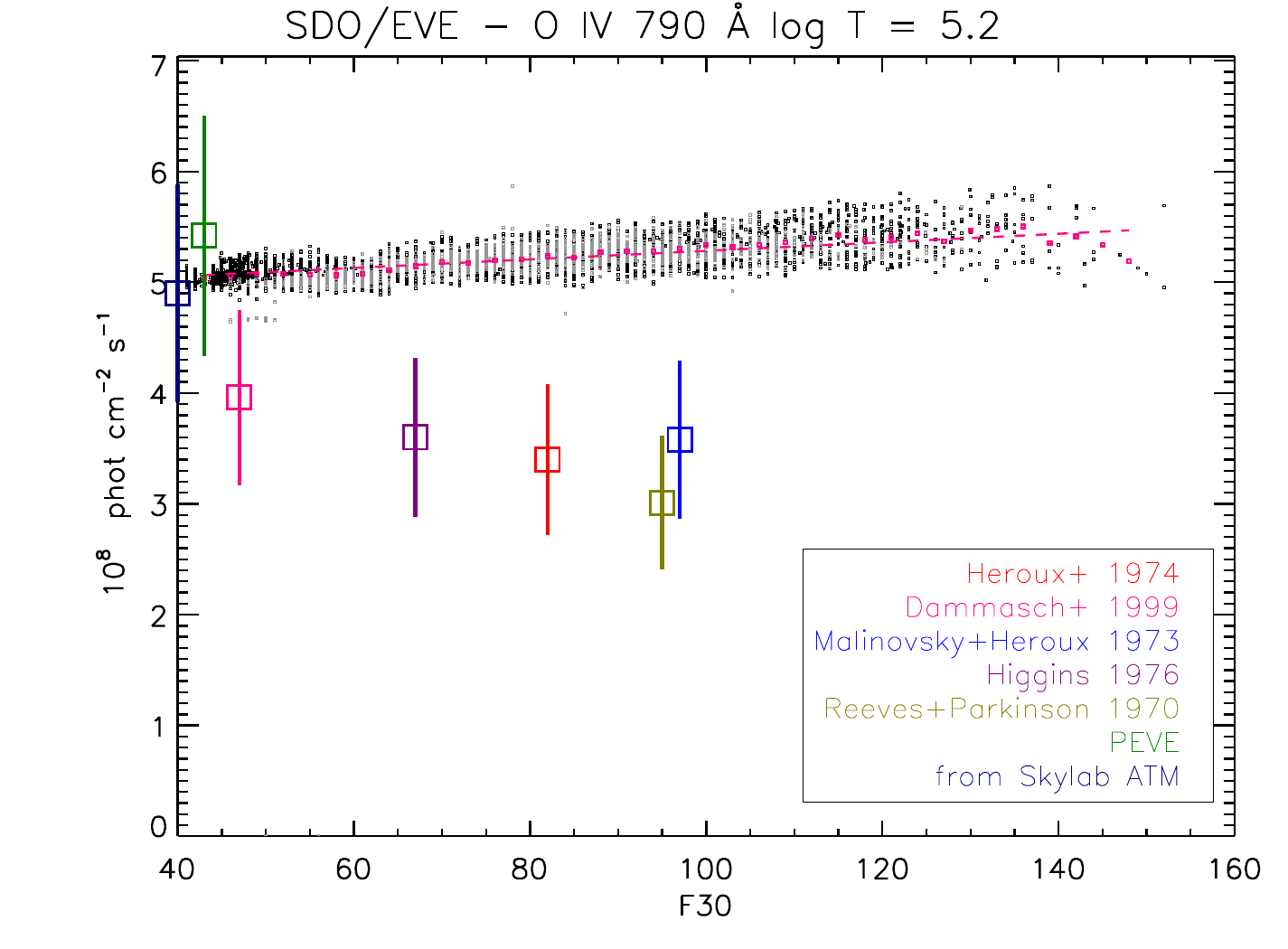}
        \includegraphics[angle=-90, width=6cm, keepaspectratio]{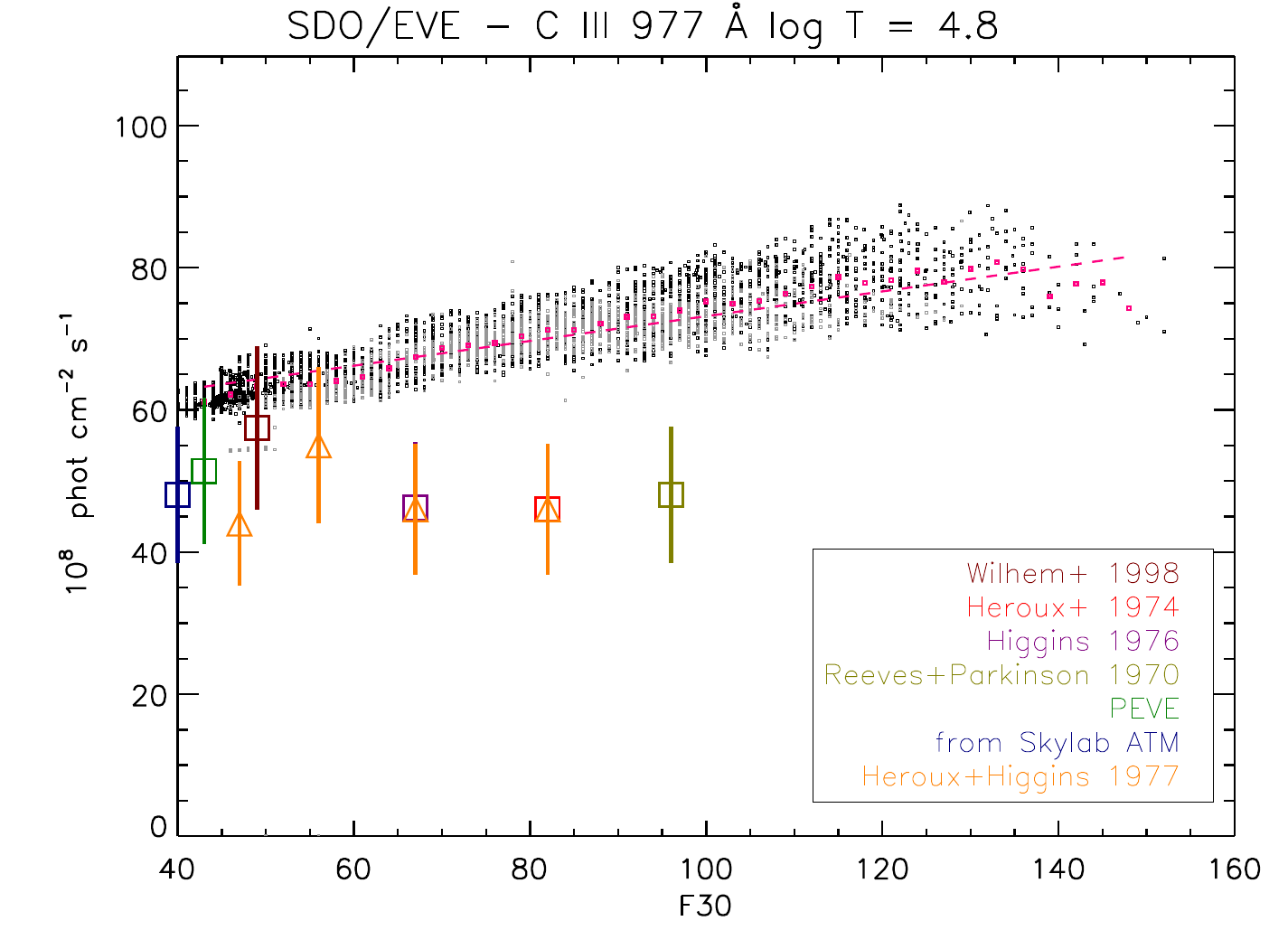}
        \includegraphics[angle=-90, width=6cm, keepaspectratio]{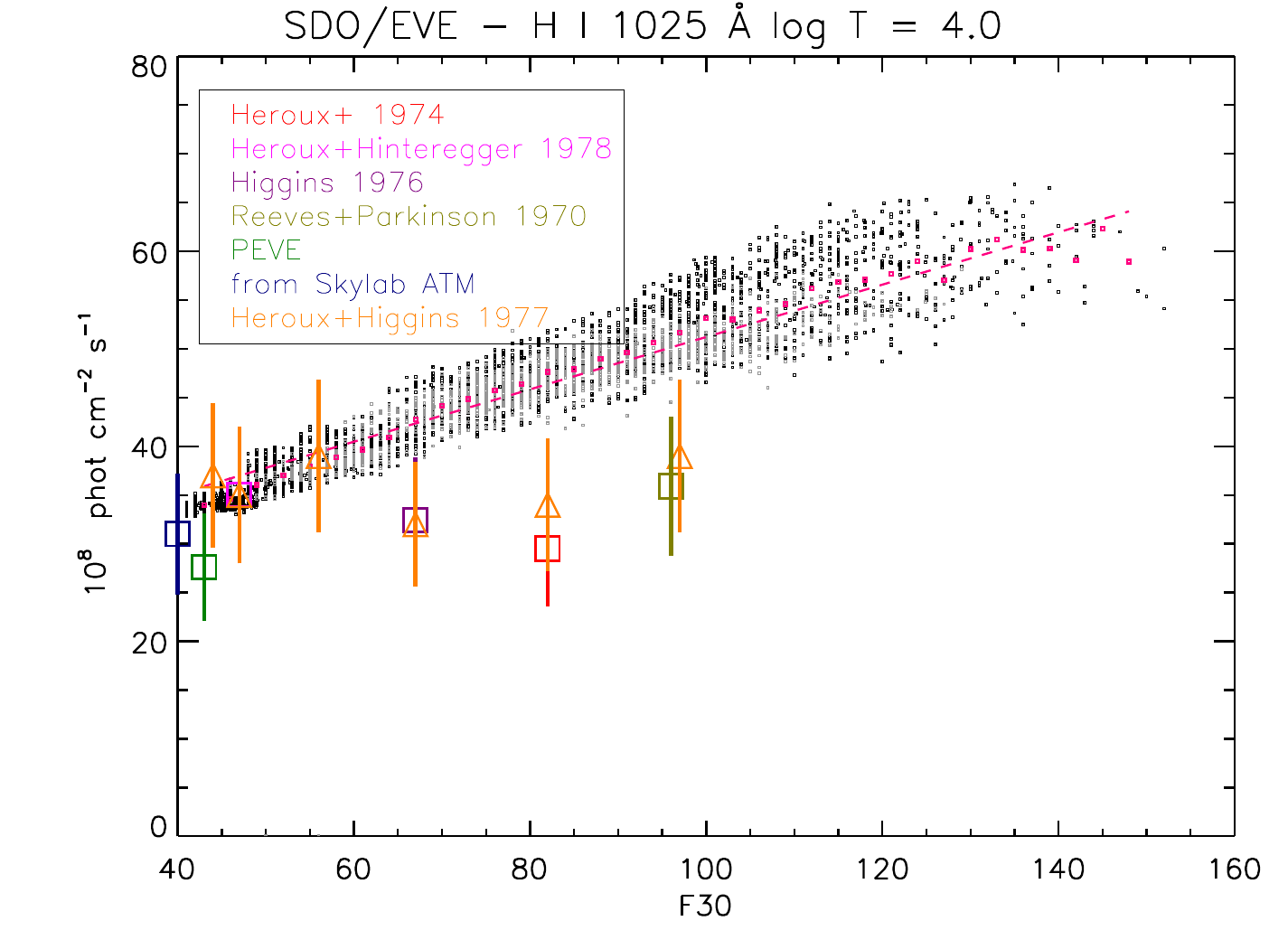}}
\centerline{
        \includegraphics[angle=-90, width=6cm, keepaspectratio]{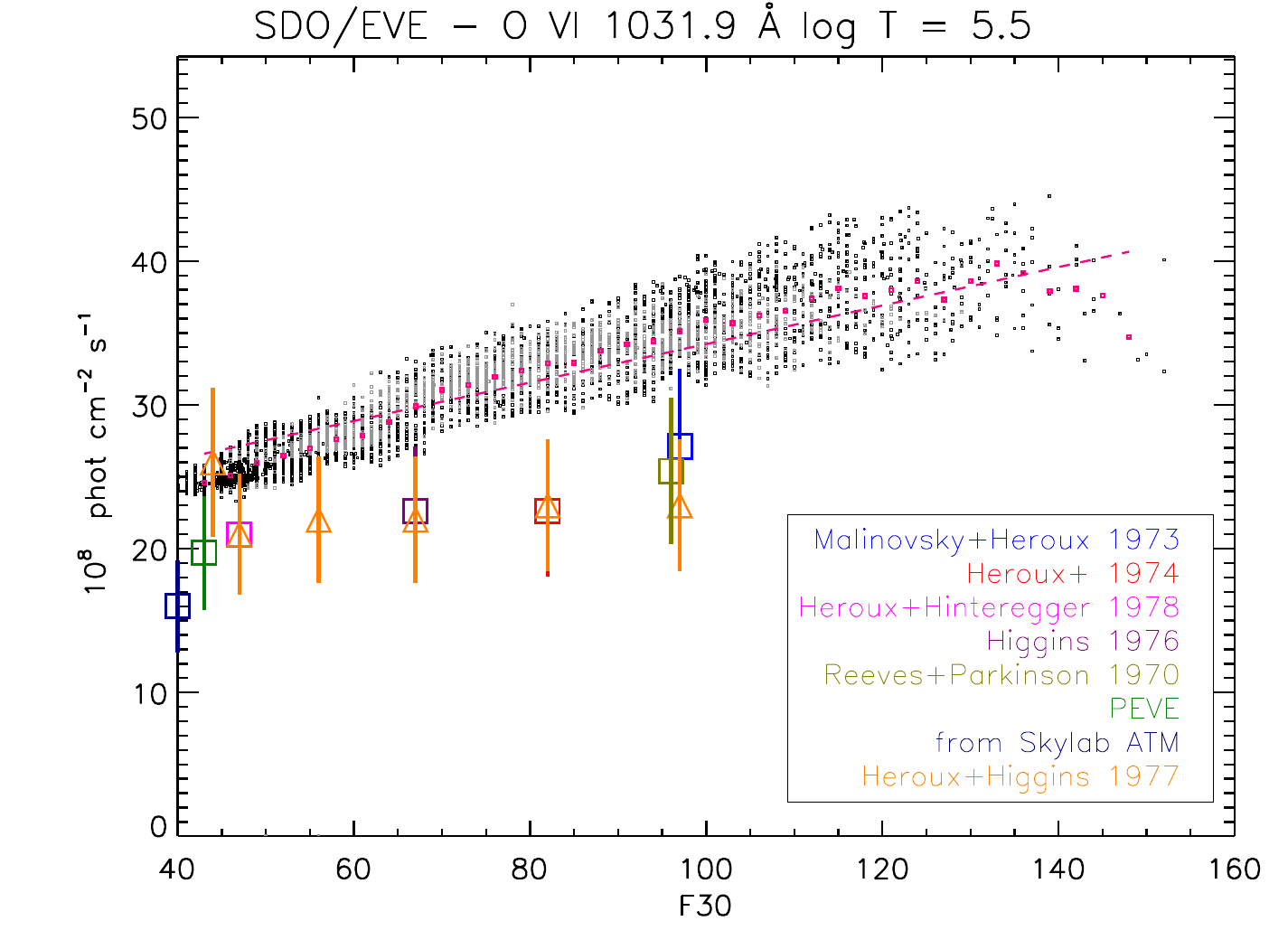}
        \includegraphics[angle=-90, width=6cm, keepaspectratio]{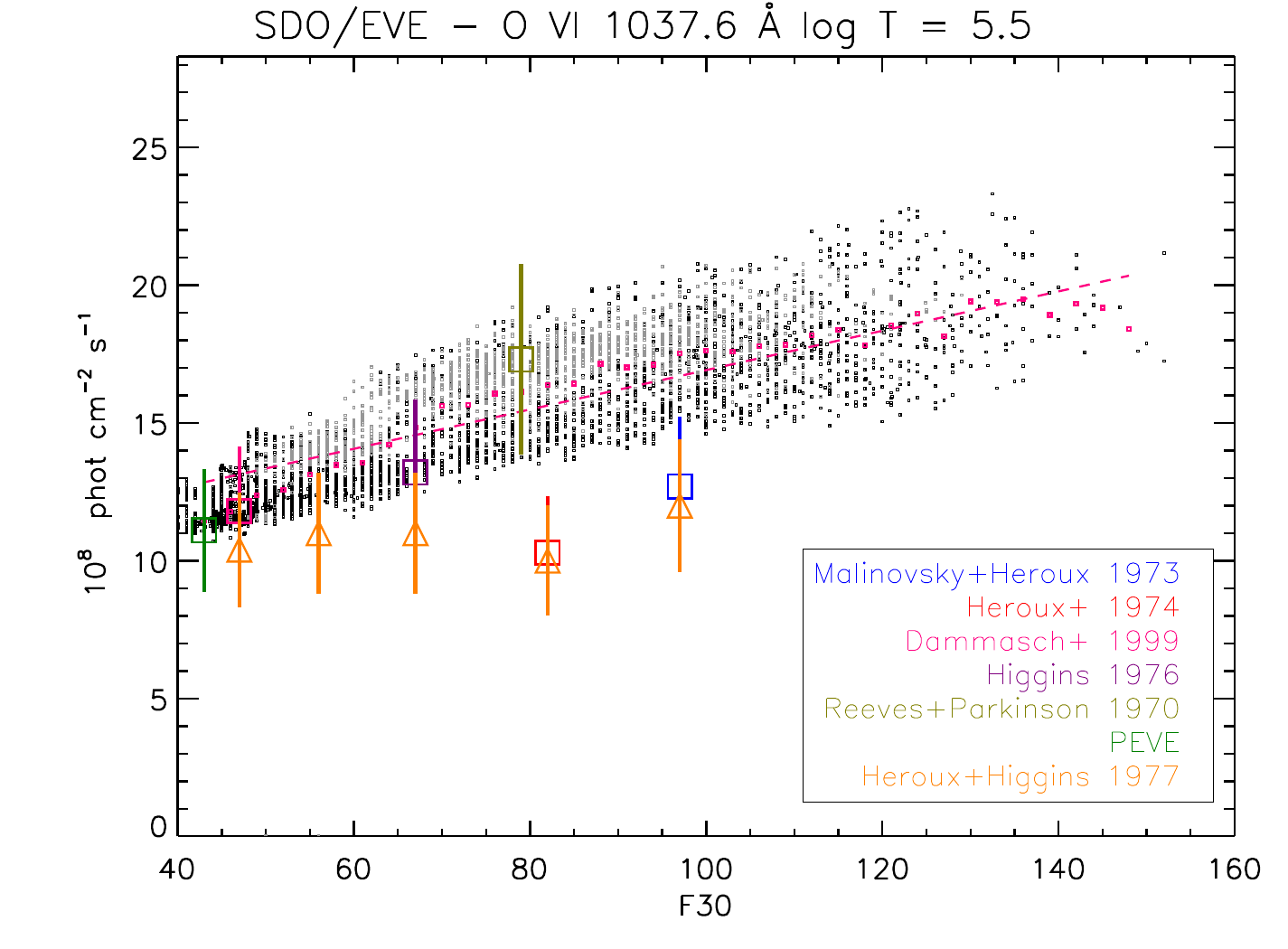}}
        \caption{Correlations of irradiances of various spectral lines measured by SDO/EVE MEGS-B and the F30 cm radio flux, with historical records from the literature and PEVE measurements from the solar minimum in 2008. Black scatter points indicate observations from 2010 until 2024 while the over-plotted grey points indicate observations from 2010 until 2014, after the failure of MEGS-A instrument. }
    \label{fig:MEGS_B_his_rec2}
\end{figure*}

This appendix presents plots of the correlations of irradiances with the F30 cm radio flux, in combination with historical records from the literature and PEVE measurements from the solar minimum in 2008. More specifically, in fig.~\ref{fig:MEGSA_his_rec} the plots are presented using data from SDO EVE MEGS-A, while fig.~\ref{fig:MEGS_B_his_rec1} and \ref{fig:MEGS_B_his_rec2} show plots with data from MEGS-B. The historical records from the literature are taken from the following papers: \cite{heroux_etal:1974}, \cite{malinovsky_analysis_1973}, \cite{heroux_hinteregger:1978}, \cite{heroux_higgins:1977}, \cite{higgins_solar_1976}, \cite{reeves_parkinson:1970a}, \cite{brekke_etal:00}, \cite{dammasch_etal:1999} and \cite{wilhelm_etal:98}. 
As noted previously, the PEVE irradiances for some strong lines are much higher than the flight EVE ones, depsite the fact that they were measured during the 
deep 2008 solar minimum, i.e. should generally be lower than any EVE values.

We have also plotted a $\pm$20\% error bar for all of the historical records (large error bars in all of the plots). 
The uncertainties in the historical records vary between different sources but
are typically 15--30\%. In some cases, even larger uncertainties are mentioned.
The errors bars are therefore shown to aid the eye and see if different  measurements agree to within $\pm$20\%.

In addition, in fig.~\ref{fig:MEGS_B_his_rec1} and \ref{fig:MEGS_B_his_rec2} we have plotted the MEGS-B data for the period of 2010-2024 as black points but we have over-plotted the period of 2010-2014 (when MEGS-A ceased to operate) as gray scatters, in order to be able to distinguish technical issues that might have been caused, such as the one discussed for the 465 \AA\, \ion{Ne}{vii} line, in fig.~\ref{fig:ne_vii_technical}. We observe a similar behaviour for the 600 \AA\, \ion{O}{iii}, the 554.5 \AA\, \ion{O}{iv} and the 765.2 \AA\, \ion{N}{iv} lines.

We find relatively good agreement between the EVE data and the \cite{malinovsky_analysis_1973} and \cite{heroux_etal:1974} records. We also found that for any line of $\lambda >$ 630 \AA\, the irradiances from EVE tend to be higher compared to the historical records or irradiances and radiances, as seen in fig.~\ref{fig:MEGS_B_his_rec2}. 

\cite{dammasch_etal:1999} has summed the 1036~\AA\ \ion{C}{ii} with the 1037~\AA\ \ion{O}{vi} while on EVE these lines are resolved separately.

\bsp	
\label{lastpage}
\end{document}